# 3D Topological Kondo Insulators

Soroush Arabi

Masterarbeit in Physik
angefertigt im Physikalischen Institut

vorgelegt der
Mathematisch-Naturwissenschaftlichen Fakultät
der
Rheinischen Friedrich-Wilhelms-Universität
Bonn

July 2017

I hereby declare that this thesis was formulated by myself and that no sources or tools other than those cited were used.

Bonn, . . . . . . . . . . . . . . . . . . . . . . . . . . . . . . . . . . . . . . . . . . . . . . . . . . .

Date Signature

1. Gutachter: Prof. Dr. Johann Kroha
2. Gutachter: Prof. Dr. Simon Trebst

# Contents

# CHAPTER 1

# Introduction

It had been long acknowleged in physics that two of the most fundamental classes of matter, conductors and insulators, can be distinguished and understood by the capability of electric conduction. Only after recent discoveries, it has been made clear that such distinction is ill-defined, and there is a new phase of matter which incorporates both of those classes. This new phase of matter shows different property at the bulk and surface; a band insulator in the bulk and a conductor in the surface [1]. According to the band theory, an ordinary or trivial insulator is described by a set of completely filled electronic bands (the valence bands) that are separated from a completely empty set of bands (the conduction bands) by an energy gap, while the Fermi energy (chemical potential) lies inside the gap. Such insulator is electronically immobile because it takes a finite energy to displace an electron. The recent theoretical and experimental researches, However, show that there is a new class of insulators, so-called topological insulators, where the energy gap between the occupied and empty bands is strongly renormalized by spin-orbit interactions [2]. Moreover, due to the coupling of the spin and orbital angular momentum of an electron, a band gap inversion happens. This means that the states located below the gap exchange with those above, and in general make a twist like electronic band structure. Such a twist is protected by a topological invariant ($\mathbb{Z}_2 = \pm 1$) of the bulk, which implies that it is not possible to remove it unless the fundamental symmetry of the bulk (e.g., time-reversal symmetry) is lifted. One exotic property of such protected twist like band structure, that does not exist in trivial band insulators, is the existence of metallic states at the surfaces of topological insulators. Due to the relativistic origin of the spin-orbit coupling, these topological surface states show a linear dispersion relation, discribed by Dirac equation, and a surface spin-texture due to a locking of spin and momentum ($\sim \boldsymbol{\sigma} \cdot \mathbf{p}$) [2].

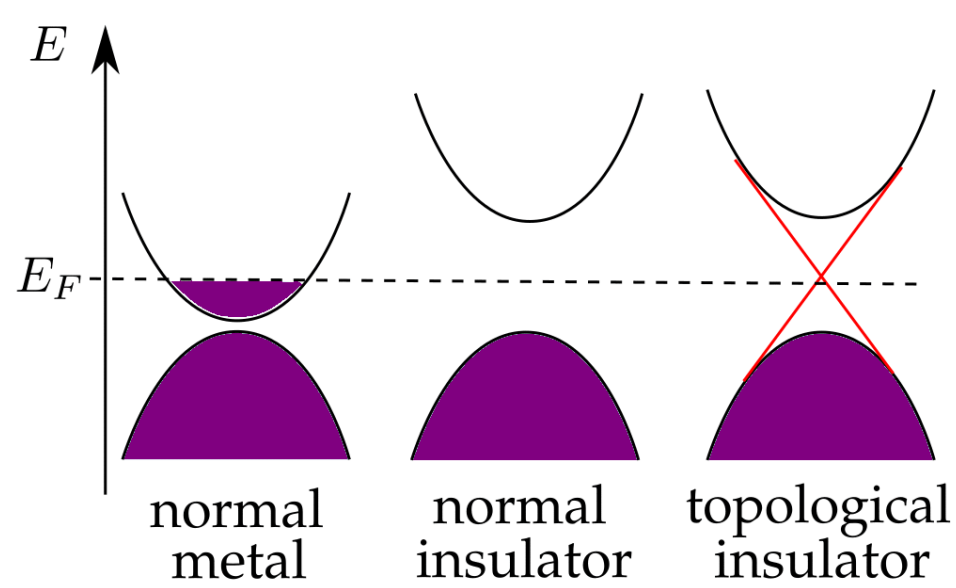

Figure 1.1: Band structure of a normal metal, normal insulator and a topological insulator with gapless metalic surface states.

The first example of a topological insulator was realized in the integer quantum Hall effect (IQHE) which shows a quantized Hall conductivity ($\sigma_{xy} = \nu\frac{e^2}{h}$) in an external perpendicular magnetic field ($B$) [3]. Subsequently, by proposing a model of spinless fermions with broken time-reversal symmetry on a two-dimensional honeycomb lattice, Haldane realized that it is possible to have the quantum Hall effect with a net zero external magnetic field [4]. In 2005 Kane and Mele proposed a new topological insulator state in two dimensional systems, the so called Quantum Spin Hall (QSH) system [5]. This proposal was based on calculations of the single-layer graphene bandstructure with intrinsic spin-orbit interactions. The quantum spin Hall state was proposed to consist of two counterpropagating edge states with opposite spin polarization and in contrast to the quantum Hall state no magnetic field is needed to create these edge states.

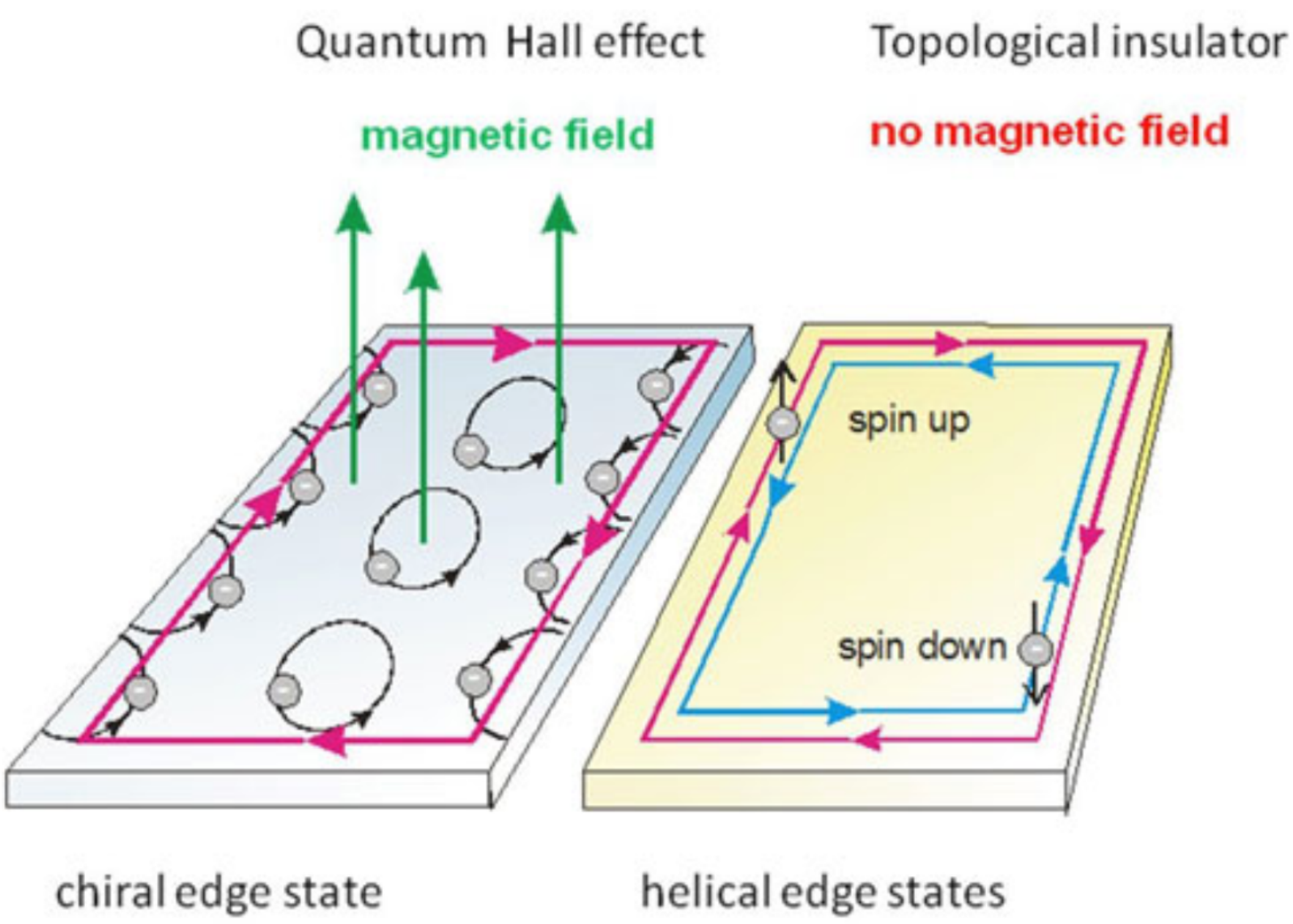

Figure 1.2: Schematic figure of the quantum spin-hall effect (right panel), which can be heuristically understood as two opposing copies of the quantum Hall effect (left panel), without an external magnetic field and with a strong spin-orbit coupling [6].

As it was mentioned earlier, symmetry-protected topological insulators (e.g, the quantum spin Hall insulator) can be explained as states of matter which has non-trivial insulating bulk and topologically protected surface states at their boundaries. As an example, the surface states of a 3D topological insulators are gapless and protected against perturbations of the Hamiltonian as long as the fundamental symmetry of the Hamiltonian, time-reversal invariance, is preserved and the bulk band gap is not closed. To characterize a 3D topological insulator, four $\mathbb{Z}_2$ invariants ($\nu_0; \nu_1, \nu_2, \nu_3$), where $\nu_0$ is the most important one, are needed. Using this classification, there are two classes of 3D topological insulators; weak and strong. The weak topological insulators made of stacked layers of 2D QSH insulators, with $\nu_0 = 0$, corresponding to unprotected helical edge states and ($\nu_1, \nu_2, \nu_3$) describe the direction of the stacking layers similar to Miller indices. The strong topological insulator cannot be interpreted in terms of the 2D QSH insulator, has $\nu_0 = 1$, where $\nu_0$ in comparision with weak topological insulator is a true topological invariant and is protected only with time-reversal symmetry.
As it was shown by Liang Fu and Charles Kane [7], in the presence of inversion symmetry, the strong topological invariant can be caluculated at eight time-reversal invariant momenta (TRIM) in the Brillouin zone, $\Gamma_i \in \{(0,0,0), ..., (\pi,\pi,\pi)\}$ as

$$(-1)^{\nu_0} = \prod_{i=1}^{8} \prod_{\gamma} \xi_{i,\gamma} \tag{1.1}$$

where the first product is over TRIM points and the second product is over all occupied Kramer's pairs and $\xi_{i,\gamma} = \pm 1$ is the well-defined parity of the Kramer's pairs $\gamma$ at $\mathbf{k} = \Gamma_i$. Similarly the weak indices $\nu_i$ are defined by product of parities on the plane $k_i = \pi$ as

$$(-1)^{\nu_i} = \prod_{i=1}^{4} \prod_{\gamma} \xi_{i,\gamma} \tag{1.2}$$

where the first product is over the $k_i = \pi$ planes and the second product is over occupied Kramers's pairs. According to the bulk–boundary correspondence, the $\mathbb{Z}_2$ invariants $(\nu_0; \nu_1, \nu_2, \nu_3)$ imply gapless boundary states on all or only certain high symmetry surfaces [8]. In a strong topological insulator, where $\nu_0 = 1$, an odd number of spin-momentum locked helical surface states are in the surface Brillouin zone.

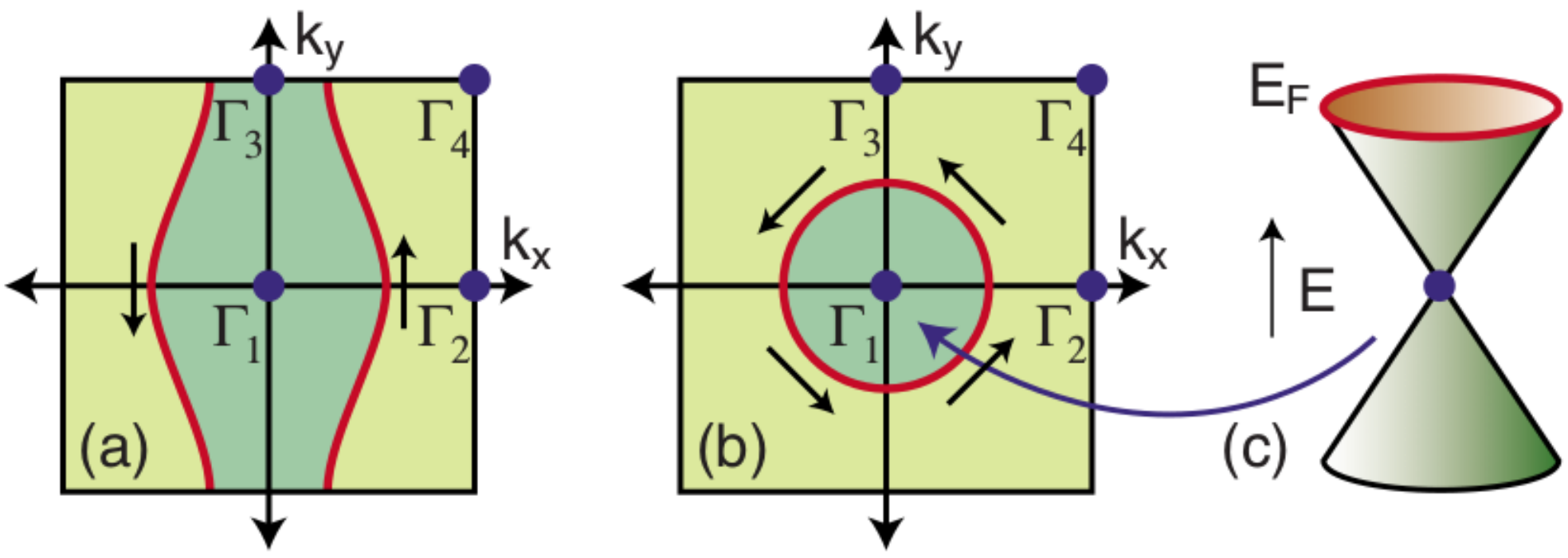

Figure 1.3: (a) A weak topological insulator with band inversions in two TRIM points $\Gamma_1$ and $\Gamma_3$ , (b) a strong topological insulators and (c) dispersion of strong topological insulator's surface states with helical spin-momentum locked polarization, encloses a single TRIM point ($\Gamma_1$) [2].

In 2010, Maxim Dzero et al., proposed a new class of topological insulators, namely **toplogical Kondo insulators**, which encorporate the interplay between strong correlations and spin-orbit coupling [9]. These topological Kondo insulators establish a platform to study both correlation and topology in the same material, and have attracted much attention from both theoretical and experimental physicists in recent years. In particular, $SmB_6$ which was discovered almost half a century ago as the first Kondo insulator, is introduced in this theory as a candidate for a topological Kondo insulator [10]. It is argued that, since Kondo insulators have inversion symmetry, the $\mathbb{Z}_2$ invariants can be calculated from the parities of occupied states at the high-symmetry points of the Brillouin zone. Due to the fact that conduction electrons (mainly $d$-orbital) and localized electrons ( mainly $f$-orbtial ) have opposite parities, a band inversion at one of the high-symmetry points should lead to a strong topological insulators. The fundamental topological properties can be studied using the infinite-$U$ Anderson lattice model with a spin-dependent hybridization form factor between nearest neighbors instead of the usual onsite hybridization that was used before [11]. This model is the main topic of this thesis.

## 1.1 Structure of the thesis

The main goal of this thesis is to realize 3D topological Kondo insulators model via a spin-orbit coupled infinite-$U$ Anderson lattice model in slave-boson representation. To achieve this aim, the thesis is structured as follows,

- In chapter 2, we start by a minimal review in some preliminiary concepts of strongly correlated condensed matter physics to be able to introduce the common infinite-$U$ Anderson model in slave-boson representation.
- In chapter 3, we discuss about the role and importance of spin-orbit coupling and crystal field effect in heavy materials. Moreover, we show how to include these effects in the Anderson model and its infinite-$U$ variant. This results in a $k$-dependent, nonlocal, odd parity, time-reversal invariant hybridization form factor, which has a central role in the presented topological Kondo insulators model Hamiltonian.
- In chapter 4, we develope a self-consistent theory to study the 3D bulk of the topological Kondo insulators at mean-field level. Namely, we perform a homogeneous slave-boson mean-field theory for a 3D bulk which has periodic boundary condition in $x, y$ and $z$ axis.
- In chapter 5, we develope a self-consistent theory to study the topological Kondo insulators in a slab geometry. By breaking periodic boundary condition in $z$-axis, we perform an inhomogeneous slave-boson mean-field theory for a multilayer system, where each layer is coupled to its nearest-neighbors by hoping and nonlocal hybridization.

CHAPTER 2

# Preliminaries

The aim of this chapter is to introduce the infinite-$U$ Anderson model [12–15], as a proposed model to describe Kondo insulator. To do so, first one need to realize the Kondo effect, which captures the concept of antiferromagnetic exchange coupling of the localized moment and conduction electrons. This results in the formation of a singlet state below the Kondo temperature and logarithmic divergence of the resistivity. We later generalize the Kondo model to the Kondo lattice model, by considering an array of impurities immersed in the conduction sea, and realizing its simplest ground state at the strong coupling limit as a Kondo insulator. In this scheme, we also introduce the Anderson model, which includes not only the spin-fluctuation but also charge-fluctuation. Thus, this model can be assumed as a more general model to study the Kondo systems. Finally, by presenting slave-boson representation, we show that how to rule out charge fluctuations from the Anderson model and modify it when there is an infinite Coulomb repulsion in the localized doubly-occupied state.

## 2.1 Kondo effect

The Kondo effect is a quantum many-body phenomenon associated with the interaction between a localized spin and free electrons. It was first discovered in metals containing small amounts of magnetic impurities. Experimentally it is observed that such impurities give anomalous contributions to many metallic properties, such as transport and thermodynamic properties. In contrast with normal metallic systems, one observes a peculiar minimum and a subsequent logarithmic divergence in the electrical resisitivity at very low temperatures (see figure (2.1)) [16]. This effect was first explained by Jun Kondo in 1964 [17], based on the so-called Kondo model, which describes a local magnetic moment spin $S$ coupled by an antiferromagnetic exchange interaction $J$ to the conduction electron spins, forming a transient singlet state, which results to the localizition of the conduction electrons around the impurities and divergence of resistivity. The seminal Hamiltoniain describing the simple physical picture and explaning the experimental results is [17]

$$\mathcal{H} = \mathcal{H}_c + \mathcal{H}_K \tag{2.1}$$

where

$$\mathcal{H}_c = \sum_{\mathbf{k}\sigma} \epsilon^c_{\mathbf{k}} c^\dagger_{\mathbf{k}\sigma} c_{\mathbf{k}\sigma}. \tag{2.2}$$

The operator $c^{\dagger}_{\mathbf{k}\sigma}$ and $c_{\mathbf{k}\sigma}$ create and annihilate conduction electrons with wave vector $\mathbf{k}$ and spin polarization $\sigma \in \{1, -1\}$. The second term of the Hamiltonian is

$$\mathcal{H}_K = J\mathbf{S}_{imp}(0) \cdot \mathbf{S_c}, \tag{2.3}$$

which is an antiferromagnetic Heisenberg exchange interaction between an impurity spin, located at the origin, with the conduction elecron. Kondo showed, using second order perturbation theory in the coupling $J$, that interaction leads to a singular scattering of the conduction electrons near the Fermi level and temperature logarithmic contribution to the resistivity [17]

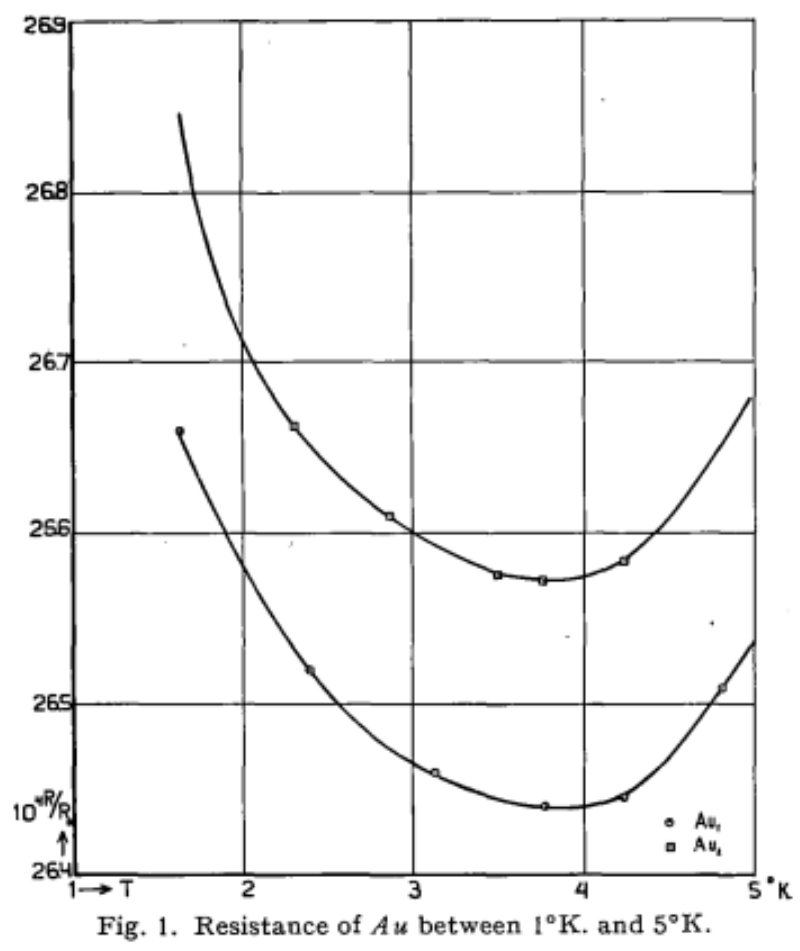

Figure 2.1: Logarithmic divergence of resistivity of iron (Fe) impurities in gold (Au) at very low temperatures due to the Kondo effect [18].

$$R = R_0 \left[ 1 - 4J\rho \log\left( \frac{k_B T}{D_0} \right) \right], \tag{2.4}$$

where $D_0$ is the band width and $\rho$ is the density of states of conduction electrons at the Fermi level.
This theory provides a good description of the phenomenon above a certain temperature ($T > T_K$), the so-called Kondo temperature, where the perturbation theory is valid, and breaks down below the Kondo temperature. This motivated many studies later for a more comprehensive theory of the Kondo problem, like the perturbative renormalization group method, commonly known as Poor Man's Scaling, which involves perturbatively eliminating excitations close to the edges of the non-interacting band. This method demonstrated that, by decreasing temperature, the effective coupling between the spins, $J$, increases unlimitedly. As this method is perturbative in $J$, it becomes invalid when J becomes large, so this method did not truly solve the Kondo problem, although it did hint at the way forward [19]. The Kondo problem was finally solved when Kenneth Wilson applied the numerical renormalization group to the Kondo model and showed that the resistivity saturates at a constant value as temperature goes to zero [20].
Generaliztion of the Kondo model to the Kondo Lattice model needs a set of Kondo impurities or a periodic lattice of magnetic impurities embedded in a sea of conduction electrons, which is described effectively in real space by the Hamiltonian [21]

$$\mathcal{H}_{KLM} = t^c \sum_{\langle i,j \rangle \sigma} (c^{\dagger}_{i\sigma} c_{j\sigma} + H.c.) + J \sum_i \mathbf{S}_{imp}(\mathbf{x_i}) \cdot \mathbf{S}_c(\mathbf{x_i}) \tag{2.5}$$

where the $t^c$ is the nearest neighbours hopping amplitude of the conduction electrons.
In the strong coupling limit, Kondo effect renormalizes the exchange coupling constant $J$ to high values,

$$\frac{t^c}{J} \to 0 \qquad \text{(strong coupling limit)}, \tag{2.6}$$

so that the conduction electron hopping term can be assumed as a perturbation to the on-site exchange interaction, resulting in the formation of singlet states of localized moments with conduction electrons at each lattice site, while the charge excitations are completely gapped.

$$\mathcal{H}_{KLM} = J \sum_i \mathbf{S}_{imp}(\mathbf{x_i}) \cdot \mathbf{S}_c(\mathbf{x_i}) + O(t) \tag{2.7}$$

In this limit (2.6), one can realize the Kondo lattice as a Kondo insulator [10].

## 2.2 Anderson impurity and lattice model

We continue the discussion with the single impurity Anderson model, and by introducing a periodic array of impurities, one can extend it to the Anderson lattice model (ALM). The single impurity Anderson model [22] is an effective model to capture the role of charge and spin fluctuations of a localized impurity in metallic environment. This impurity could be for example the $d$- or $f$-level of a transition metal atoms embedded in a nonmagnetic metal. The Hamiltonian of the SIAM consists of three parts

$$\mathcal{H} = \mathcal{H}_{imp} + \mathcal{H}_c + \mathcal{H}_{hyb} \tag{2.8}$$

where $\mathcal{H}_{imp}$ describes the interacting impurity electron level and is given by

$$\mathcal{H}_{imp} = \sum_\sigma \epsilon^d d^\dagger_\sigma d_\sigma + U n^d_\uparrow n^d_\downarrow. \tag{2.9}$$

The operators $d^\dagger_\sigma$ and $d_\sigma$ create and annihilate impurity electrons with spin component $\sigma \in \{1, -1\}$, $\epsilon^d$ is the impurity heavy flat-band and $n^d_\sigma = d^\dagger_\sigma d_\sigma$ is the particle number operator. Occupying the impurity level with two electrons costs a repulsive interaction energy $U > 0$ caused by the Coulomb Interaction.

$$U = \int d\mathbf{r} d\mathbf{r}' \phi^*_d(\mathbf{r})\phi^*_d(\mathbf{r}') \frac{e^2}{|\mathbf{r} - \mathbf{r}'|} \phi_d(\mathbf{r})\phi_d(\mathbf{r}') \tag{2.10}$$

where $\phi_d(\mathbf{r})$ is the impurity wavefunction. In atomic d shells this Coulomb interation can be very large, of the order of 30 eV [16].
The conduction electrons are modeled by a non-interacting electron gas as in equation (2.2). The coupling of the impurity and the conduction electron levels due to hybridization is described by the Hamiltonian

$$\mathcal{H}_{hyb} = \sum_{\mathbf{k}\sigma} \left( V_\mathbf{k} c^\dagger_{\mathbf{k}\sigma} d_\sigma + H.c. \right) \tag{2.11}$$

where $V_\mathbf{k} = \langle R = 0, \sigma | V | \mathbf{k}, \sigma \rangle$ is the on-site hybridization potential.
As an initial step to understand the concept of local moment and the interplay between charge and spin fluctuations through this model, one can first turn off the hybridization which mixes the conduction electrons with the impurity and realize different possible energy configurations of the impurity state: (i) empty state with total energy $E_0 = 0$; (ii) singly-occupied state with spin polarization $\sigma$ and total energy of $E^1_\sigma = \epsilon^d$; (iii) doubly-occupied state with antiparallel spin configuration and total energy $E^2 = 2\epsilon^d + U$ [16].
In the atomic limit, where the d- or f-electrons are localized deep inside the atomic shells with negative energies, they contribute to the valence band of the solid with $\epsilon^d < \epsilon_F$. In this limit, the singly-occupied state is the magnetic ground-state of the model with a two-fold degenracy of $S = \frac{1}{2}$ state. The second and third configurations, empty and doubly-occupied state; are non-magentic exited state of the impurity [16].
The generalization of the Anderson impurity model to a periodic lattice of Kondo impurities or localized moments, has attracted the attention of both theorists and experimentalists of heavy-fermion materials in recent years. The effective Hamiltonian of the Anderson lattice model is [13]

$$\mathcal{H} = \sum_{\mathbf{k}\sigma} \epsilon^c_{\mathbf{k}} c^\dagger_{\mathbf{k}\sigma} c_{\mathbf{k}\sigma} + \sum_{i\sigma} \left( \epsilon^d d^\dagger_{i\sigma} d_{i\sigma} + U n^d_{i\uparrow} n^d_{i\downarrow} \right) + \sum_{\mathbf{k}i\sigma} \left( V_{\mathbf{k}} e^{i\mathbf{k}\cdot\mathbf{R}_i} c^\dagger_{\mathbf{k}\sigma} d_{i\sigma} + H.c. \right) \tag{2.12}$$

where the index $i$ is the impurity lattice index.

## 2.3 Slave-boson representation and infinite-$U$ Anderson model

It has been known that a full study of the properties of the Anderson model (2.12) is a very complicated and demanding problem [23]. The obstacle in solving this model resides in the fact that the usual many-body perturbation theory does not work in this case. This breakdown is due to exceeding of the on-site Coulomb repulsion $U$ from the conduction electron band width $D_0$. Consider the Anderson model with large on-site repulsion $U$, explained in the last section, where each impurity state can either be empty, singly occupied, or doubly occupied. The electron behaves completely different on each of these states. For large $U$, the doubly occupied states will have very high energy and only contribute through virtual processes at low temperatures. Therefore, one can project the original Hilbert space onto a smaller subspace, where the double occupancy is excluded. It has been acknowleged that such a projection is a challenging task within the conventional quantum many-body theory.
To do the projection onto the Hilbert subspace one can use the well-stablished slave-particle formalism [14]. In this formalism, the electron operator is expressed in terms of psuedo-fermions and slave-bosons. For instance,

$$\begin{cases} a^\dagger |vac\rangle = |2\rangle \\ b^\dagger |vac\rangle = |0\rangle \\ f^\dagger_\sigma |vac\rangle = |\sigma\rangle \end{cases} \tag{2.13}$$

where $a$ is heavy slave-boson and $b$ is a light slave-boson that creates doubly-occupied and empty impurity states out of vacuum. The Fermi character of the real electrons enforces that the single-occupation operator $f_\sigma$ to be fermionic. The bosons created by $b^\dagger$ are also called holons and are supposed to carry the electron's charge. On the other hand, fermions created by $f^\dagger_\sigma$ are called spinons and are supposed to carry the electron's spin but carry no charge.
The connection of these auxilary operators $(a, b, f)$ to the physical operator $d$ is

$$d^\dagger_\sigma = f^\dagger_\sigma b + \eta_\sigma a^\dagger f_{-\sigma}, \tag{2.14}$$

where $\eta_\uparrow = 1$ and $\eta_\downarrow = -1$ are the phase factors needed to be faithful to the anticommutation relation of $d$ and $d^\dagger$ operators. To implement the projection to the physical Hilbert subspace, one needs to impose the

constraint

$$Q = \sum_{\sigma} f^{\dagger}_{\sigma} f_{\sigma} + b^{\dagger} b + a^{\dagger} a = 1. \tag{2.15}$$

Basically, this constraint imposes the conservation of auxilary particles number and restricts it to 1.
In this formalism, it is then straightforward to do the projection to the physical Hilbert subspace for $U \to \infty$, where the double-occupancy is completely gapped out. Therefore, the operator $a$ will vanish from the equation (2.15) and one obtains

$$Q = \sum_{\sigma} f^{\dagger}_{\sigma} f_{\sigma} + b^{\dagger} b = 1. \tag{2.16}$$

One should take into account that the conservation of auxilary particle number guarantees the existence of a local gauge symmetry $U(1)$ of the Hamiltoninan.
Now one should satisfy equation (2.16) by projecting out ($Q > 1$) sectors of the Hilbert space. To do this, we consider the grand canonical density operator

$$\rho_G = \frac{1}{Z_G} e^{-\beta(H+\lambda Q)}, \tag{2.17}$$

where $Z_G = tr\{e^{-\beta(\mathcal{H}+\lambda Q)}\}$ is the grand canonnical partition function and $\lambda$ is the auxilary particle chemical potential. Since the norm of the operator $Q$ is one, one can use the following trick to rewrite the partition function as follows

$$Z_G = tr\{Q e^{-\beta(\mathcal{H}+\lambda Q)}\}. \tag{2.18}$$

In the grand canonical ensemble, the expectation value of a physical operator $O$ is defined as

$$\langle O \rangle = tr\{\rho_G O\} = \lim_{\lambda\to\infty} \frac{tr\{O e^{-\beta(\mathcal{H}+\lambda Q)}\}}{tr\{Q e^{-\beta(\mathcal{H}+\lambda Q)}\}} \tag{2.19}$$

where the limit ($\lambda \to \infty$) restricts the evaluation of the expectation value to the $Q = 1$ sector of the Hilbert space as follows,

$$\langle O \rangle = \lim_{\lambda\to\infty} \frac{\langle Q=0|O e^{-\beta(\mathcal{H}+\lambda Q)}|Q=0\rangle + \langle Q=1|O e^{-\beta(\mathcal{H}+\lambda Q)}|Q=1\rangle + \langle Q=2|O e^{-\beta(\mathcal{H}+\lambda Q)}|Q=2\rangle + ...}{\langle Q=0|Q e^{-\beta(\mathcal{H}+\lambda Q)}|Q=0\rangle + \langle Q=1|Q e^{-\beta(\mathcal{H}+\lambda Q)}|Q=1\rangle + \langle Q=2|Q e^{-\beta(\mathcal{H}+\lambda Q)}|Q=2\rangle + ...}. \tag{2.20}$$

Considering the fact that the expectation value of a physical operator is vanishing in $Q = 0$ sector; $O|Q = 0\rangle = 0$ and $Q|Q = 0\rangle = 0$. Moreover, the terms with $Q > 1$ drops faster than $Q = 1$,

$$\to \langle O \rangle = \frac{\langle Q=1|O e^{-\beta(\mathcal{H}+\lambda Q)}|Q=1\rangle}{\langle Q=1|Q e^{-\beta(\mathcal{H}+\lambda Q)}|Q=1\rangle} = \frac{\frac{1}{Z_G^{Q=1}}\langle Q=1|O e^{-\beta(\mathcal{H}+\lambda Q)}|Q=1\rangle}{\frac{1}{Z_G^{Q=1}}\langle Q=1|Q e^{-\beta(\mathcal{H}+\lambda Q)}|Q=1\rangle} = \frac{\langle O\rangle_{Q=1}}{\langle Q\rangle_{Q=1}}. \tag{2.21}$$

Therefore, in this projection method, one should calculate the expectation value of a physical operator from a grand canonical operator.
It is now straightforward to rewrite the Anderson model (2.12) with infinite Coulomb repulsion $U$ in the slave-boson representaion as [14]

$$\mathcal{H} = \sum_{\mathbf{k}\sigma} \epsilon^{c}_{\mathbf{k}} c^{\dagger}_{\mathbf{k}\sigma} c_{\mathbf{k}\sigma} + \sum_{\sigma} \epsilon^{f} f^{\dagger}_{\sigma} f_{\sigma} + \sum_{\mathbf{k}\sigma} \left( V_{\mathbf{k}} c^{\dagger}_{\mathbf{k}\sigma} f_{\sigma} b^{\dagger} + H.c. \right) + \lambda(Q-1), \tag{2.22}$$

with the following vertices

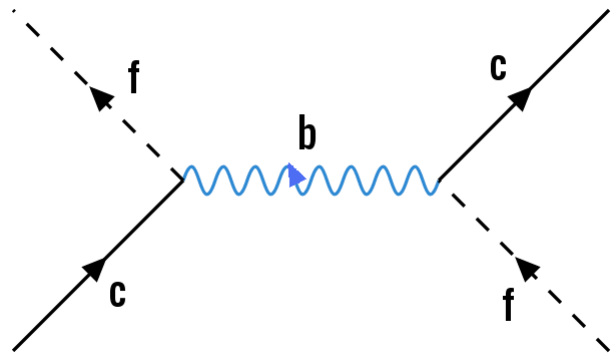

Figure 2.2: Slave-particle vertices of the infinite-$U$ Anderson model. Slave-boson $b$ appears as an exchange particle that mediate the interaction.

In 1979, Martin and Allen proposed that in terms of adiabaticity, the low-temperature behavior of a Kondo lattice is reasonably well-described in a low-energy limit derivative of Anderson lattice model, or better to say infinite-$U$ Anderson lattice model [24]. Following this idea, we will later realize Kondo insulator through infinite-$U$ Anderson lattice model by applying slave-boson mean-field theory.

CHAPTER 3

# Model of topological Kondo insulators

In this chapter, the proposed model for topological Kondo insulators (TKI), namely spin-orbit coupled infinite-$U$ Anderson lattice model [9, 25], which will be used for the rest of this thesis, is introduced. To have an insight for building the model, first the physical backgrounds and conventions that are used, such as the properties of quantum mechanical states and the role of spin-orbit coupling, are presented. Then, time-reversal symmetry and its impact on the quantum states, in terms of Kramers theorem, is discussed. Following this strategy, effects of spin-orbit coupling and crystal field splitting in heavy materials such as rare-earth elements and $SmB_6$ as one of the candidates of TKIs are reviewed. Finally, by including the effect of strong spin-orbit coupling and crystal field splitting in Anderson model and its infinite-$U$ variant, we obtain a $k$-dependent, nonlocal, odd parity, time-reversal invariant hybridization form factor, which has a central role in the topological Kondo insulators model Hamiltonian.

## 3.1 Quantum mechanical state

In quantum mechanics, the state of an electron in a hydrogen atom is determined by a set of four quantum numbers; the principal quantum number $n$, the azimuthal quantum number $l$, the magnetic quantum number $m_l$ and the spin quantum number $m_s = \sigma$. The wavefunction of an electron in real space can be written as a product of a radial part $R_{nl}$, an angular part $Y_{lm_l}$ and a spin part $S_\sigma$. In spherical coordinates, $\mathbf{r} = (r, \theta, \phi)$, it can be written as

$$\Psi_{nlm_l\sigma} = R_{nl}(r)Y_{lm_l}(\theta, \phi)S_\sigma, \tag{3.1}$$

where the angular component $Y_{lm_l}$ is a spherical harmonic and the radial component $R_{nl}$ is generally an exponential function decaying with distance $r$ (see figure (3.1)). The radial together with the angular part of these wavefunctions, $R_{nl}(r)Y_{lm_l}(\theta, \phi)$, forms the so-called atomic orbitals, which give the probability amplitude for an electron to be found in a specific location around a nucleus of the atom. The principal quantum number $n$ can take integer values $n \in \{1, 2, 3, ...\}$ and it determines the different atomic shell energies. The azimuthal quantum number $l$ determines the orbital angular momentum of the electron. For a given energy level $n$, the group of orbitals corresponding to different values of $l$ are labelled as $s, p, d, f, ...$; referring to $l = 0, 1, 2, 3, ...$ , respectively. The atomic subshell determined by the two quantum numbers $n$ and $l$ is described by the magnetic quantum number $m_l \in \{-l, -l + 1, ...l\}$. Therefore, describing a specific atomic orbital requires three quantum numbers $n$, $l$ and $m_l$. Due to the Pauli exclusion principle for Fermions, each orbital can host at most two electrons of different spin quantum number, and thus the state of a single electron is entirely described by the four quantum numbers [26].

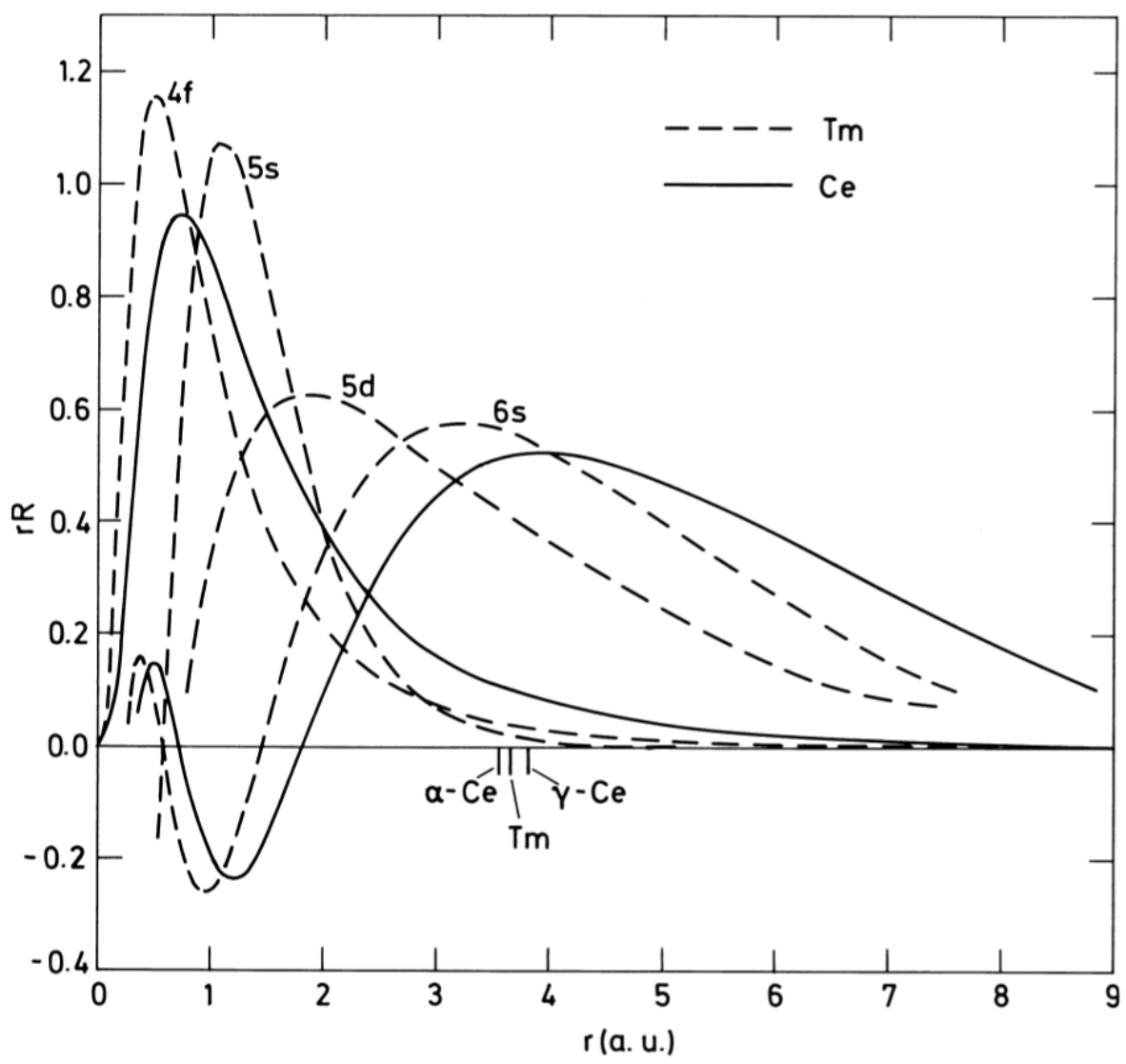

Figure 3.1: The radial components of atomic wavefunctions for $Ce$, with one $4f$ electron, and $Tm$, with 13 $4f$-electrons, or one $4f$-hole [27].

In a spherical potential $V(r)$, the total energy of an atomic state can be determined by two quantum numbers $n$ and $l$, independent of the value of the magnetic quantum number $m_l$. The degenracy over the values of $m_l$ can be lifted by internal or external effects, such as spin-orbit interaction, which is considerable for heavy nucleus atoms, and crystal field effect which is a static electric field produced by a surrounding charge distribution. In the later calculations, we will use the spherical harmonics $Y_{lm_l}$ obtained for the hydrogen atom, for the atoms with more than one electrons.

## 3.2 Spin-orbit interaction

One can explain the spin-orbit interaction by considering an electron orbiting a nucleus with positive amount of charge ($+Ze$), where $Z$ is the atomic number and $e = 1.602 \times 10^{-19} C$ is the elementary electric charge. Since the electron moves with speeds close to the speed of light, relativistic effects become considerable. These effects can be simply realized in special relativity by changing from the laboratory frame of reference to that of the electron; the moving nucleus induces a magnetic field which then interacts with the spin of the electron. Apart from this simple intuitive picture, one can start from the relativistic Dirac equation in a spherically symmetric potential field with minimal substitution as

$$(i\,\hbar \not{D} - m_e c)\Psi = 0, \tag{3.2}$$

where $\not{D} = \gamma_\mu D^\mu = \gamma_\mu(\partial^\mu + i\frac{e}{\hbar c}A^\mu)$, with $A^\mu = (V, \mathbf{A})$ being the electromagnetic four-potential.
The low energy limit of the Dirac equation (3.2) can be found by applying perturbation theory up to the leading order of $(\frac{p}{m_e c})^2$ [28]

$$\mathcal{H} \approx \underbrace{\frac{\mathbf{p}^2}{2m_e} + V}_{\text{non-relativistic term}} - \underbrace{\frac{\mathbf{p}^4}{8m_e^3c^2}}_{\text{kinetic correction term}} + \underbrace{\frac{\hbar^2}{8m_e^2c^2}\nabla^2 V}_{\text{Darwin term}} + \underbrace{\frac{\hbar^2}{2m_e^2c^2 r}\frac{dV}{dr}\mathbf{L}.\mathbf{S}}_{\text{spin-orbit interaction}}\ , \tag{3.3}$$

where $\mathbf{L} = (L_x, L_y, L_z)$ and $\mathbf{S} = (S_x, S_y, S_z)$ is the spin of the electron. The first and second term of the expansion describe the non-relativistic part of the Hamiltonian (3.3). The third term, which is a correction to the kinetic energy due to the increase of mass with velocity, reduces the energy of all states by an amount which is determined by $l$. The fourth term, is the so-called 'Darwin' term, which merely increases the energy of $s$ states. These effects may both be incorporated into the central field ($V$), but the last term couples together the spin and orbital motion in a way that has far-reaching consequences for the magnetic properties [27].
When dealing with different sources of angular momentum, like $\mathbf{L}$ and $\mathbf{S}$, there are two obvious bases that one might choose to work in. The first is called the uncoupled basis, which is the common eigenbasis of the operators $\mathbf{L}^2, \mathbf{L}_z, \mathbf{S}^2$ and $\mathbf{S}_z$, that simultaneously commute with Hamiltonian operator, with the following eigenvalues

$$\begin{aligned} \mathbf{L}^2|l, m_l; s, \sigma\rangle &= \hbar^2 l(l+1)|l, m_l; s, \sigma\rangle \\ \mathbf{L}_z|l, m_l; s, \sigma\rangle &= \hbar m_l|l, m_l; s, \sigma\rangle \\ \mathbf{S}^2|l, m_l; s, \sigma\rangle &= \hbar^2 s(s+1)|l, m_l; s, \sigma\rangle \\ \mathbf{S}_z|l, m_l; s, \sigma\rangle &= \hbar m_s|l, m_l; s, \sigma\rangle \end{aligned}, \tag{3.4}$$

where the quantum numbers $l, m_l, s$ and $\sigma$ are needed to identify the eigenstates, as explained in the section 3.1. In case of having a spin-orbit coupling term in the Hamiltonian, these operators do not commute with the Hamiltonian simultanously; this is the consequence of spin $s$ and angular momentum $l$ not being good quantum numbers. In this case, the total angular momentum can be defined as

$$\mathbf{J} = \mathbf{L} + \mathbf{S}, \tag{3.5}$$

and one can show that the Hamiltonian can simultaneously commute with the operators $\mathbf{J}^2, \mathbf{L}^2, \mathbf{S}^2$ and $\mathbf{J}_z$, which means that the good quantum numbers needed to identify the eigenstates are $j, l, s$ and $m_j$, respectively. In this case, the one-electron wavefunction of the atomic subshell may be written as a superposition of the uncoupled wavefunctions

$$\Psi(l, s; j, m_j) = \sum_{m_l, \sigma} C_{ls}(m_l, \sigma; j, m_j)\Psi(l, m_l; s, \sigma), \tag{3.6}$$

where $C_{ls}(m_l, \sigma; j, m_j)$ is the Clebsch–Gordan coefficient. Without loss of generality, they are assumed to be real numbers [28]. It is convenient to write the expansion (3.6) in a representation independent form, in terms of state vectors

$$|l, s; j, m_j\rangle = \sum_{m_l, m_s} \langle l, m_l; s, \sigma|l, s; j, m_j\rangle|l, m_l; s, \sigma\rangle. \tag{3.7}$$

Therefore, the problem of finding the coupled wavefunction or eigenbasis of the Hamitonian which includes spin-orbit coupling, reduces to finding the corresponding Clebsch–Gordan coefficients.

To observe the importance of spin-orbit coupling for heavy elements, one can try to find the spin-orbit coupling constant $\lambda$. The energy shift due to the spin-orbit contribution can be written as

$$\Delta E = \lambda\langle \mathbf{L}.\mathbf{S}\rangle = \frac{\lambda}{2}\left(\langle \mathbf{J}\rangle^2 - \langle \mathbf{L}\rangle^2 - \langle \mathbf{S}\rangle^2\right) = \frac{\lambda}{2}\left(j(j+1) - l(l+1) - s(s+1)\right). \tag{3.8}$$

From the expansion of the Dirac equation (3.3), one finds that

$$\lambda = \frac{\hbar^2}{2m_e^2c^2r}\frac{dV}{dr}, \tag{3.9}$$

where $r$ is the distance between electron and nucleus. The potential $V(r)$ created by a nucleus with $Z$ protons is given by

$$V(r) = -\frac{Ze^2}{4\pi\epsilon_0}\frac{1}{r} \quad . \tag{3.10}$$

Besides, one can find [29]

$$\langle\frac{1}{r^3}\rangle_{nl} = \frac{Z^3}{l(l+\frac{1}{2})(l+1)n^3a_B}, \tag{3.11}$$

where $a_B = \frac{4\pi\epsilon_0\hbar^2}{m_e e^2}$ is the Bohr radius.
Using equations (3.9), (3.10) and (3.11); one obtains that the coupling constant $\lambda$ is proportional to [26, 29]

$$\lambda \sim Z^4, \tag{3.12}$$

which means that the spin-orbit interaction is proportional to the fourth power of the atomic number and becomes considerable in heavy materials.

## 3.3 Time-reversal symmetry and Kramers' theorem

A symmetry which is obeyed by many classical and quantum systems is the so-called time-reversal symmetry (TRS), which means that the physical properties of the system are invariant under switch of time arrow, $t \to -t$. The time-reversal implies that quantities like linear momentum $\mathbf{p}$ and spin $\mathbf{s} = \frac{\hbar}{2}\boldsymbol{\sigma}$ change direction at the same position $\mathbf{r}$ [30],

$$\begin{cases} \mathbb{T}\mathbf{r}\mathbb{T}^{-1} = \mathbf{r} \\ \mathbb{T}\mathbf{p}\mathbb{T}^{-1} = -\mathbf{p} \\ \mathbb{T}\mathbf{s}\mathbb{T}^{-1} = -\mathbf{s} \end{cases} \tag{3.13}$$

where $\mathbb{T}$ is the time-reversal operator.
A system is called time-reversal invariant, if the Hamiltonian of the system $\mathcal{H}$ does not change under transformations (3.13) or, in other words; it commutes with the time-reversal operator

$$\mathbb{T}\mathcal{H}\mathbb{T}^{-1} = \mathcal{H} \iff \mathbb{T}\mathcal{H} = \mathcal{H}\mathbb{T} \iff [\mathbb{T}, \mathcal{H}] = 0 \quad . \tag{3.14}$$

Therefore, a typical interaction which is odd on both spin and momentum is invariant under time reversal,

$$\mathbb{T}\mathbf{s}\cdot\mathbf{p}\mathbb{T}^{-1} = (\mathbb{T}\mathbf{s}\mathbb{T}^{-1})\cdot(\mathbb{T}\mathbf{p}\mathbb{T}^{-1}) = (-\mathbf{s})\cdot(-\mathbf{p}) = \mathbf{s}\cdot\mathbf{p} \quad . \tag{3.15}$$

Without loss of generality, one can show that the representation of the time-reversal operator ($\mathbb{T}$), compatible with the equations (3.13), is [31]

$$\mathbb{T} = i\sigma_y C, \tag{3.16}$$

where $C$ is the complex conjugation operator.
From equation (3.16), it is obvious that the time-reversal operator of a spin-$\frac{1}{2}$ fermion, is an anti-unitary operator with the property,

$$\mathbb{T}^2 = -1 \iff \mathbb{T}^{-1} = -\mathbb{T}, \tag{3.17}$$

which maps the Hilbert space onto itself and preserves the norm of the states. It differs from a unitary operator by an extra complex conjugation as

$$\langle Tx|Ty\rangle = \overline{\langle x|y\rangle} = \langle y|x\rangle. \tag{3.18}$$

One of the fundamental consequences of the time-reversal symmetry for a quantum mechanical system, is the so-called Kramers' theorem. According to this theorem, each eigenvalue of a time-reversal invariant Hamiltonian with odd number of spin-$\frac{1}{2}$ fermions is at least doubly degenerate.
For the proof, consider two eigenstates, $|E^+\rangle$ and $|E^-\rangle$, of the Hamiltonian, which are related to each other by time reversal operation as

$$|E^+\rangle = \mathbb{T}|E^-\rangle \qquad \text{or} \qquad |E^-\rangle = \mathbb{T}^{-1}|E^+\rangle. \tag{3.19}$$

Since the Hamiltonian has the time-reversal symmetry, these two eigenstates have the same energy eigenvalue. Now one should show that these two state are independent of each other, or in the other words; they are orthogonal

$$\langle E^+|E^-\rangle = 0. \tag{3.20}$$

Using equations (3.17), (3.18) and (3.19), one finds

$$\begin{aligned}&\langle E^+|E^-\rangle = \langle E^-|\mathbb{T}^\dagger\mathbb{T}^{-1}|E^+\rangle = -\langle E^-|\mathbb{T}^\dagger\mathbb{T}|E^+\rangle = -\langle \mathbb{T}E^-|\mathbb{T}E^+\rangle = -\overline{\langle E^-|E^+\rangle} = -\langle E^+|E^-\rangle \\ &\rightarrow\ 2\langle E^+|E^-\rangle = 0\ .\end{aligned} \tag{3.21}$$

Thus, the theorem is proved.

## 3.4 Rare earth elements

One group of heavy materials where the spin-orbit interaction is of considerable importance, is the so-called rare earth elements. The rare earth elements are the 15 lanthanide elements ($Z = 57$ through 71) and yttrium ($Z = 39$), which are filling the $4f$ electron shell. Denoting that lanthanum is considered a $d$-block element, but is included in the rare earth elements due to its chemical similarities with the other fourteen [27]. The $4f$ electrons are located deep inside the atomic shells, and covered by the $5s$ and $5p$ states. The $5d$ and $6s$ electrons form the conduction bands in the metals. The partial screening of the increasing nuclear charge along the rare earth elements, contracts the wavefunctions of the lanthanide, which is reflected in their ionic and atomic radii [27]. For example, as illustrated in (3.1), the $4f$ wavefunction contracts significantly between Ce, which has one $4f$ electron, and Tm, which has one $4f$ hole in the atom, while both of them are in the metallic state. Generally, when a large number of rare earth atoms are gathered together to form a solid, the $4f$ electron shells remain highly localized, so that their magnetic properties closely resemble to those in the free atoms. The $5d$ and $6s$ electrons, which are located at the external shells; on the other hand, become delocalized into Bloch states, spreading throughout the metal and creating a conduction-electron gas [27].

## 3.5 Spin-orbit interaction and crystal field effect in $SmB_6$

An old candidate of mixed-valent heavy-fermion Kondo insulator is $SmB_6$, which belongs to the group of rare-earth hexaborides with a cubic crystal symmetry (lattice constant $a \approx 4.13$ Å) [32–34] and hybridization gap of around 20 $meV$ [35]. In $SmB_6$, each Sm ion is surrounded by B cages located in each corner of the cubic cell, and consequently electrons in the Sm ion feel the cubic crystal field and a strong spin-orbit interaction due to its heavy nucleus.

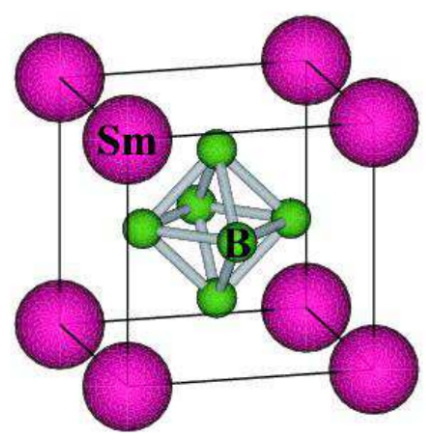

Figure 3.2: $SmB_6$ crystal structure [36].

The first intuition to build a tight-binding model for $SmB_6$ is coming from the first principle studies of Yanase, Harima [37] and Antonov [38]. Their studies show that samarium $4f$ orbitals hybridize with the samarium $5d$ orbitals. In this scheme, one can consider the effect of the crystal field splitting and spin-orbit interaction on the $5d$ and $4f$ electrons. In case of $5d$ electrons, crystal field splitting has a more stronger effect than the spin-orbit coupling, in comparison with the $f$-electrons. This is because the wavefunction of the $d$-electrons are spatially more spread than the localized $f$-electrons. So that first crystal field effect lifts the degenracy of the $d$ orbitals ($m_l = 0, \pm1, \pm2$), leading to a low-lying $e_g$ quartet and a higher energy $t_{2g}$ multiplet at the Γ point (figure 3.4). Away from the Γ point, due to the spin-orbit interaction $e_g$ orbitals splits into a $\Gamma_8$ quartet and $t_{2g}$ splitts into a $\Gamma_7$ doublet and $\Gamma_8$ quartet [36]. Thus, for the $5d$ electrons only $\{\Gamma_8^{(1)}, \Gamma_8^{(2)}\}$ doublets of $e_g$ are close to the Fermi energy [39].

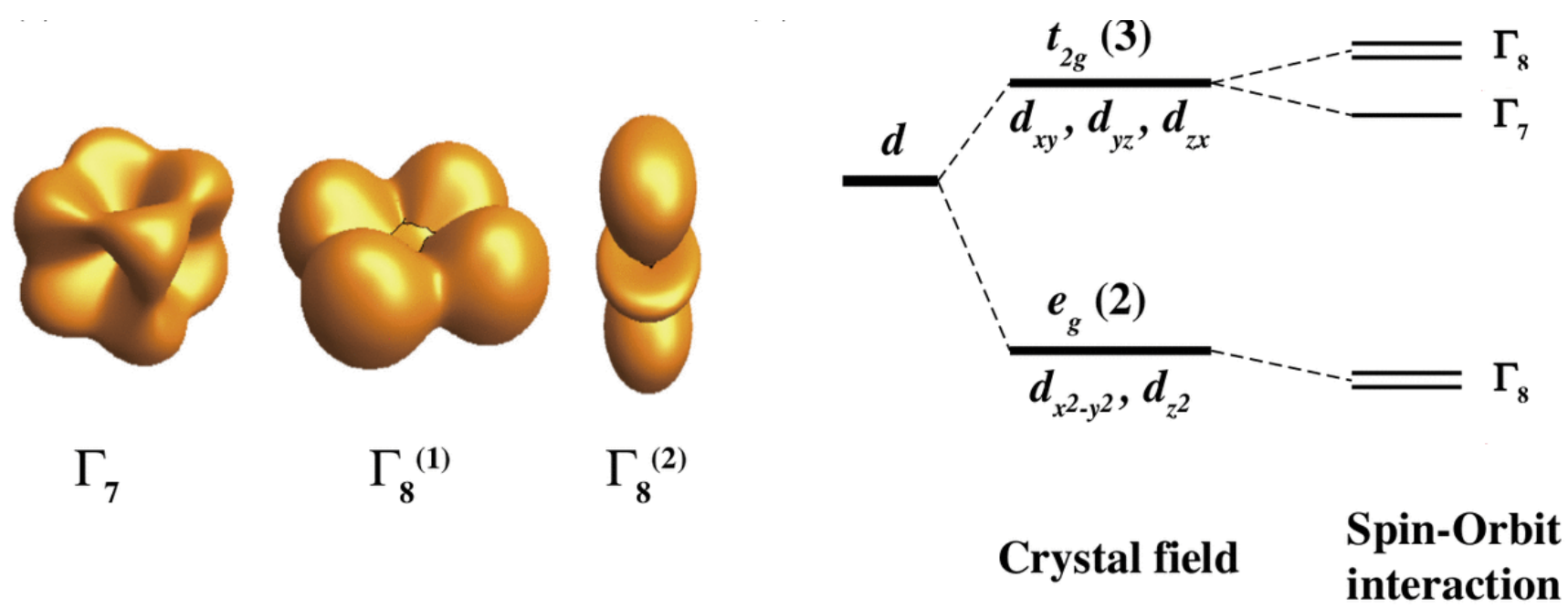

Figure 3.3: Crystal field and the spin-orbit splittings for Sm $5d$ levels [36].

In case of $4f$ electron's state, first spin-orbit coupling splits it into $j = \frac{7}{2}$ and $j = \frac{5}{2}$ multiplets, where $j = \frac{7}{2}$ states can be ignored, since *ab initio* methods show that their energies are far away from the Fermi level [36, 40]. Then crystal field effect splits the $j = \frac{5}{2}$ multiplets, with degenracy of $2(\frac{5}{2}) + 1 = 6$, into the $\Gamma_7$ doublet and $\Gamma_8$ quartet, where $\Gamma_7$ can also be ignored since it is far away from Fermi energy. Away from the Γ point, the $\Gamma_8$ quartet further splits into two doublets $\Gamma_8^{(1)}$ and $\Gamma_8^{(2)}$. These wavefunctions can be written as [36]

$$\begin{aligned} |\Gamma_7^f\rangle &= \sqrt{\frac{5}{6}}\ |m_j = \pm\frac{3}{2}\rangle - \sqrt{\frac{1}{6}}\ |m_j = \mp\frac{5}{2}\rangle \\ |\Gamma_8^{f(1)}\rangle &= \sqrt{\frac{1}{6}}\ |m_j = \pm\frac{3}{2}\rangle + \sqrt{\frac{5}{6}}\ |m_j = \mp\frac{5}{2}\rangle \ . \\ |\Gamma_8^{f(2)}\rangle &= |m_j = \pm\frac{1}{2}\rangle \end{aligned} \tag{3.22}$$

Thus, for the $4f$ electrons only $\{\Gamma_8^{(1)}, \Gamma_8^{(2)}\}$ doublets of $j = \frac{5}{2}$ are close to the Fermi energy.

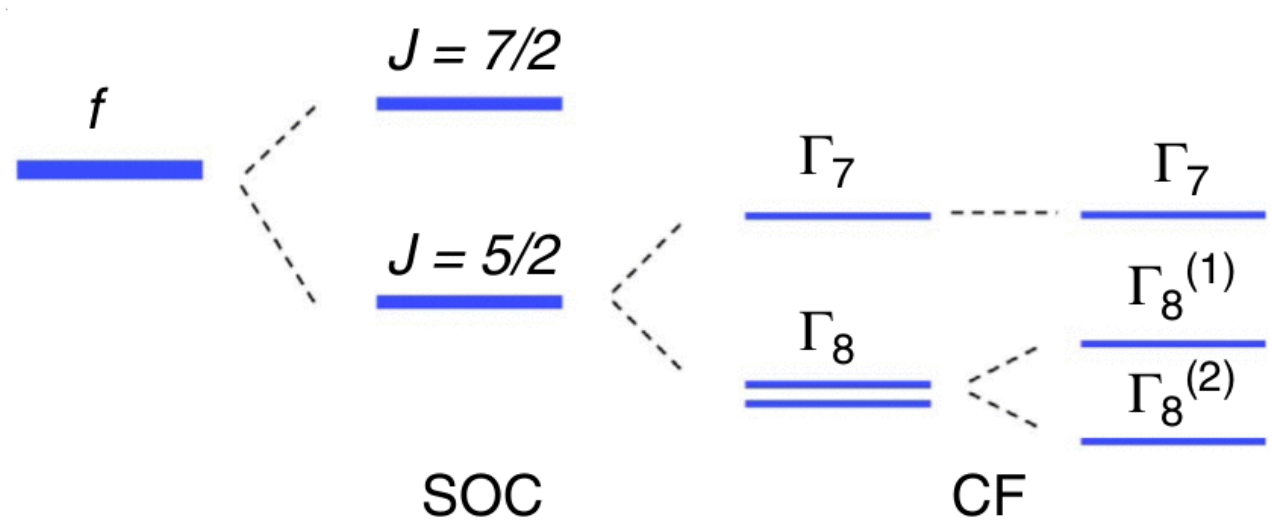

Figure 3.4: Crystal field and the spin-orbit splittings for Sm $5d$ level [41].

Altogether, one can conclude that effectively in a Kondo insulator the $\Gamma_8$ quartet of the $f$ state hybridizes with the $e_g$ quartet of the $5d$ conduction state [10].

## 3.6 Single impurity Anderson model with spin-orbit coupling

Having insight from the first principle studies and experimental results about $SmB_6$ in mind, we can present a toy model for topological Kondo insulators introduced by Maxim Dzero and Piers Coleman in 2010 [9, 25]. In this toy model, it is assumed that the conduction band consists of Bloch electrons and the heavy band consists of only the lowest-lying Kramers doublet with respect to the Fermi energy. We start with the single impurity Anderson model written in terms of the fermionic operators, associated with the crystal field symmetry and spin-orbit coupling as [9]

$$\mathcal{H} = \sum_{\mathbf{k}\sigma} \epsilon^c_{\mathbf{k}} c^\dagger_{\mathbf{k}\sigma} c_{\mathbf{k}\sigma} + \sum_{m_j} \epsilon^f f^\dagger_{m_j} f_{m_j} + U n^f_\uparrow n^f_\downarrow + \sum_{\mathbf{k}} \sum_{\sigma m_j} \Big( \langle \mathbf{k}, \sigma | V | \Gamma^f : \mathbf{R} = 0, m_j \rangle c^\dagger_{\mathbf{k}\sigma} f_{m_j} + H.c. \Big), \tag{3.23}$$

where the fermionic operator $c^\dagger_{\mathbf{k}\sigma}$ creates a Bloch state $|\mathbf{k}, \sigma\rangle$ with $\mathbf{k}$ and $\sigma \in \{\uparrow, \downarrow\}$ being momentum and real spin quantum numbers, respectively. The fermionic operator $f^\dagger_{m_j}$ creates a localized Kramers doublet state $|\Gamma^f : \mathbf{R} = 0, m_j\rangle$ associated with the crystal field effect and spin-orbit coupling, where $m_j \in \{\Uparrow, \Downarrow\}$ is a pseudospin quantum number, corresponding to the two possible values of the projection of total angular momentum in the lowest-lying Kramers doublet. For example, in case of $SmB_6$ the lowest-lying Kramers doublet is $|\Gamma_8^{f(2)}\rangle = |m_j = \pm\frac{1}{2}\rangle$.

In the following, we try to simplify the hybridization matrix elements $\langle \mathbf{k}, \sigma | V | \Gamma^f : \mathbf{R} = 0, m_j\rangle$, using

certain assumptions. In this model, it is implied that the localized states can be reasonbaly well described by the wavefunctions of the Hydrogen atom. Moreover, it is assumed that the hybridization operator ($V$) has a spherical symmetry in the vicinity of the heavy atom. At the first step, we expand the Bloch wave $\langle \mathbf{k}, \sigma|$ of $s = \frac{1}{2}$ fermion in terms of spherical harmonics,

$$Y_{l',m_{l'}}(\hat{\mathbf{k}}) = \langle \theta', \phi' | l', m_{l'} \rangle, \tag{3.24}$$

and partial waves, $\langle k; l', m_{l'}; s, \sigma|$ (or uncoupled basis), as

$$\langle \mathbf{k}, \sigma| = \sum_{l',m_{l'}} \langle k; l', m_{l'}; s, \sigma| Y_{l',m_{l'}}(\hat{\mathbf{k}}) = \sum_{l',m_{l'}} \langle k| \otimes \langle l', m_{l'}; s, \sigma| Y_{l',m_{l'}}(\hat{\mathbf{k}}). \tag{3.25}$$

Besides, the lowest-lying Kramers doublet state can be can be decomposed into radial and angular part as

$$|\Gamma^f : \mathbf{R} = 0, m_j\rangle = |R = 0; l, s; j, m_j\rangle = |R = 0\rangle \otimes |l, s; j, m_j\rangle. \tag{3.26}$$

For fixed value of $l$ and $s = \frac{1}{2}$,

$$m_l = m_j - \sigma. \tag{3.27}$$

Now one can rewrite the hybridization matrix elements by using equations (3.25), (3.26), (3.27) as

$$\begin{aligned} \langle \mathbf{k}, \sigma|V|\Gamma^f : \mathbf{R} = 0, m_j\rangle &= \sum_{l',m_{l'}} \langle k|V|R = 0\rangle \underbrace{\langle l', m_{l'}; s, \sigma|l, s; j, m_j\rangle}_{\sim \delta_{l,l'} \delta_{m_j-\sigma, m_{l'}}} Y_{l,m_l}(\hat{\mathbf{k}}) \\ &= \langle k|V|R = 0\rangle \langle l, m_j - \sigma; s, \sigma|l, s; j, m_j\rangle Y_{l,m_j-\sigma}(\hat{\mathbf{k}}), \end{aligned} \tag{3.28}$$

where $\langle l, m_j - \sigma; s, \sigma|l, s; j, m_j\rangle$ is the Clebsch-Gordan coefficient. A relevant simplification for the hybridization matrix elements is to assume a constant hybridization amplitude as

$$\langle k|V|R = 0\rangle \sim V_0. \tag{3.29}$$

Thus, the equation (3.28) becomes

$$\langle \mathbf{k}, \sigma|V|\Gamma^f : \mathbf{R} = 0, m_j\rangle = V_0 C_{ls}(j, m_j; m_j - \sigma, \sigma) Y_{l,m_j-\sigma}(\hat{\mathbf{k}}), \tag{3.30}$$

with

$$\begin{cases} \sigma \in \{-\frac{1}{2}, \frac{1}{2}\} \\ j \in \{l - \frac{1}{2}, l + \frac{1}{2}\} \\ m_j \in \{-j, -j + 1, ..., j\} \end{cases}. \tag{3.31}$$

The Clebsh-Gordan coefficient for $s = \frac{1}{2}$ is [28]

$$C_{ls}(j, m_j; m_j - \sigma, \sigma) = \begin{cases} \sqrt{\frac{1}{2} + \frac{2m_j\sigma}{2l+1}}, & j = l + \frac{1}{2} \\ \mathrm{sgn}(\sigma)\sqrt{\frac{1}{2} - \frac{2m_j\sigma}{2l+1}}, & j = l - \frac{1}{2} \end{cases}. \tag{3.32}$$

By defining spin-orbit coupled spherical harmonics as

$$\mathcal{Y}_{\sigma m_j}(\hat{\mathbf{k}}) = C_{ls}(j, m_j; m_j - \sigma, \sigma) Y_{l, m_j - \sigma}(\hat{\mathbf{k}}), \tag{3.33}$$

one can finally write the hybridization matrix element as

$$\Phi_{\sigma m_j}(\hat{\mathbf{k}}) = \langle \mathbf{k}, \sigma | V | \Gamma^f : \mathbf{R} = 0, m_j \rangle = V_0 \mathcal{Y}_{\sigma m_j}(\hat{\mathbf{k}}). \tag{3.34}$$

From now on, we replace the symbol of the pseudospin quantum number $m_j$ with $\alpha$ for simplicity. Now, one can calculate the hybridization matrix for two interesting cases,

$$\begin{cases} \text{(i)} \;\; l = 1, s = \frac{1}{2}, j = \frac{1}{2} \;\text{ and }\; m_j = \pm\frac{1}{2} \\ \text{(ii)} \;\; l = 3, s = \frac{1}{2}, j = \frac{5}{2} \;\text{ and }\; m_j = \pm\frac{1}{2} \end{cases} \tag{3.35}$$

For the case (i) [25],

$$\Phi(\hat{\mathbf{k}}) = V_0 \begin{bmatrix} \mathcal{Y}_{\frac{1}{2},\frac{1}{2}}(\hat{\mathbf{k}}) & \mathcal{Y}_{\frac{1}{2},-\frac{1}{2}}(\hat{\mathbf{k}}) \\ \mathcal{Y}_{-\frac{1}{2},\frac{1}{2}}(\hat{\mathbf{k}}) & \mathcal{Y}_{-\frac{1}{2},-\frac{1}{2}}(\hat{\mathbf{k}}) \end{bmatrix} = \begin{bmatrix} C_{1,\frac{1}{2}}(\frac{1}{2},\frac{1}{2};0,\frac{1}{2})Y_{1,0}(\hat{\mathbf{k}}) & C_{1,\frac{1}{2}}(\frac{1}{2},-\frac{1}{2};-1,\frac{1}{2})Y_{1,-1}(\hat{\mathbf{k}}) \\ C_{1,\frac{1}{2}}(\frac{1}{2},\frac{1}{2};1,-\frac{1}{2})Y_{1,1}(\hat{\mathbf{k}}) & C_{1,\frac{1}{2}}(\frac{1}{2},-\frac{1}{2};0,-\frac{1}{2})Y_{1,0}(\hat{\mathbf{k}}) \end{bmatrix}, \tag{3.36}$$

with

$$\begin{aligned} \mathcal{Y}_{\frac{1}{2},\frac{1}{2}}(\hat{\mathbf{k}}) &= C_{1,\frac{1}{2}}(\frac{1}{2},\frac{1}{2};0,\frac{1}{2})Y_{1,0}(\hat{\mathbf{k}}) = \sqrt{\frac{1}{3}}\left(\sqrt{\frac{3}{4\pi}}\hat{k}_z\right) = \sqrt{\frac{1}{4\pi}}\hat{k}_z \\ \mathcal{Y}_{\frac{1}{2},-\frac{1}{2}}(\hat{\mathbf{k}}) &= C_{1,\frac{1}{2}}(\frac{1}{2},-\frac{1}{2};-1,\frac{1}{2})Y_{1,-1}(\hat{\mathbf{k}}) = \sqrt{\frac{2}{3}}\left(\sqrt{\frac{3}{4\pi}}\frac{\hat{k}_x - i\hat{k}_y}{\sqrt{2}}\right) = \sqrt{\frac{1}{4\pi}}(\hat{k}_x - i\hat{k}_y) \\ \mathcal{Y}_{-\frac{1}{2},\frac{1}{2}}(\hat{\mathbf{k}}) &= C_{1,\frac{1}{2}}(\frac{1}{2},\frac{1}{2};1,-\frac{1}{2})Y_{1,1}(\hat{\mathbf{k}}) = -\sqrt{\frac{2}{3}}\left(-\sqrt{\frac{3}{4\pi}}\frac{\hat{k}_x + i\hat{k}_y}{\sqrt{2}}\right) = \sqrt{\frac{1}{4\pi}}(\hat{k}_x + i\hat{k}_y) \\ \mathcal{Y}_{-\frac{1}{2},-\frac{1}{2}}(\hat{\mathbf{k}}) &= C_{1,\frac{1}{2}}(\frac{1}{2},-\frac{1}{2};0,-\frac{1}{2})Y_{1,0}(\hat{\mathbf{k}}) = -\sqrt{\frac{1}{3}}\left(\sqrt{\frac{3}{4\pi}}\hat{k}_z\right) = -\sqrt{\frac{1}{4\pi}}\hat{k}_z \end{aligned} . \tag{3.37}$$

Thus, it becomes

$$\Phi(\hat{\mathbf{k}}) = \frac{V_0}{\sqrt{4\pi}} \begin{bmatrix} \hat{k}_z & \hat{k}_x - i\hat{k}_y \\ \hat{k}_x + i\hat{k}_y & -\hat{k}_z \end{bmatrix}. \tag{3.38}$$

Rewriting equation (3.38) in terms of Pauli matrices and absorbing the prefactor into the hybridization amplitude, one finds

$$\Phi(\hat{\mathbf{k}}) = V_0 \hat{\mathbf{k}} \cdot \boldsymbol{\sigma} \;\; . \tag{3.39}$$

As it was proved before (3.15), since the form factor is proportional to a product of an odd number of spin and odd number of momentum operators, it is time-reversal invariant.
The hybridization form factor of the case (ii), which is relevant for $SmB_6$ is [9]

$$\Phi(\hat{\mathbf{k}}) = V_0 \begin{bmatrix} \mathcal{Y}_{\frac{1}{2},\frac{1}{2}}(\hat{\mathbf{k}}) & \mathcal{Y}_{\frac{1}{2},-\frac{1}{2}}(\hat{\mathbf{k}}) \\ \mathcal{Y}_{-\frac{1}{2},\frac{1}{2}}(\hat{\mathbf{k}}) & \mathcal{Y}_{-\frac{1}{2},-\frac{1}{2}}(\hat{\mathbf{k}}) \end{bmatrix} = \begin{bmatrix} C_{3,\frac{1}{2}}(\frac{5}{2},\frac{1}{2};0,\frac{1}{2})Y_{3,0}(\hat{\mathbf{k}}) & C_{3,\frac{1}{2}}(\frac{5}{2},-\frac{1}{2};-1,\frac{1}{2})Y_{3,-1}(\hat{\mathbf{k}}) \\ C_{3,\frac{1}{2}}(\frac{5}{2},\frac{1}{2};1,-\frac{1}{2})Y_{3,1}(\hat{\mathbf{k}}) & C_{3,\frac{1}{2}}(\frac{5}{2},-\frac{1}{2};0,-\frac{1}{2})Y_{3,0}(\hat{\mathbf{k}}) \end{bmatrix}, \tag{3.40}$$

which results in

$$\Phi(\hat{\mathbf{k}}) \propto V_0 \frac{k^3}{\sqrt{7}} \begin{bmatrix} \sqrt{3}Y_{3,0}(\hat{\mathbf{k}}) & -2Y_{3,-1}(\hat{\mathbf{k}}) \\ 2Y_{3,1}(\hat{\mathbf{k}}) & -\sqrt{3}Y_{3,0}(\hat{\mathbf{k}}) \end{bmatrix}. \tag{3.41}$$

In general, the form factor of a single impurity Anderson model can be written as [9]

$$\Phi(\hat{\mathbf{k}}) = V_0 \mathbf{S}_k \cdot \boldsymbol{\sigma}, \tag{3.42}$$

with $S_k$ being a vector function of momentum.
Up to here, the hybridization form factor of a single impurity Anderson model has been derived. Now, one need to generalize it to a lattice case. The form factor of a lattice should be periodic in momentum and also reduces to its simplest form ($\propto \hat{\mathbf{k}} \cdot \boldsymbol{\sigma}$) for $k \to 0$. Thus, the simplistic model form factor of the lattice can be proposed as [11]

$$\Phi(\mathbf{k}) = V_0 \mathbf{S}_k \cdot \boldsymbol{\sigma}, \tag{3.43}$$

with

$$\begin{cases} \mathbf{S}_k = (sin(k_x), sin(k_y), sin(k_z)) \\ \boldsymbol{\sigma} = (\sigma_x, \sigma_y, \sigma_z) \end{cases} . \tag{3.44}$$

One can see this model form factor is nonlocal, vanishing at high-symmetry points ($k = 0, \pm\pi$) and has odd parity

$$\Phi(-\mathbf{k}) = -\Phi(\mathbf{k}). \tag{3.45}$$

## 3.7 Model Hamiltonian of topological Kondo insulators

The proposed model Hamiltonian for topological Kondo insulators is the spin-orbit coupled infinite-$U$ Anderson lattice model [9, 25]. This model Hamiltonian is a variant of the Anderson model (3.23) described in the last section. It incorporates the effect of the spin-orbit interaction and crystal field splitting of the $f$ states through the hybridization form factor. Moreover, it includes infinite Coulomb repulsion which is neccessary to have the localized moments. The slave-boson representaion of this model Hamiltonian in real space is [25]

$$\begin{aligned} \mathcal{H} &= \mathcal{H}_c + \mathcal{H}_f + \mathcal{H}_{hyb} + \mathcal{H}_Q \,; \\ \mathcal{H}_c &= -t^c \sum_{\langle i,j \rangle} \sum_{\sigma=\uparrow,\downarrow} [c^\dagger_{i\sigma} c_{j\sigma} + H.c.] \\ \mathcal{H}_f &= -t^f \sum_{\langle i,j \rangle} \sum_{\alpha=\uparrow,\downarrow} [(f^\dagger_{i\alpha} b_i)(b^\dagger_j f_{j\alpha}) + H.c.] \\ \mathcal{H}_{hyb} &= V_0 \sum_{\langle i,j \rangle} \sum_{\sigma,\alpha=\uparrow,\downarrow} [c^\dagger_{i\sigma} \Phi_{\sigma\alpha}(\mathbf{R}_i - \mathbf{R}_j) b^\dagger_j f_{j\alpha} + H.c.] \\ \mathcal{H}_Q &= \sum_i \lambda_i (Q_i - 1), \end{aligned} \tag{3.46}$$

where $\mathcal{H}_c$ is a tight-binding model for the conduction electrons. $\mathcal{H}_f$ is a slave-boson representation of nearest-neighbor hoping of the $f$ electrons. $\mathcal{H}_{hyb}$ describes the hybridization of the spin-orbit coupled $f$ electrons with the conduction electrons via a nonlocal hybridization form factor $\Phi_{\sigma\alpha}(\mathbf{R}_i - \mathbf{R}_j)$. $\mathcal{H}_Q$ describes the constraint on the psuedoparticles, imposed at each lattice site in the infinite Coulomb repulsion.
In the next chapters, we will study the topological Kondo insulators via this model Hamiltonian.

CHAPTER 4

# Homogenous slave-boson mean-field theory of 3D topological Kondo insulators

In this chapter we develope a self-consistent theory to study the properties of the 3D bulk of the TKI with the model Hamiltonian introduced in chapter 3. For this purpose we use slave-boson mean-field theory which allows the study of low-energy regime of a Kondo system with a quadratic single-particle Hamiltonian. On the other hand, at higher temperatures, it produces an spurious first-order transition to the local-moment regime rather than a continuous crossover. Therefore, this mean-field theory is safe for low enough temperatures ($\frac{T}{D_0} \leq 10^{-4}$) to analyse Kondo systems. In this approximation, it is straightforward to calculate spectral functions and local density of states (LDOS) as well as the band structure and eigenstates of the Hamiltonian through exact diagonaliztion. In order to do so, however, the mean-field parameters of the model are first found by solving self-consistency equations, which can be obtained by minimization of the free energy.

## 4.1 Saddle-point approximation to the slave-boson representation

Taking the simplest form, saddle point approximation or slave-boson mean-field of the infinite-$U$ Anderson lattice model consists of replacing slave-boson field ($b_i$) at each lattice site by the modulus of its expectation value [15]. Besides, assuming translational invariance in the bulk, slave-boson field ($b_i$) and auxilary chemical potential ($\lambda_i$) can be constant throughout the lattice. Thus, one can write

$$\begin{cases} b_i = \langle b_i \rangle = b \\ b_i^\dagger = \langle b_i^\dagger \rangle = b \end{cases} , \quad \lambda_i = \lambda$$

where without loss of generality we assumed that $b$ is real-valued.
Therefore, the mean-field Hamiltonian of TKI may be expressed in momentum space as

$$\overline{\mathcal{H}} = \sum_{\mathbf{k},\sigma} \epsilon^c_{\mathbf{k}} c^\dagger_{\mathbf{k}\sigma} c_{\mathbf{k}\sigma} + \sum_{\mathbf{k},\alpha} \left(\tilde{\epsilon}^f_{\mathbf{k}} + \lambda\right) f^\dagger_{\mathbf{k}\alpha} f_{\mathbf{k}\alpha} + V_0 b \sum_{\mathbf{k},\sigma,\alpha} \left(c^\dagger_{\mathbf{k}\sigma} \Phi_{\sigma\alpha}(\mathbf{k}) f_{\mathbf{k}\alpha} + H.c.\right) + N_f \lambda \left(b^2 - 1\right), \tag{4.1}$$

where $N_f$ is the number of lattice sites for $f$ electrons and

$$\begin{cases} \epsilon^c_{\mathbf{k}} = -2t^c \sum_{l=x,y,z} \cos(k_l) \\ \tilde{\epsilon}^f_{\mathbf{k}} = -2t^f b^2 \sum_{l=x,y,z} \cos(k_l) = b^2 \epsilon^f_{\mathbf{k}} \end{cases} \tag{4.2}$$

are the dispersion of $c$- and $f$-electrons, respectively. The dispersion of the $f$-electron is renormalized by $\lambda$ and its hopping amplitude $t^f$ by slave-boson $b$. Moreover, the hybridization amplitude is also renormalized by the $b$. One should also note that $\Phi(\mathbf{k})^\dagger = \Phi(\mathbf{k})$, since the form factor is a Hermitian operator.

## 4.2 Mean-field equations of the topological Kondo insulators in the bulk

To study the system at the mean-field level, we propose the grand canonical free energy of the TKI as follows

$$\Omega = \langle\overline{\mathcal{H}}\rangle - TS - \mu\Big((N_c + N_f) - N_i\Big) - \eta(N_c - N_f), \tag{4.3}$$

where $\langle\overline{\mathcal{H}}\rangle$ is the thermal average of the TKI mean-field Hamiltonian. $S$ is the entropy of the system at finite temperature. Third term is the constraint that fixes the total particle number ($N = N_c + N_f$). Fourth term is the constraint that enforces Kondo insulator with equal conduction and $f$ electrons ($N_c = N_f$). One should note that the temperature range that slave-boson mean-theory is valid, limited to very low values ($T \to 0$). Therefore, the contribution of entropy to free energy can be ignored in this mean-field. Moreover, we make a convention and absorb all Lagrange multipliers into the mean-field Hamiltonian. In this way, the Hamiltonian thermal average becomes

$$\langle\overline{\mathcal{H}}\rangle = \sum_{\mathbf{k},\sigma} \xi^c_{\mathbf{k}} \langle c^\dagger_{\mathbf{k}\sigma} c_{\mathbf{k}\sigma}\rangle + \sum_{\mathbf{k},\alpha} \Big(\xi^f_{\mathbf{k}} + \lambda\Big) \langle f^\dagger_{\mathbf{k}\alpha} f_{\mathbf{k}\alpha}\rangle + V_0 b \sum_{\mathbf{k},\sigma,\alpha} \Big(\Phi(\mathbf{k})_{\sigma\alpha} \langle c^\dagger_{\mathbf{k}\sigma} f_{\mathbf{k}\alpha}\rangle + H.c.\Big) + N_f \lambda \Big(b^2 - 1\Big), \tag{4.4}$$

where

$$\begin{cases} \xi^c_{\mathbf{k}} = -2t^c \sum_{l=x,y,z} \cos(k_l) - \mu - \eta, \\ \xi^f_{\mathbf{k}} = -2t^f b^2 \sum_{l=x,y,z} \cos(k_l) - \mu + \eta \end{cases} . \tag{4.5}$$

The mean-field equations can be obtained by minimizing the thermal average of the mean-field Hamiltonian with respect to the mean-field parameters. In the slave-boson mean-field theory of the TKI, mean-field parameters are slave-boson ($b$) and a set of Lagrange multipliers, involving auxilary chemical potentials ($\lambda$) and ($\eta$). So that

$$\frac{\partial\langle\overline{\mathcal{H}}\rangle}{\partial b} = 0, \qquad \frac{\partial\langle\overline{\mathcal{H}}\rangle}{\partial \lambda} = 0, \qquad \frac{\partial\langle\overline{\mathcal{H}}\rangle}{\partial \eta} = 0. \tag{4.6}$$

In this mean-field theory, we do not take the physical chemical potential ($\mu$) as a mean-field parameter, but it is treated as a free parameter. Therefore, the mean-field equations are

$$\boxed{\frac{1}{N_f} \sum_{\mathbf{k},\sigma,\alpha} \Phi_{\sigma\alpha}(\mathbf{k}) \Big(\langle c^\dagger_{\mathbf{k}\sigma} f_{\mathbf{k}\alpha}\rangle + \langle f^\dagger_{\mathbf{k}\alpha} c_{\mathbf{k}\sigma}\rangle\Big) - \frac{1}{N_f} \sum_{\mathbf{k},\alpha} 2b\epsilon^f_{\mathbf{k}} \langle f^\dagger_{\mathbf{k}\alpha} f_{\mathbf{k}\alpha}\rangle + 2\lambda b = 0} \qquad \text{(b-eq.)}, \tag{4.7}$$

similarly for $\lambda$

$$\boxed{\frac{1}{N_f} \sum_{\mathbf{k},\alpha} \langle f^\dagger_{\mathbf{k}\alpha} f_{\mathbf{k}\alpha}\rangle + b^2 - 1 = 0} \qquad (\lambda\text{-eq.}), \tag{4.8}$$

and for $\eta$

$$\boxed{\sum_{\mathbf{k},\alpha}\langle f^{\dagger}_{\mathbf{k}\alpha} f_{\mathbf{k}\alpha}\rangle - \sum_{\mathbf{k},\sigma}\langle c^{\dagger}_{\mathbf{k}\sigma} c_{\mathbf{k}\sigma}\rangle = 0} \qquad (\eta\text{-eq.}). \tag{4.9}$$

Basically, the equation for $\lambda$ is the holonomic constraint ($Q = n_b + n_f = 1$) on the infinite-$U$ Anderson lattice model that enforces single occupancy of the impurity (valence) band. This equation can be interpreted as fluctuation/deviation of the $f$-band from perfect half-filling by the amount of $n_b$ (slave-boson particle number).
In order to have a Kondo insulator, formation of singlet states between $c$- and $f$- electrons is needed at each lattice site. This means that the number of $c$- and $f$-electrons are equal on average. The equation over $\eta$ imposesm this condition to the system. It shifts all bands to the position of physical chemical potential ($\mu$), in such a way that it lies inside the gap.

## 4.3 The equation of motion method

To solve the self-consistency equations for $(b, \lambda, \eta)$ one needs to know about certain thermal averages like $\langle c^{\dagger}_{\mathbf{k}\sigma} c_{\mathbf{k}\sigma}\rangle$ , $\langle f^{\dagger}_{\mathbf{k}\alpha} f_{\mathbf{k}\alpha}\rangle$ , $\langle c^{\dagger}_{\mathbf{k}\sigma} f_{\mathbf{k}\alpha}\rangle$ and $\langle f^{\dagger}_{\mathbf{k}\alpha} c_{\mathbf{k}\sigma}\rangle$ that appear in the equations. It is straightforward to show that these thermal averages are average occupation numbers, and can be expressed in terms of Green's functions as follows,

$$\begin{cases} \langle c^{\dagger}_{\mathbf{k}\sigma} c_{\mathbf{k}\sigma}\rangle = \int_{-\infty}^{+\infty} d\omega \;\; A_{cc}(\mathbf{k},\omega;\sigma,\sigma) n_F(\omega) = \frac{1}{\pi}\int_{-\infty}^{+\infty} d\omega \;\; Im\left(G^{A}_{cc}(\mathbf{k},\omega;\sigma,\sigma)\right) n_F(\omega) \\ \langle f^{\dagger}_{\mathbf{k}\alpha} f_{\mathbf{k}\alpha}\rangle = \int_{-\infty}^{+\infty} d\omega \;\; A_{ff}(\mathbf{k},\omega;\alpha,\alpha) n_F(\omega) = \frac{1}{\pi}\int_{-\infty}^{+\infty} d\omega \;\; Im\left(G^{A}_{ff}(\mathbf{k},\omega;\alpha,\alpha)\right) n_F(\omega) \\ \langle c^{\dagger}_{\mathbf{k}\sigma} f_{\mathbf{k}\alpha}\rangle = \int_{-\infty}^{+\infty} d\omega \;\; A_{fc}(\mathbf{k},\omega;\alpha,\sigma) n_F(\omega) = \frac{1}{\pi}\int_{-\infty}^{+\infty} d\omega \;\; Im\left(G^{A}_{fc}(\mathbf{k},\omega;\alpha,\sigma)\right) n_F(\omega) \\ \langle f^{\dagger}_{\mathbf{k}\alpha} c_{\mathbf{k}\sigma}\rangle = \int_{-\infty}^{+\infty} d\omega \;\; A_{cf}(\mathbf{k},\omega;\sigma,\alpha) n_F(\omega) = \frac{1}{\pi}\int_{-\infty}^{+\infty} d\omega \;\; Im\left(G^{A}_{cf}(\mathbf{k},\omega;\sigma,\alpha)\right) n_F(\omega) \end{cases} , \tag{4.10}$$

where $n_F(\omega)$ is the Fermi distribution function [42]. Therefore, one needs to find the advanced Green's function, which can be done analytically, first by calculating the thermal Green's function in the imaginary time domain, then Fourier transforming to the frequency domain (Matsubara space) and eventually performing the analytic continuation ($i\omega_n \to \omega - i\delta$) to the complex plane, where $\delta$ is an infinitesimal number. Following this strategy, the time-translational invariant thermal Green's function in the basis of $\Psi_{\mathbf{k}} = \left(c_{\mathbf{k}\uparrow}, c_{\mathbf{k}\downarrow}, f_{\mathbf{k}\uparrow}, f_{\mathbf{k}\downarrow}\right)$ is

$$\begin{aligned} \mathcal{G}(\mathbf{k},\tau) &= -\langle \hat{\mathcal{T}} \Psi_{\mathbf{k}}(\tau) \Psi^{\dagger}_{\mathbf{k}}(0)\rangle \\ &= -\langle \hat{\mathcal{T}} \begin{bmatrix} c_{\mathbf{k}\uparrow}(\tau) \\ c_{\mathbf{k}\downarrow}(\tau) \\ f_{\mathbf{k}\uparrow}(\tau) \\ f_{\mathbf{k}\downarrow}(\tau) \end{bmatrix} \left[ c^{\dagger}_{\mathbf{k}\uparrow}(0)\, c^{\dagger}_{\mathbf{k}\downarrow}(0)\, f^{\dagger}_{\mathbf{k}\uparrow}(0)\, f^{\dagger}_{\mathbf{k}\downarrow}(0) \right]\rangle \\ &= \left[\begin{array}{cc|cc} \mathcal{G}_{cc}(\mathbf{k},\tau;\uparrow\uparrow) & \mathcal{G}_{cc}(\mathbf{k},\tau;\uparrow\downarrow) & \mathcal{G}_{cf}(\mathbf{k},\tau;\uparrow\uparrow) & \mathcal{G}_{cf}(\mathbf{k},\tau;\uparrow\downarrow) \\ \mathcal{G}_{cc}(\mathbf{k},\tau;\downarrow\uparrow) & \mathcal{G}_{cc}(\mathbf{k},\tau;\downarrow\downarrow) & \mathcal{G}_{cf}(\mathbf{k},\tau;\downarrow\uparrow) & \mathcal{G}_{cf}(\mathbf{k},\tau;\downarrow\downarrow) \\ \hline \mathcal{G}_{fc}(\mathbf{k},\tau;\uparrow\uparrow) & \mathcal{G}_{fc}(\mathbf{k},\tau;\uparrow\downarrow) & \mathcal{G}_{ff}(\mathbf{k},\tau;\uparrow\uparrow) & \mathcal{G}_{ff}(\mathbf{k},\tau;\uparrow\downarrow) \\ \mathcal{G}_{fc}(\mathbf{k},\tau;\downarrow\uparrow) & \mathcal{G}_{fc}(\mathbf{k},\tau;\downarrow\downarrow) & \mathcal{G}_{ff}(\mathbf{k},\tau;\downarrow\uparrow) & \mathcal{G}_{ff}(\mathbf{k},\tau;\downarrow\downarrow) \end{array}\right], \end{aligned} \tag{4.11}$$

where $\tau$ is the imaginary time and $\hat{\mathcal{T}}$ is the imaginary time-ordering operator. It is straightforward to check that for the given Hamiltonian, there shouldn't be any spin flip for the ($cc$) and ($ff$) electrons in the corresponding Green's function. Therefore, the full Green's fuction (4.11) can be simplified to

$$\mathcal{G}(\mathbf{k},\tau)=\left[\begin{array}{c|c}\mathcal{G}^{(2\times2)}_{cc}(\mathbf{k},\tau) & \mathcal{G}^{(2\times2)}_{cf}(\mathbf{k},\tau)\\ \hline \mathcal{G}^{(2\times2)}_{fc}(\mathbf{k},\tau) & \mathcal{G}^{(2\times2)}_{ff}(\mathbf{k},\tau)\end{array}\right]. \tag{4.12}$$

By Fourier transforming (4.12) to the frequency space, one would obtain

$$\mathcal{G}(\mathbf{k},i\omega_n)=\frac{1}{\beta}\sum_{\tau}e^{-i\omega_n\tau}\mathcal{G}(\mathbf{k},\tau)=\left[\begin{array}{c|c}\mathcal{G}^{(2\times2)}_{cc}(\mathbf{k},i\omega_n) & \mathcal{G}^{(2\times2)}_{cf}(\mathbf{k},i\omega_n)\\ \hline \mathcal{G}^{(2\times2)}_{fc}(\mathbf{k},i\omega_n) & \mathcal{G}^{(2\times2)}_{ff}(\mathbf{k},i\omega_n)\end{array}\right], \tag{4.13}$$

where $\frac{1}{\beta}=k_BT$ is the inverse thermal energy and $\omega_n=\frac{\pi}{\beta}(2n+1)$ with $n\in\mathbb{Z}$ is the fermionic Matsubara frequency. Now one should find the corresponding Matsubara Green's functions through the equation of motion method. To do so, we start from the matrix representation of the mean-field Hamiltonian in momentum space, including all chemical potentials, written in the given basis

$$\overline{\mathcal{H}}(\mathbf{k})=\sum_{\mathbf{k}}\begin{bmatrix}c^\dagger_{\mathbf{k}\uparrow}\\ c^\dagger_{\mathbf{k}\downarrow}\\ f^\dagger_{\mathbf{k}\uparrow}\\ f^\dagger_{\mathbf{k}\downarrow}\end{bmatrix}^T\left[\begin{array}{cc|cc}\xi^c_{\mathbf{k}} & 0 & V_0bS^z_k & V_0b\left(S^x_k-iS^y_k\right)\\ 0 & \xi^c_{\mathbf{k}} & V_0b\left(S^x_k+iS^y_k\right) & -V_0bS^z_k\\ \hline V_0bS^z_k & V_0b\left(S^x_k-iS^y_k\right) & \xi^f_{\mathbf{k}}+\lambda & 0\\ V_0b\left(S^x_k+iS^y_k\right) & -V_0bS^z_k & 0 & \xi^f_{\mathbf{k}}+\lambda\end{array}\right]\begin{bmatrix}c_{\mathbf{k}\uparrow}\\ c_{\mathbf{k}\downarrow}\\ f_{\mathbf{k}\uparrow}\\ f_{\mathbf{k}\downarrow}\end{bmatrix}+N_f\lambda\left(b^2-1\right), \tag{4.14}$$

where the constant term $N_f\lambda\left(b^2-1\right)$ can be absorbed into the formalism as a constant shift of the ground state energy. One can see that $\mathcal{H}(\mathbf{k})$ is a $4\times4$ matrix with $2\times2$ blocks. Consequently, the eigenvalues of $\mathcal{H}(\mathbf{k})$ are

$$E^{\pm}=\frac{1}{2}(\xi^c_{\mathbf{k}}+\xi^f_{\mathbf{k}}+\lambda)\pm\sqrt{\left(\frac{\xi^c_{\mathbf{k}}-\xi^f_{\mathbf{k}}-\lambda}{2}\right)^2+3V_0^2b^2\Phi(\mathbf{k})^2}, \tag{4.15}$$

which means that the matrix $\mathcal{H}(\mathbf{k})$ has four eigenvalues with two degenerate ones. Moreover, one can show that $\Phi(\mathbf{k})^2$ is scalar, like the following

$$\Phi(\mathbf{k})^2=(\vec{S}_{\mathbf{k}}.\vec{\sigma})(\vec{S}_{\mathbf{k}}.\vec{\sigma})=(\vec{S}_{\mathbf{k}}.\vec{S}_{\mathbf{k}})+\underbrace{(\vec{S}_{\mathbf{k}}\times\vec{S}_{\mathbf{k}})}_{=0}.\vec{\sigma}=|S_{\mathbf{k}}|^2\mathbb{1}_{2\times2}. \tag{4.16}$$

The equation of motion in Matsubara space is

$$(i\omega_n\mathbb{1}_{4\times4}-\mathcal{H}(\mathbf{k}))\,\mathcal{G}(\mathbf{k},i\omega_n)=\mathbb{1}_{4\times4}\ , \tag{4.17}$$

where $\mathcal{H}(\mathbf{k})$ is the Hamiltonian matrix. Writting this equation in a matrix representation as

$$\left(\begin{bmatrix}i\omega_n\mathbb{1}_{2\times2} & 0\\ 0 & i\omega_n\mathbb{1}_{2\times2}\end{bmatrix}-\begin{bmatrix}\xi^c_{\mathbf{k}}\mathbb{1}_{2\times2} & V_0b\Phi(\mathbf{k})\\ V_0b\Phi(\mathbf{k}) & \left(\xi^f_{\mathbf{k}}+\lambda\right)\mathbb{1}_{2\times2}\end{bmatrix}\right)\begin{bmatrix}\mathcal{G}^{(2\times2)}_{cc}(\mathbf{k},i\omega_n) & \mathcal{G}^{(2\times2)}_{cf}(\mathbf{k},i\omega_n)\\ \mathcal{G}^{(2\times2)}_{fc}(\mathbf{k},i\omega_n) & \mathcal{G}^{(2\times2)}_{ff}(\mathbf{k},i\omega_n)\end{bmatrix}=\begin{bmatrix}1 & 0\\ 0 & 1\end{bmatrix}. \tag{4.18}$$

The matrix equation (4.18) can be expanded more elaborately in terms of four algebraic equations

$$\begin{cases}(i\omega_n - \xi^c_{\mathbf{k}})\,\mathbb{1}_{2\times2}\mathcal{G}^{(2\times2)}_{cc}(\mathbf{k},i\omega_n) - V_0 b\Phi(\mathbf{k})\mathcal{G}^{(2\times2)}_{fc}(\mathbf{k},i\omega_n) = 1\\ -V_0 b\Phi(\mathbf{k})\mathcal{G}^{(2\times2)}_{cc}(\mathbf{k},i\omega_n) + \left[i\omega_n - \left(\xi^f_{\mathbf{k}} + \lambda\right)\right]\mathbb{1}_{2\times2}\mathcal{G}^{(2\times2)}_{fc}(\mathbf{k},i\omega_n) = 0\end{cases} \tag{4.19}$$

and

$$\begin{cases}\left[i\omega_n - \left(\xi^f_{\mathbf{k}} + \lambda\right)\right]\mathbb{1}_{2\times2}\mathcal{G}^{(2\times2)}_{ff}(\mathbf{k},i\omega_n) - V_0 b\Phi(\mathbf{k})\mathcal{G}^{(2\times2)}_{cf}(\mathbf{k},i\omega_n) = 1\\ -V_0 b\Phi(\mathbf{k})\mathcal{G}^{(2\times2)}_{ff}(\mathbf{k},i\omega_n) + (i\omega_n - \xi^c_{\mathbf{k}})\,\mathbb{1}_{2\times2}\mathcal{G}^{(2\times2)}_{cf}(\mathbf{k},i\omega_n) = 0\end{cases} \tag{4.20}$$

where it is obvious that each of these two sets of equations is independent of the other; so that the question of finding different Matsubara Green's function reduces to solving equation set (4.19) and (4.20) algebraically as follows

$$\begin{cases}\mathcal{G}^{(2\times2)}_{cc}(\mathbf{k},i\omega_n) = \dfrac{1}{i\omega_n - \xi^c_{\mathbf{k}} - V_0^2 b^2 |S_{\mathbf{k}}|^2 \left(i\omega_n - (\xi^f_{\mathbf{k}}+\lambda)\right)^{-1}}\\ \mathcal{G}^{(2\times2)}_{fc}(\mathbf{k},i\omega_n) = \dfrac{V_0 b\Phi(\mathbf{k})}{\left(i\omega_n - (\xi^f_{\mathbf{k}}+\lambda)\right)\left(i\omega_n - \xi^c_{\mathbf{k}}\right) - V_0^2 b^2 |S_{\mathbf{k}}|^2}\\ \mathcal{G}^{(2\times2)}_{cf}(\mathbf{k},i\omega_n) = \dfrac{V_0 b\Phi(\mathbf{k})}{\left(i\omega_n - (\xi^f_{\mathbf{k}}+\lambda)\right)\left(i\omega_n - \xi^c_{\mathbf{k}}\right) - V_0^2 b^2 |S_{\mathbf{k}}|^2}\\ \mathcal{G}^{(2\times2)}_{ff}(\mathbf{k},i\omega_n) = \dfrac{1}{i\omega_n - (\xi^f_{\mathbf{k}}+\lambda) - V_0^2 b^2 |S_{\mathbf{k}}|^2 \left(i\omega_n - \xi^c_{\mathbf{k}}\right)^{-1}}\end{cases}. \tag{4.21}$$

It is obvious that $\mathcal{G}_{cf}$ is equal to $\mathcal{G}_{fc}$. Besides, by taking into account that all Green's functions are $(2\times2)$ matrices, the superscript $(2\times2)$ will be dropped from here on.
The corresponding self-energies of the $\mathcal{G}_{cc}$ and $\mathcal{G}_{ff}$ in (4.21) can be written in terms of the bare $c$ and $f$ Green's functions

$$\begin{cases}\Sigma_{cc}(\mathbf{k},i\omega_n) = V_0^2 b^2 \Phi(\mathbf{k})^2 \left(i\omega_n - (\xi^f_{\mathbf{k}}+\lambda)\right)^{-1} = V_0^2 b^2 |S_{\mathbf{k}}|^2 \mathcal{G}^{(0)}_{ff}(\mathbf{k},i\omega_n)\\ \Sigma_{ff}(\mathbf{k},i\omega_n) = V_0^2 b^2 \Phi(\mathbf{k})^2 \left(i\omega_n - \xi^c_{\mathbf{k}}\right)^{-1} = V_0^2 b^2 |S_{\mathbf{k}}|^2 \mathcal{G}^{(0)}_{cc}(\mathbf{k},i\omega_n)\end{cases}. \tag{4.22}$$

One can see that $\mathcal{G}_{cc}$ and $\mathcal{G}_{ff}$ have a simple pole, and $\mathcal{G}_{fc}$ has a double pole in the complex plane. To avoid further complexities in the analytic continuation, it is always recommended to work with simple-poled functions as far as possible; hence before performing the prescription ($i\omega_n \to \omega - i\delta$), one can decompose the double pole of $\mathcal{G}_{fc}$ into two simple pole, using partial fraction decomposition method as follows

$$\mathcal{G}_{fc}(\mathbf{k},i\omega_n) = \frac{V_0 b\Phi(\mathbf{k})}{\left(i\omega_n - (\xi^f_{\mathbf{k}}+\lambda)\right)\left(i\omega_n - \xi^c_{\mathbf{k}}\right) - V_0^2 b^2 |S_{\mathbf{k}}|^2} = \frac{V_0 b\Phi(\mathbf{k})}{\left(i\omega_n - \mathcal{W}^+_{\mathbf{k}}\right)\left(i\omega_n - \mathcal{W}^-_{\mathbf{k}}\right)}, \tag{4.23}$$

where

$$\mathcal{W}^{\pm}_{\mathbf{k}} = \frac{1}{2}\left[\left(\xi^c_{\mathbf{k}} + \xi^f_{\mathbf{k}} + \lambda\right) \pm \sqrt{\left(\xi^c_{\mathbf{k}} - \xi^f_{\mathbf{k}} - \lambda\right)^2 + 4b^2 V_0^2 |S_{\mathbf{k}}|^2}\right]. \tag{4.24}$$

Now by using the identity

$$\frac{1}{(x-a)(x-b)} = \frac{1}{a-b}\left(\frac{1}{x-a} - \frac{1}{x-b}\right), \tag{4.25}$$

the equation (4.23) becomes

$$\mathcal{G}_{fc}(\mathbf{k},i\omega_n) = \frac{V_0 b\Phi(\mathbf{k})}{\mathcal{W}_{\mathbf{k}}{}^+ - \mathcal{W}_{\mathbf{k}}{}^-}\left(\frac{1}{i\omega_n - \mathcal{W}_{\mathbf{k}}{}^+} - \frac{1}{i\omega_n - \mathcal{W}_{\mathbf{k}}{}^-}\right). \tag{4.26}$$

Having all Green's functions in a convenient form, one can apply the prescription first on the self-energies

$$\begin{cases}\Sigma_{cc}(\mathbf{k}, i\omega_n) \\ \Sigma_{ff}(\mathbf{k}, i\omega_n)\end{cases} \rightarrow \begin{cases}\Sigma^A_{cc}(\mathbf{k},\omega) = V_0^2 b^2 |S_\mathbf{k}|^2 \left(\omega - (\xi^f_\mathbf{k} + \lambda) - i\delta\right)^{-1} = V_0^2 b^2 |S_\mathbf{k}|^2 G^{(0)A}_{ff}(\mathbf{k},\omega) \\ \Sigma^A_{ff}(\mathbf{k},\omega) = V_0^2 b^2 |S_\mathbf{k}|^2 (\omega - \xi^c_\mathbf{k} - i\delta)^{-1} = V_0^2 b^2 |S_\mathbf{k}|^2 G^{(0)A}_{cc}(\mathbf{k},\omega)\end{cases}, \tag{4.27}$$

with the real and imaginary part of

$$\begin{cases} Re\,\Sigma^A_{cc}(\mathbf{k},\omega) = \frac{V_0^2 b^2 |S_\mathbf{k}|^2 \left(\omega - (\xi^f_\mathbf{k}+\lambda)\right)}{\left(\omega-(\xi^f_\mathbf{k}+\lambda)\right)^2+\delta^2} \\ Im\,\Sigma^A_{cc}(\mathbf{k},\omega) = \frac{V_0^2 b^2 |S_\mathbf{k}|^2 \delta}{\left(\omega-(\xi^f_\mathbf{k}+\lambda)\right)^2+\delta^2}\end{cases}, \begin{cases} Re\,\Sigma^A_{ff}(\mathbf{k},\omega) = \frac{V_0^2 b^2 |S_\mathbf{k}|^2 (\omega-\xi^c_\mathbf{k})}{(\omega-\xi^c_\mathbf{k})^2+\delta^2} \\ Im\,\Sigma^A_{ff}(\mathbf{k},\omega) = \frac{V_0^2 b^2 |S_\mathbf{k}|^2 \delta}{(\omega-\xi^c_\mathbf{k})^2+\delta^2}\end{cases}; \tag{4.28}$$

then for the Green's functions

$$\begin{cases} G^A_{cc}(\mathbf{k},\omega) = \frac{1}{\omega-\xi^c_\mathbf{k}-\Sigma^A_{cc}(\mathbf{k},\omega)} \\ G^A_{fc}(\mathbf{k},\omega) = \frac{V_0 b \Phi(\mathbf{k})}{\mathcal{W}_\mathbf{k}^+ - \mathcal{W}_\mathbf{k}^-}\left(\frac{1}{\omega-\mathcal{W}_\mathbf{k}^+-i\delta} - \frac{1}{\omega-\mathcal{W}_\mathbf{k}^- - i\delta}\right) \\ G^A_{ff}(\mathbf{k},\omega) = \frac{1}{\omega-(\xi^f_\mathbf{k}+\lambda)-\Sigma^A_{ff}(\mathbf{k},\omega)}\end{cases}, \tag{4.29}$$

where one should note that the imaginary infinitesimal number ($i\delta$) has been not added to the denominator of the $G^A_{cc}$ and $G^A_{ff}$, since their self-energies are already complex numbers, which guarantee that the poles of the advanced Green's function are located at the upper-half plane.

Having an explicit form for $G^A_{fc}$ makes it possible to simplify more the mean-field equation for $b$ (4.7), which impacts the numerics performance dramatically. This extra calculation is done in the appendix (A) of the thesis.
To sum up, the self-consistency equations loop in terms of Green's functions are

$$\boxed{\begin{aligned} &\frac{1}{N_f}\sum_\mathbf{k} |S_\mathbf{k}|^2 \frac{n_F(\mathcal{W}_\mathbf{k}^+) - n_F(\mathcal{W}_\mathbf{k}^-)}{\mathcal{W}_\mathbf{k}^+ - \mathcal{W}_\mathbf{k}^-} - \frac{1}{\pi N_f}\sum_{\mathbf{k},\alpha}\xi^f_\mathbf{k}\int_{-\infty}^{+\infty} d\omega\ Im\left(G^A_{ff}(\mathbf{k},\omega;\alpha,\alpha)\right) n_F(\omega) + \lambda = 0 \\ &\frac{1}{\pi N_f}\sum_{\mathbf{k},\alpha}\int_{-\infty}^{+\infty} d\omega\ Im\left(G^A_{ff}(\mathbf{k},\omega;\alpha,\alpha)\right) n_F(\omega) + b^2 - 1 = 0 \\ &\frac{1}{\pi}\sum_{\mathbf{k},\alpha}\int_{-\infty}^{+\infty} d\omega\ Im\left(G^A_{ff}(\mathbf{k},\omega;\alpha,\alpha)\right) n_F(\omega) - \frac{1}{\pi}\sum_{\mathbf{k},\sigma}\int_{-\infty}^{+\infty} d\omega\ Im\left(G^A_{cc}(\mathbf{k},\omega;\sigma,\sigma)\right) n_F(\omega) = 0 \\ &\begin{cases} G^A_{cc}(\mathbf{k},\omega;\sigma,\sigma) = \frac{1}{\omega-\xi^c_\mathbf{k}-\Sigma^A_{cc}(\mathbf{k},\omega)} \\ G^A_{ff}(\mathbf{k},\omega;\alpha,\alpha) = \frac{1}{\omega-(\xi^f_\mathbf{k}+\lambda)-\Sigma^A_{ff}(\mathbf{k},\omega)}\end{cases} \\ &\begin{cases} Re\,\Sigma^A_{cc}(\mathbf{k},\omega) = \frac{V_0^2 b^2 |S_\mathbf{k}|^2 \left(\omega - (\xi^f_\mathbf{k}+\lambda)\right)}{\left(\omega-(\xi^f_\mathbf{k}+\lambda)\right)^2+\delta^2} \\ Im\,\Sigma^A_{cc}(\mathbf{k},\omega) = \frac{V_0^2 b^2 |S_\mathbf{k}|^2 \delta}{\left(\omega-(\xi^f_\mathbf{k}+\lambda)\right)^2+\delta^2}\end{cases} \begin{cases} Re\,\Sigma^A_{ff}(\mathbf{k},\omega) = \frac{V_0^2 b^2 |S_\mathbf{k}|^2 (\omega-\xi^c_\mathbf{k})}{(\omega-\xi^c_\mathbf{k})^2+\delta^2} \\ Im\,\Sigma^A_{ff}(\mathbf{k},\omega) = \frac{V_0^2 b^2 |S_\mathbf{k}|^2 \delta}{(\omega-\xi^c_\mathbf{k})^2+\delta^2}\end{cases} \\ &\mathcal{W}^\pm_\mathbf{k} = \frac{1}{2}\left[\left(\xi^c_\mathbf{k} + \xi^f_\mathbf{k} + \lambda\right) \pm \sqrt{\left(\xi^c_\mathbf{k} - \xi^f_\mathbf{k} - \lambda\right)^2 + 4b^2V_0^2|S_\mathbf{k}|^2}\right]\end{aligned}}. \tag{4.30}$$

## 4.4 Numerical notes

To find the corresponding values of the mean-field parameters $(b, \lambda, \eta)$, one should solve the self-consistency integral equations (4.30), which consist a triple integral over momentum and a single integral over frequency, iteratively. Obviously, it is numerically too costly and impractical in an appropriate time extent if one does the job with a brute-force method. Therefore, some approximations should be implemented. We assume an isotropic form factor $\Phi(\mathbf{k})$ and dispersions for the bulk, which is a factual assumption, since we have a cubic lattice with translational invariance for all dimensions.
One can also try to write the discrete summation over momentum as an integral, with the following normalization factors

$$\frac{1}{N}\sum_{\mathbf{k}} \to \frac{1}{\mathcal{V}_{BZ}}\int_{BZ} d^3k, \tag{4.31}$$

where $\mathcal{V}_{BZ}$ is the volume of the Brillouin zone.
Having this assumption, the momentum integration becomes (4.31)

$$\frac{1}{\mathcal{V}_{BZ}}\int_{BZ} d^3k = \frac{1}{\mathcal{V}_{BZ}}\underbrace{\int_{\Omega} d\Omega}_{=4\pi}\int_0^{k_{BZ}} k^2 dk \tag{4.32}$$

and by defining $\kappa := \frac{k}{k_{BZ}}$, one can make the integral to be normalized to 1 as follows

$$\frac{1}{\mathcal{V}_{BZ}}\int_{BZ} d^3k = \underbrace{\frac{4\pi k_{BZ}^3}{\mathcal{V}_{BZ}}}_{=3}\int_0^1 \kappa^2 d\kappa, \tag{4.33}$$

where the volume of the Brillouin zone in spherical space is assumed to be $V_{BZ} = \frac{4}{3}\pi k_{BZ}^3$. Besides, one should note if we had used Cartesian integration, the volume of Brillouin zone would be $k_{BZ}^3 = (\frac{2\pi}{a})^3$, with $a$ being the lattice constant.

## 4.5 Results

### 4.5.1 Bulk dispersion

Renormalized band structure of the TKI in the bulk can be obtained from diagonalization of the Hamiltonian. The result shows an insulating bulk with a very narrow hybridization gap. One can see that due to the hybridization, lower band takes the character of the conduction electrons and upper band takes the character of $f$ electrons, which are localized.

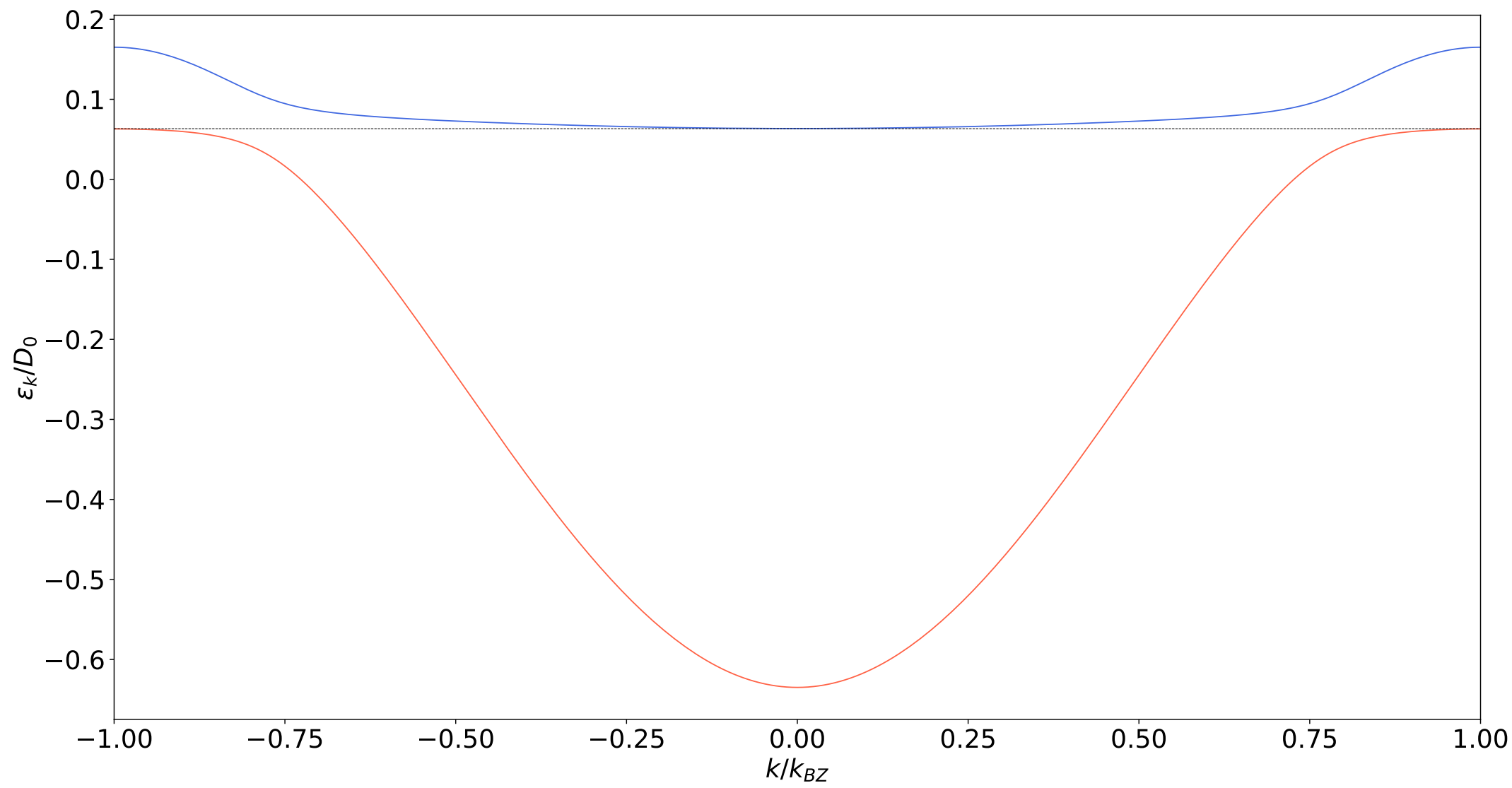

Figure 4.1: Bulk dispersion of the TKI slave-boson mean-field theory at temperature $T/D_0 = 10^{-4}$ with mean-field parameters $b = 0.463$, $\lambda/D_0 = 0.162$ and $\eta/D_0 = 0.127$. The parameters chosen are $V_0/D_0 = 0.1$, $E_0^f/D_0 = -0.2$, $t^c/D_0 = 0.2$, $t^f/D_0 = -0.001$, $\mu/D_0 = 0.063$ and $\delta = 10^{-4}$.

### 4.5.2 Local density of states (LDOS)

The renormalized local density of state can be obtained via

$$\rho(\omega) = \frac{1}{\pi N}\sum_{\mathbf{k}} ImG(\mathbf{k},\omega)e^{i\mathbf{k}\cdot\mathbf{0}} = \frac{1}{\pi N}\sum_{\mathbf{k}} ImG(\mathbf{k},\omega). \tag{4.34}$$

The result shows a very sharp Kondo peak and a very narrow hybridization gap arround ($\omega = 0$). Incomplete opening of the gap, specially in $\rho_f(\omega)$, is a numerical problem, due to the finite width ($\delta = 10^{-4}$) given to the Green's function.

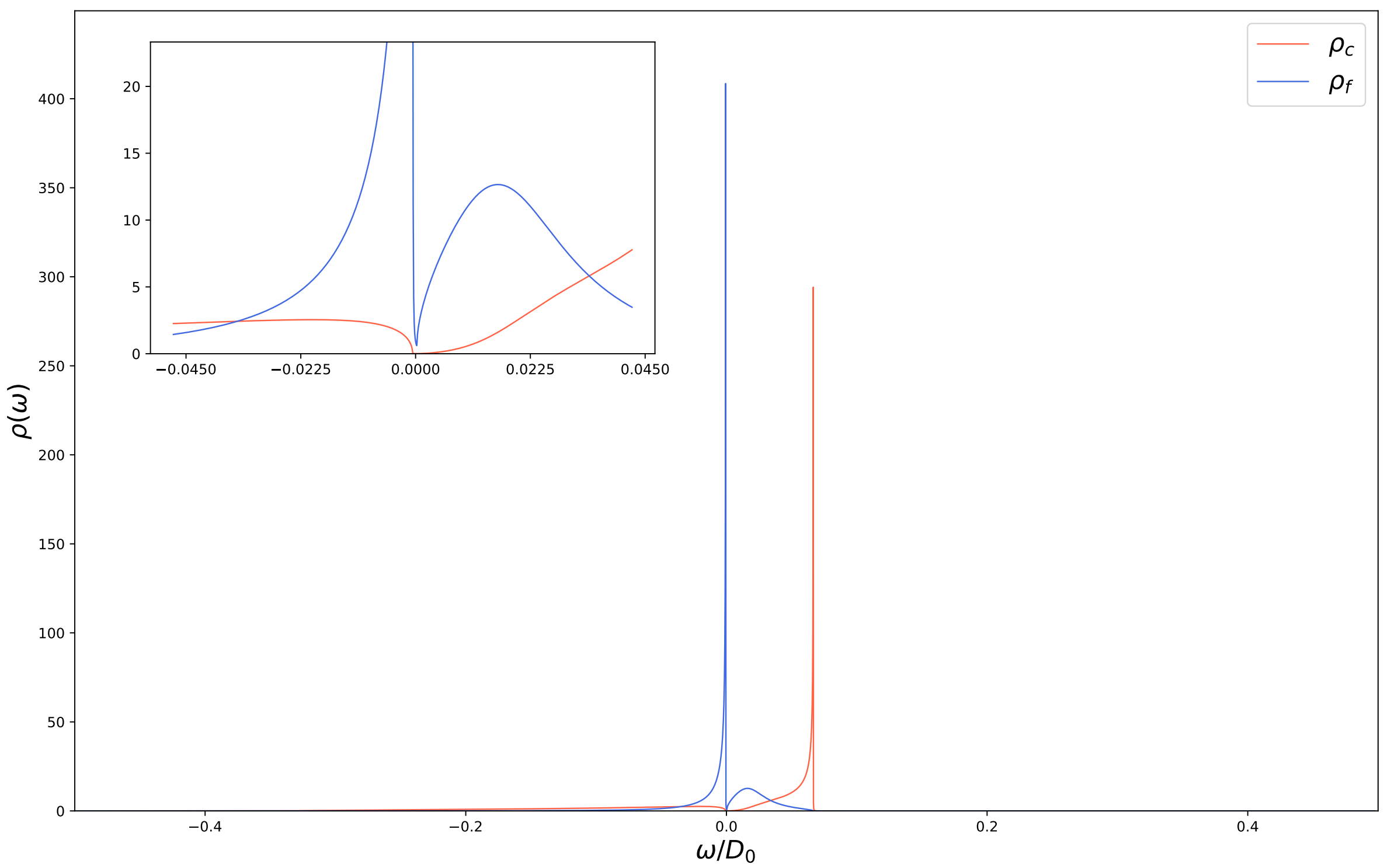

Figure 4.2: Renormalized LDOS of $c$ and $f$ electrons of the TKI at temperature $T/D_0 = 10^{-4}$ with mean-field parameters $b = 0.463$, $\lambda/D_0 = 0.162$ and $\eta/D_0 = 0.127$. The parameters chosen are $V_0/D_0 = 0.1$, $E_0^f/D_0 = -0.2$, $t^c/D_0 = 0.2$, $t^f/D_0 = -0.001$, $\mu/D_0 = 0.063$ and $\delta = 10^{-4}$.

CHAPTER 5

# Inhomogeneous slave-boson mean-field theory of a 3D TKI in a slab geometry

In this chapter, we develop a self-consistent theory for a 3D TKI in a slab geometry with broken periodic boundary condition (PBC) in $z-$axis and periodic in $xy-$plane. In this multilayer system, each layer is coupled to its neighboring layers by hoping and nonlocal hybridization form factor. This means that we should perform an inhomogeneous slave-boson mean-field theory. In comparison to the bulk case, where the mean-field parameters $(b, \lambda)$ are constant throughout the lattice, in such a layered-dependent system, mean-field parameters changes layer by layer. Consequently, one needs to adjust the bulk mean-field Hamiltonian in such a way that it incorporates a new eigenbasis and layer-dependent parameters.

## 5.1 Multilayer Hamiltonian of the bulk

One can imagine a multilayer system as a number of atomic monolayers extended into the $x-$ and $y-$direction that are stacked onto each other in $z-$direction. The $z-$direction is called perpendicular and any direction in the $xy$-planes is called parallel. We split up the lattice vectors into parallel and perpendicular directions $\mathbf{R} = (\mathbf{R}_{\|i}, n)$, where $n$ is the layer index. Likewise, the wavevectors read $\mathbf{k} = (\mathbf{k}_{\|i}, q)$. So one should note that the momentum $k$ is only a good quantum number in parallel plane and not in perpendicular plane. Therefore, a suitable eignbasis would be a direct product of a parallel momentum eigenstate and localized parallel state

$$|\mathbf{k}\rangle \to |\mathbf{k}_{\|}, n\rangle.$$

We start our analysis by Fourier transforming mean-field Hamiltonian of the bulk from momentum to real space in $z-$axis. The bulk Hamiltonian is

$$\overline{\mathcal{H}} = \sum_{\mathbf{k},\sigma} \xi^c_{\mathbf{k}} c^\dagger_{\mathbf{k}\sigma} c_{\mathbf{k}\sigma} + \sum_{\mathbf{k},\alpha} \left(\xi^f_{\mathbf{k}} + \lambda\right) f^\dagger_{\mathbf{k}\alpha} f_{\mathbf{k}\alpha} + \sum_{\mathbf{k},\sigma,\alpha} V_0 b \left(\Phi(\mathbf{k})_{\sigma\alpha} c^\dagger_{\mathbf{k}\sigma} f_{\mathbf{k}\alpha} + H.c.\right) + N_f \lambda \left(b^2 - 1\right),$$

where

$$\begin{cases} \xi^c_{\mathbf{k}} = -2t^c \sum_{l=x,y,z} cos(k_l) - \mu - \eta = \epsilon^c_{\mathbf{k}} - \mu - \eta \\ \xi^f_{\mathbf{k}} = -2b^2 t^f \sum_{l=x,y,z} cos(k_l) - \mu + \eta = \tilde{\epsilon}^f_{\mathbf{k}} - \mu + \eta = b^2 \epsilon^f_{\mathbf{k}} - \mu + \eta \\ \Phi(\mathbf{k}) = \sum_{l=x,y,z} S^l_k . \sigma^l \end{cases}.$$

One can write the matrix representation of the Hamiltonian in the space of spins ($\sigma = \{\uparrow, \downarrow\}, \alpha = \{\uparrow, \downarrow\}$) and species $(c, f)$ with the basis $(\psi^c_{\mathbf{k}}, \psi^f_{\mathbf{k}}) = (c_{\mathbf{k}\uparrow}, c_{\mathbf{k}\downarrow}, f_{\mathbf{k}\uparrow}, f_{\mathbf{k}\downarrow})$ as

$$\overline{\mathcal{H}} = \sum_{\mathbf{k}} \begin{bmatrix} c^\dagger_{\mathbf{k}\uparrow} \\ c^\dagger_{\mathbf{k}\downarrow} \\ f^\dagger_{\mathbf{k}\uparrow} \\ f^\dagger_{\mathbf{k}\downarrow} \end{bmatrix}^T \left[\begin{array}{c|c} \overline{\mathcal{H}}^{(2\times2)}_{cc}(k_x, k_y, k_z) & \overline{\mathcal{H}}^{(2\times2)}_{cf}(k_x, k_y, k_z) \\ \hline \overline{\mathcal{H}}^{(2\times2)}_{cf}(k_x, k_y, k_z) & \overline{\mathcal{H}}^{(2\times2)}_{ff}(k_x, k_y, k_z) \end{array}\right] \begin{bmatrix} c_{\mathbf{k}\uparrow} \\ c_{\mathbf{k}\downarrow} \\ f_{\mathbf{k}\uparrow} \\ f_{\mathbf{k}\downarrow} \end{bmatrix}, \tag{5.1}$$

where one should take into account that $\overline{\mathcal{H}}^{(2\times2)}_{cf}(k_x, k_y, k_z)$ is equal to $\overline{\mathcal{H}}^{(2\times2)}_{fc}(k_x, k_y, k_z)$, since the form factor $\Phi(\mathbf{k})$ consists of pauli matrices, which are Hermitian.
More elaborately, the $2 \times 2$ blocks take the following form

$$= \sum_{\mathbf{k}} \begin{bmatrix} c^\dagger_{\mathbf{k}\uparrow} \\ c^\dagger_{\mathbf{k}\downarrow} \\ f^\dagger_{\mathbf{k}\uparrow} \\ f^\dagger_{\mathbf{k}\downarrow} \end{bmatrix}^T \left[\begin{array}{cc|cc} \xi^c_{\mathbf{k}} & 0 & V_0 b S^z_k & V_0 b\left(S^x_k - iS^y_k\right) \\ 0 & \xi^c_{\mathbf{k}} & V_0 b\left(S^x_k + iS^y_k\right) & -V_0 b S^z_k \\ \hline V_0 b S^z_k & V_0 b\left(S^x_k - iS^y_k\right) & \xi^f_{\mathbf{k}} + \lambda & 0 \\ V_0 b\left(S^x_k + iS^y_k\right) & -V_0 b S^z_k & 0 & \xi^f_{\mathbf{k}} + \lambda \end{array}\right] \begin{bmatrix} c_{\mathbf{k}\uparrow} \\ c_{\mathbf{k}\downarrow} \\ f_{\mathbf{k}\uparrow} \\ f_{\mathbf{k}\downarrow} \end{bmatrix}. \tag{5.2}$$

Now we can Fourier transform the Hamiltonian matrix elements in $z-$axis as

$$\overline{\mathcal{H}}_{IJ}(k_x, k_y) = \frac{1}{N_z}\sum_{k_z} \langle k_z | \langle I | \overline{\mathcal{H}}(k_x, k_y, k_z) | J \rangle | k_z \rangle = \frac{1}{N_z}\sum_{k_z} \langle I | k_z \rangle \overline{\mathcal{H}}(k_x, k_y, k_z) \langle k_z | J \rangle \tag{5.3}$$

and by considering $\langle I | k_z \rangle = e^{-ik_z z_I}$ and $\langle k_z | J \rangle = e^{ik_z z_J}$, one can obtain

$$\overline{\mathcal{H}}_{IJ}(k_x, k_y) = \frac{1}{N_z}\sum_{k_z} e^{ik_z(z_J - z_I)}\overline{\mathcal{H}}(k_x, k_y, k_z) \tag{5.4}$$

where $N_z$ is the number of the lattice sites and $|n = I\rangle$ is the local state eigenbasis in $z-$axis. Having equation (5.4), it is then straightforward to Fourier transform each block of the Hamiltonian matrix (5.1). First decompose $\overline{\mathcal{H}}^{(2\times2)}_{cc}(k_x, k_y, k_z)$ into parallel and perpendicular parts

$$\overline{\mathcal{H}}^{(2\times2)}_{cc}(k_x, k_y, k_z) = \begin{bmatrix} \xi^c_{\mathbf{k}} & 0 \\ 0 & \xi^c_{\mathbf{k}} \end{bmatrix} = \begin{bmatrix} -2t^c\left(cos(k_x) + cos(k_y) + cos(k_z)\right) - \mu - \eta & 0 \\ 0 & -2t^c\left(cos(k_x) + cos(k_y) + cos(k_z)\right) - \mu - \eta \end{bmatrix} \tag{5.5}$$

$$= \begin{bmatrix} -2t^c\left(cos(k_x) + cos(k_y)\right) - \mu - \eta & 0 \\ 0 & -2t^c\left(\left(cos(k_x) + cos(k_y)\right) - \mu - \eta\right. \end{bmatrix} + \begin{bmatrix} -2t^c cos(k_z) & 0 \\ 0 & -2t^c cos(k_z) \end{bmatrix}. \tag{5.6}$$

Using $cos(k_z) = \frac{e^{ik_z}+e^{-ik_z}}{2}$ and defining the parallel dispersion $\xi^c_{\mathbf{k}_\parallel} = -2t^c\left(cos(k_x) + cos(k_y)\right) - \mu - \eta$, one can write equation (5.5) as

$$\overline{\mathcal{H}}^{(2\times2)}_{cc}(k_x, k_y, k_z) = \begin{bmatrix} \xi^c_{\mathbf{k}_\parallel} & 0 \\ 0 & \xi^c_{\mathbf{k}_\parallel} \end{bmatrix} + \begin{bmatrix} -t^c & 0 \\ 0 & -t^c \end{bmatrix} e^{ik_z} + \begin{bmatrix} -t^c & 0 \\ 0 & -t^c \end{bmatrix} e^{-ik_z}.$$

Now by using equation (5.4), Fourier transform reads

$$\overline{\mathcal{H}}^{IJ}_{cc}(k_x, k_y) = \begin{bmatrix} \xi^c_{\mathbf{k}_\parallel} & 0 \\ 0 & \xi^c_{\mathbf{k}_\parallel} \end{bmatrix} \delta_{I,J} + \begin{bmatrix} -t^c & 0 \\ 0 & -t^c \end{bmatrix} \delta_{I,J+1} + \begin{bmatrix} -t^c & 0 \\ 0 & -t^c \end{bmatrix} \delta_{I,J-1} \quad . \tag{5.7}$$

According to this equation, conduction electron Hamiltonian splits into dispersion in parallel plane and nearest neighbors hoping in the $z-$axis, with blocks of the form

$$a_c = \begin{bmatrix} \xi^c_{\mathbf{k}_\parallel} & 0 \\ 0 & \xi^c_{\mathbf{k}_\parallel} \end{bmatrix}, \quad b_c = \begin{bmatrix} -t^c & 0 \\ 0 & -t^c \end{bmatrix}. \tag{5.8}$$

which simplifies $\overline{\mathcal{H}}^{IJ}_{cc}(k_x, k_y)$ as follows

$$\overline{\mathcal{H}}^{IJ}_{cc}(k_x, k_y) = a_c\delta_{I,J} + b_c\delta_{I,J+1} + b_c\delta_{I,J-1} \quad . \tag{5.9}$$

For a lattice with $N_z$ sites in $z-$axis, the full matrix representation of $\overline{\mathcal{H}}_{cc}(k_x, k_y)$ would be

$$\overline{\mathcal{H}}_{cc}(k_x, k_y) = \begin{bmatrix} a_c & b_c & 0 & 0 & \cdots & b_c \\ b_c & a_c & b_c & 0 & \cdots & 0 \\ 0 & b_c & a_c & b_c & \cdots & 0 \\ \vdots & \ddots & \ddots & \ddots & \ddots & \\ \vdots & & \ddots & \ddots & \ddots & b_c \\ b_c & & & 0 & b_c & a_c \end{bmatrix}_{2N_z\times2N_z} , \tag{5.10}$$

where the first and last site is enforced to be equal to each other due to the periodic boundary conditions

$$\begin{cases} \overline{\mathcal{H}}^{21}_{cc}(k_x, k_y) = \overline{\mathcal{H}}^{1N_z}_{cc}(k_x, k_y) \\ \overline{\mathcal{H}}^{12}_{cc}(k_x, k_y) = \overline{\mathcal{H}}^{N_z1}_{cc}(k_x, k_y) \end{cases} . \tag{5.11}$$

Similarly for $\overline{\mathcal{H}}_{ff}(k_x, k_y)$, one can obtain

$$\overline{\mathcal{H}}^{IJ}_{ff}(k_x, k_y) = \begin{bmatrix} \xi^f_{\mathbf{k}_\parallel} + \lambda & 0 \\ 0 & \xi^f_{\mathbf{k}_\parallel} + \lambda \end{bmatrix} \delta_{I,J} + \begin{bmatrix} -t^f b^2 & 0 \\ 0 & -t^f b^2 \end{bmatrix} \delta_{I,J+1} + \begin{bmatrix} -t^f b^2 & 0 \\ 0 & -t^f b^2 \end{bmatrix} \delta_{I,J-1} \ , \tag{5.12}$$

with a similar blocks

$$a_f = \begin{bmatrix} \xi^f_{\mathbf{k}_\parallel} + \lambda & 0 \\ 0 & \xi^f_{\mathbf{k}_\parallel} + \lambda \end{bmatrix}, \quad b_f = \begin{bmatrix} -t^f b^2 & 0 \\ 0 & -t^f b^2 \end{bmatrix} \tag{5.13}$$

where $\xi^{f}_{\mathbf{k}_{\parallel}} = -2t_f b^2 \sum_{l=x,y} cos(k_l) - \mu + \eta$. This can be written in a compact form as

$$\overline{\mathcal{H}}^{IJ}_{ff}(k_x, k_y) = a_f \delta_{I,J} + b_f \delta_{I,J+1} + b_f \delta_{I,J-1} \quad . \tag{5.14}$$

So the full matrix representation of $H_{ff}(k_x, k_y)$ for a lattice with $N_z$ sites in $z-$axis would be

$$\overline{\mathcal{H}}_{ff}(k_x, k_y) = \begin{bmatrix} a_f & b_f & 0 & 0 & \cdots & b_f \\ b_f & a_f & b_f & 0 & \cdots & 0 \\ 0 & b_f & a_f & b_f & \cdots & 0 \\ \vdots & \ddots & \ddots & \ddots & \ddots & \\ \vdots & & \ddots & \ddots & \ddots & b_f \\ b_f & & & 0 & b_f & a_f \end{bmatrix}_{2N_z \times 2N_z} , \tag{5.15}$$

with the following periodic boundary condition

$$\begin{cases} \overline{\mathcal{H}}^{21}_{ff}(k_x, k_y) = \overline{\mathcal{H}}^{1N_z}_{ff}(k_x, k_y) \\ \overline{\mathcal{H}}^{12}_{ff}(k_x, k_y) = \overline{\mathcal{H}}^{N_z 1}_{ff}(k_x, k_y). \end{cases} . \tag{5.16}$$

Now one need to find the relevant expressions for $H_{cf}(k_x, k_y)$ and $H_{fc}(k_x, k_y)$. Like before, we decompose it into parallel and perpendicular parts as

$$\overline{\mathcal{H}}^{(2\times 2)}_{cf}(k_x, k_y, k_z) = \begin{bmatrix} V_0 b S^z_k & V_0 b \left(S^x_k - iS^y_k\right) \\ V_0 b \left(S^x_k + iS^y_k\right) & -V_0 b S^z_k \end{bmatrix} = \begin{bmatrix} V_0 b sin(k_z) & V_0 b \left(sin(k_x) - isin(k_y)\right) \\ V_0 b \left(sin(k_x) + isin(k_y)\right) & -V_0 b sin(k_z) \end{bmatrix}$$
$$= \begin{bmatrix} V_0 b sin(k_z) & 0 \\ 0 & -V_0 b sin(k_z) \end{bmatrix} + \begin{bmatrix} 0 & V_0 b \left(sin(k_x) - isin(k_y)\right) \\ V_0 b \left(sin(k_x) + isin(k_y)\right) & 0 \end{bmatrix}.$$

Using $sin(k_z) = \frac{e^{ik_z} - e^{-ik_z}}{2i}$, one can obtain

$$\overline{\mathcal{H}}^{(2\times 2)}_{cf}(k_x, k_y, k_z) = \begin{bmatrix} V_0 b & 0 \\ 0 & -V_0 b \end{bmatrix} \frac{e^{ik_z}}{2i} - \begin{bmatrix} V_0 b & 0 \\ 0 & -V_0 b \end{bmatrix} \frac{e^{-ik_z}}{2i} + \begin{bmatrix} 0 & V_0 b \left(sin(k_x) - isin(k_y)\right) \\ V_0 b \left(sin(k_x) + isin(k_y)\right) & 0 \end{bmatrix}.$$

By using equation (5.4), Fourier transform reads

$$\overline{\mathcal{H}}^{IJ}_{cf}(k_x, k_y, k_z) = \frac{1}{2i} \begin{bmatrix} V_0 b & 0 \\ 0 & -V_0 b \end{bmatrix} \delta_{I,J+1} - \frac{1}{2i} \begin{bmatrix} V_0 b & 0 \\ 0 & -V_0 b \end{bmatrix} \delta_{I,J-1} + \begin{bmatrix} 0 & V_0 b \left(sin(k_x) - isin(k_y)\right) \\ V_0 b \left(sin(k_x) + isin(k_y)\right) & 0 \end{bmatrix} \delta_{I,J} \; .$$

According to this equation, hybridization Hamiltonian splits into sinusoidal form factor in parallel plane and constant hybridization between nearest neighbors in perpendicular plane, with blocks of the form

$$a_{cf} = \begin{bmatrix} 0 & V_0 b\left(sin(k_x) - isin(k_y)\right) \\ V_0 b\left(sin(k_x) + isin(k_y)\right) & 0 \end{bmatrix}, \quad b_{cf} = \frac{1}{2i}\begin{bmatrix} V_0 b & 0 \\ 0 & -V_0 b \end{bmatrix} \tag{5.17}$$

and a compact form of

$$\overline{\mathcal{H}}^{IJ}_{cf}(k_x, k_y) = a_{cf}\delta_{I,J} + b_{cf}\delta_{I,J+1} - b_{cf}\delta_{I,J-1}. \tag{5.18}$$

So the full matrix representation of $H_{cf}(k_x, k_y)$ for a lattice with $N_z$ sites in $z-$axis would be

$$\overline{\mathcal{H}}_{cf}(k_x, k_y) = \begin{bmatrix} a_{cf} & -b_{cf} & 0 & 0 & \cdots & b_{cf} \\ b_{cf} & a_{cf} & -b_{cf} & 0 & \cdots & 0 \\ 0 & b_{cf} & a_{cf} & -b_{cf} & \cdots & 0 \\ \vdots & \ddots & \ddots & \ddots & \ddots & \\ \vdots & & \ddots & \ddots & \ddots & -b_{cf} \\ -b_{cf} & & & 0 & b_{cf} & a_{cf} \end{bmatrix}_{2N_z \times 2N_z}, \tag{5.19}$$

with the following periodic boundary condition

$$\begin{cases} \overline{\mathcal{H}}^{21}_{cf}(k_x, k_y) = \overline{\mathcal{H}}^{1N_z}_{cf}(k_x, k_y) \\ \overline{\mathcal{H}}^{12}_{cf}(k_x, k_y) = \overline{\mathcal{H}}^{N_z 1}_{cf}(k_x, k_y) \end{cases}. \tag{5.20}$$

Finally, we have found all the block matrices need to reconstruct the full Hamiltonian matrix of a bulk with $N_z$ layers in $z-$axis

$$\overline{\mathcal{H}} = \sum_{\mathbf{k}_\parallel} \begin{bmatrix} c^\dagger_{\mathbf{k}_\parallel 1\uparrow} \\ c^\dagger_{\mathbf{k}_\parallel 1\downarrow} \\ \vdots \\ c^\dagger_{\mathbf{k}_\parallel N_z\uparrow} \\ c^\dagger_{\mathbf{k}_\parallel N_z\downarrow} \\ \hline f^\dagger_{\mathbf{k}_\parallel 1\uparrow} \\ f^\dagger_{\mathbf{k}_\parallel 1\downarrow} \\ \vdots \\ f^\dagger_{\mathbf{k}_\parallel N_z\uparrow} \\ f^\dagger_{\mathbf{k}_\parallel N_z\downarrow} \end{bmatrix}^T \left[\begin{array}{c:c} \overline{\mathcal{H}}_{cc}(k_x, k_y) & \overline{\mathcal{H}}_{cf}(k_x, k_y) \\ \hdashline \overline{\mathcal{H}}_{cf}(k_x, k_y) & \overline{\mathcal{H}}_{ff}(k_x, k_y) \end{array}\right]_{4N_z \times 4N_z} \begin{bmatrix} c_{\mathbf{k}_\parallel 1\uparrow} \\ c_{\mathbf{k}_\parallel 1\downarrow} \\ \vdots \\ c_{\mathbf{k}_\parallel N_z\uparrow} \\ c_{\mathbf{k}_\parallel N_z\downarrow} \\ \hline f_{\mathbf{k}_\parallel 1\uparrow} \\ f_{\mathbf{k}_\parallel 1\downarrow} \\ \vdots \\ f_{\mathbf{k}_\parallel N_z\uparrow} \\ f_{\mathbf{k}_\parallel N_z\downarrow} \end{bmatrix}. \tag{5.21}$$

## 5.2 Multilayer Hamiltonian without periodic boundary condition

To find the mean-field Hamiltonian with broken periodic boundary condition in the $z-$axis, one should remove all linking elements (5.11), (5.16) and (5.20) of each block, which makes them tri-banded. Moreover, one should note that the mean-field parameters will be layer-dependent by removing the PBC in the $z-$axis

$$(b, \lambda, \eta) \rightarrow (b_n, \lambda_n, \eta)\,. \tag{5.22}$$

where since the physical chemical ($\mu$) is a thermodynamic macroscopic quantity, it is the same for all layers. Consequntly, the auxilary chemical potential ($\eta$) which shifts all bands to the position of $\mu$, do not carry layer index and is constant.

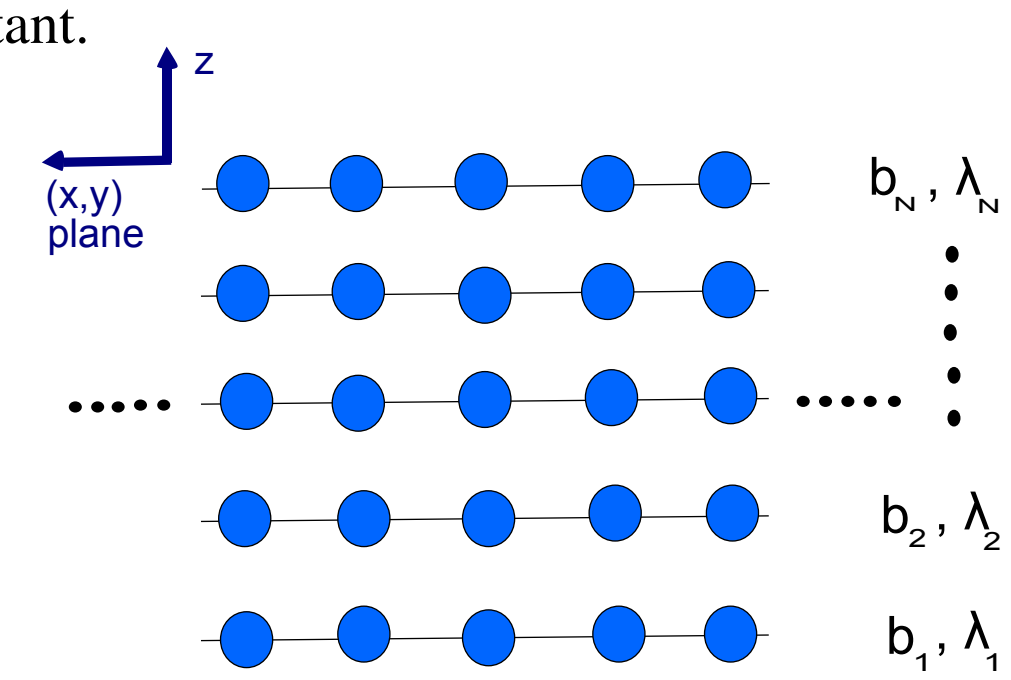

Figure 5.1: Schematic figure of a multilayer system with broken PBC in $z-$axis and periodic in $xy$-plane.

As a consequence, all the building blocks become layer-dependent

$$\begin{cases} a_{c/f/cf} \rightarrow a^n_{c/f/cf} \\ b_{c/f/cf} \rightarrow b^n_{c/f/cf} \end{cases} . \tag{5.23}$$

Appllying all these changes to (5.10), (5.15) and (5.19), one can find the new matrix elements of the Hamiltonian as

$$\overline{\mathcal{H}}^{ML}_{c/f}(k_x, k_y) = \begin{bmatrix} a^1_{c/f} & b^2_{c/f} & 0 & 0 & \cdots & 0 \\ b^1_{c/f} & a^2_{c/f} & b^3_{c/f} & 0 & \cdots & 0 \\ 0 & b^2_{c/f} & a^3_{c/f} & b^4_{c/f} & \cdots & 0 \\ \vdots & \ddots & \ddots & \ddots & \ddots & \\ \vdots & & \ddots & \ddots & \ddots & b^{N_z}_{c/f} \\ 0 & & & 0 & b^{N_z-1}_{c/f} & a^{N_z}_{c/f} \end{bmatrix}_{2N_z\times 2N_z} \tag{5.24}$$

and

$$\overline{\mathcal{H}}^{ML}_{cf}(k_x, k_y) = \begin{bmatrix} a^1_{cf} & -b^2_{cf} & 0 & 0 & \cdots & 0 \\ b^1_{cf} & a^2_{cf} & -b^3_{cf} & 0 & \cdots & 0 \\ 0 & b^2_{cf} & a^3_{cf} & -b^4_{cf} & \cdots & 0 \\ \vdots & \ddots & \ddots & \ddots & \ddots & \\ \vdots & & \ddots & \ddots & \ddots & -b^{N_z}_{cf} \\ 0 & & & 0 & b^{N_z-1}_{cf} & a^{N_z}_{cf} \end{bmatrix}_{2N_z\times 2N_z} , \tag{5.25}$$

where the superscript "ML" stands for a multilayer system with open boundary condition in $z-$axis. To find the mean-field equation, it is more convenient to have the inhomogenous Hamiltonian in an operator form, $\left(\psi^c_{\mathbf{k}_\parallel n}, \psi^f_{\mathbf{k}_\parallel n}\right) = \left(c_{\mathbf{k}_\parallel n\uparrow}, c_{\mathbf{k}_\parallel n\downarrow}, f_{\mathbf{k}_\parallel n\uparrow}, f_{\mathbf{k}_\parallel n\downarrow}\right)$, like the following

$$\begin{aligned}
\overline{\mathcal{H}}^{ML}_{c/f} &= \sum_{\mathbf{k}_\parallel} \{ \left[\psi^{c/f\dagger}_{\mathbf{k}_\parallel 1} a^1_{c/f} \psi^{c/f}_{\mathbf{k}_\parallel 1} + ... + \psi^{c/f\dagger}_{\mathbf{k}_\parallel N_z} a^{N_z}_{c/f} \psi^{c/f}_{\mathbf{k}_\parallel N_z}\right] \\
&+ \left[\psi^{c/f\dagger}_{\mathbf{k}_\parallel 1} b^2_{c/f} \psi^{c/f}_{\mathbf{k}_\parallel 2} + ... + \psi^{c/f\dagger}_{\mathbf{k}_\parallel N_z-1} b^{N_z}_{c/f} \psi^{c/f}_{\mathbf{k}_\parallel N_z}\right] \\
&+ \left[\psi^{c/f\dagger}_{\mathbf{k}_\parallel 2} b^1_{c/f} \psi^{c/f}_{\mathbf{k}_\parallel 1} + ... + \psi^{c/f\dagger}_{\mathbf{k}_\parallel N_z} b^{N_z-1}_{c/f} \psi^{c/f}_{\mathbf{k}_\parallel N_z-1}\right] \} \\
&= \sum_{\mathbf{k}_\parallel} \left[ \sum_{n=1}^{N_z} \psi^{c/f\dagger}_{\mathbf{k}_\parallel n} a^n_{c/f} \psi^{c/f}_{\mathbf{k}_\parallel n} + \sum_{n=1}^{N_z-1} \left( \psi^{c/f\dagger}_{\mathbf{k}_\parallel n+1} b^n_{c/f} \psi^{c/f}_{\mathbf{k}_\parallel n} + \psi^{c/f\dagger}_{\mathbf{k}_\parallel n} b^{n+1}_{c/f} \psi^{c/f}_{\mathbf{k}_\parallel n+1} \right) \right].
\end{aligned} \tag{5.26}$$

Writing (5.26) in terms of $c$ and $f$ operators, and the building blocks $a^n_{c/f}$ and $b^n_{c/f}$ elementwise, one finds

$$\rightarrow \quad \overline{\mathcal{H}}^{ML}_{cc} = \sum_{\mathbf{k}_\parallel,\sigma} \left[ \sum_{n=1}^{N_z} \xi^c_{\mathbf{k}_\parallel} c^\dagger_{\mathbf{k}_\parallel n\sigma} c_{\mathbf{k}_\parallel n\sigma} - t^c \sum_{n=1}^{N_z-1} \left( c^\dagger_{\mathbf{k}_\parallel n+1\sigma} c_{\mathbf{k}_\parallel n\sigma} + c^\dagger_{\mathbf{k}_\parallel n\sigma} c_{\mathbf{k}_\parallel n+1\sigma} \right) \right] \tag{5.27}$$

and

$$\rightarrow \quad \overline{\mathcal{H}}^{ML}_{ff} = \sum_{\mathbf{k}_\parallel,\alpha} \left[ \sum_{n=1}^{N_z} \left( \xi^f_{n\mathbf{k}_\parallel} + \lambda_n \right) f^\dagger_{\mathbf{k}_\parallel n\alpha} f_{\mathbf{k}_\parallel n\alpha} - t^f \sum_{n=1}^{N_z-1} \left( b^2_n f^\dagger_{\mathbf{k}_\parallel n+1\alpha} f_{\mathbf{k}_\parallel n\alpha} + b^2_{n+1} f^\dagger_{\mathbf{k}_\parallel n\alpha} f_{\mathbf{k}_\parallel n+1\alpha} \right) \right], \tag{5.28}$$

where $\xi^f_{n\mathbf{k}_\parallel} = -2t^f b^2_n \sum_{l=x,y} cos(k_l) - \mu + \eta$. Moreover, it is assumed that, for a cubic lattice, interlayer and intralayer hoping are equal. The same calculation can be done for $H_{cf}$ with slight changes like the following

$$\begin{aligned}
\overline{\mathcal{H}}^{ML}_{cf} &= \sum_{\mathbf{k}_\parallel} \{ \left[\psi^{cf\dagger}_{\mathbf{k}_\parallel 1} a^1_{cf} \psi^{cf}_{\mathbf{k}_\parallel 1} + ... + \psi^{cf\dagger}_{\mathbf{k}_\parallel N_z} a^{N_z}_{cf} \psi^{cf}_{\mathbf{k}_\parallel N_z}\right] \\
&+ \left[\psi^{cf\dagger}_{\mathbf{k}_\parallel 1} (-b^2_{cf}) \psi^{cf}_{\mathbf{k}_\parallel 2} + ... + \psi^{cf\dagger}_{\mathbf{k}_\parallel N_z-1} (-b^{N_z}_{cf}) \psi^{cf}_{\mathbf{k}_\parallel N_z}\right] \\
&+ \left[\psi^{cf\dagger}_{\mathbf{k}_\parallel 2} b^1_{cf} \psi^{cf}_{\mathbf{k}_\parallel 1} + ... + \psi^{cf\dagger}_{\mathbf{k}_\parallel N_z} b^{N_z-1}_{cf} \psi^{cf}_{\mathbf{k}_\parallel N_z-1}\right] \} \\
&= \sum_{\mathbf{k}_\parallel} \left[ \sum_{n=1}^{N_z} \psi^{cf\dagger}_{\mathbf{k}_\parallel n} a^n_{cf} \psi^{cf}_{\mathbf{k}_\parallel n} + \sum_{n=1}^{N_z-1} \left( \psi^{cf\dagger}_{\mathbf{k}_\parallel n+1} b^n_{cf} \psi^{cf}_{\mathbf{k}_\parallel n} - \psi^{cf\dagger}_{\mathbf{k}_\parallel n} b^{n+1}_{cf} \psi^{cf}_{\mathbf{k}_\parallel n+1} \right) \right].
\end{aligned} \tag{5.29}$$

Then by writing equation (5.29) in terms of $c$ and $f$ operators, and the building blocks $a^n_{cf}$ and $b^n_{cf}$ elementwise, one can find

$$\rightarrow \quad \overline{\mathcal{H}}^{ML}_{cf} = \sum_{\mathbf{k}_\parallel,\sigma,\alpha} \left[ \sum_{n=1}^{N_z} V_0 b_n \Phi_{\sigma\alpha}(\mathbf{k}_\parallel) c^\dagger_{\mathbf{k}_\parallel n\sigma} f_{\mathbf{k}_\parallel n\alpha} + \sum_{n=1}^{N_z-1} V_0 \Phi^z_{\sigma\alpha} \left( c^\dagger_{\mathbf{k}_\parallel n+1\sigma} b_n f_{\mathbf{k}_\parallel n\alpha} - c^\dagger_{\mathbf{k}_\parallel n\sigma} b_{n+1} f_{\mathbf{k}_\parallel n+1\alpha} \right) \right] \tag{5.30}$$

where

$$\begin{cases}\Phi(\mathbf{k}_{\|}) = \sum_{l=x,y} S^{l}_{k}\sigma_{l} \\ \Phi^{z} = \frac{1}{2i}\sigma_{z}\end{cases} . \tag{5.31}$$

By Hermition conjugating of $\overline{\mathcal{H}}^{ML}_{cf}$, one can find $\overline{\mathcal{H}}^{ML}_{fc}$ as follows

$$\rightarrow \quad \overline{\mathcal{H}}^{ML}_{fc} = \sum_{\mathbf{k}_{\|},\sigma,\alpha}\left[\sum_{n=1}^{N_z} V_0 b_n \Phi_{\sigma\alpha}(\mathbf{k}_{\|}) f^{\dagger}_{\mathbf{k}_{\|}n\alpha} c_{\mathbf{k}_{\|}n\sigma} + \sum_{n=1}^{N_z-1} V_0 \Phi^{z*}_{\sigma\alpha}\left(f^{\dagger}_{\mathbf{k}_{\|}n\alpha} b_n c_{\mathbf{k}_{\|}n+1\sigma} - f^{\dagger}_{\mathbf{k}_{\|}n+1\alpha} b_{n+1} c_{\mathbf{k}_{\|}n\sigma}\right)\right] \tag{5.32}$$

Finally, by collecting all terms we can reconstruct the mean-field Hamiltonian of an open multilayer system as follows

$$\begin{aligned}\overline{\mathcal{H}}^{ML} =& \sum_{\mathbf{k}_{\|},\sigma}\left[\sum_{n=1}^{N_z} \xi^{c}_{\mathbf{k}_{\|}} c^{\dagger}_{\mathbf{k}_{\|}n\sigma} c_{\mathbf{k}_{\|}n\sigma} - t^{c} \sum_{n=1}^{N_z-1}\left(c^{\dagger}_{\mathbf{k}_{\|}n+1\sigma} c_{\mathbf{k}_{\|}n\sigma} + c^{\dagger}_{\mathbf{k}_{\|}n\sigma} c_{\mathbf{k}_{\|}n+1\sigma}\right)\right] \\ &+ \sum_{\mathbf{k}_{\|},\alpha}\left[\sum_{n=1}^{N_z}\left(\xi^{f}_{n\mathbf{k}_{\|}} + \lambda_n\right) f^{\dagger}_{\mathbf{k}_{\|}n\alpha} f_{\mathbf{k}_{\|}n\alpha} - t^{f} \sum_{n=1}^{N_z-1}\left(b_n^2 f^{\dagger}_{\mathbf{k}_{\|}n+1\alpha} f_{\mathbf{k}_{\|}n\alpha} + b_{n+1}^2 f^{\dagger}_{\mathbf{k}_{\|}n\alpha} f_{\mathbf{k}_{\|}n+1\alpha}\right)\right] \\ &+ V_0 \sum_{\mathbf{k}_{\|},\sigma,\alpha}\left\{\sum_{n=1}^{N_z}\left(b_n \Phi_{\sigma\alpha}(\mathbf{k}_{\|}) c^{\dagger}_{\mathbf{k}_{\|}n\sigma} f_{\mathbf{k}_{\|}n\alpha} + H.c.\right) + \sum_{n=1}^{N_z-1}\left[\Phi^{z}_{\sigma\alpha}\left(c^{\dagger}_{\mathbf{k}_{\|}n+1\sigma} b_n f_{\mathbf{k}_{\|}n\alpha} - c^{\dagger}_{\mathbf{k}_{\|}n\sigma} b_{n+1} f_{\mathbf{k}_{\|}n+1\alpha}\right) + H.c.\right]\right\} \\ &+ N_{f\|} \sum_{n} \lambda_n \left(b_n^2 - 1\right)\end{aligned} \tag{5.33}$$

## 5.3 Mean-field equations of the multilayer system

Following the calculations of the mean-field theory of the bulk in chapter 4, one needs to minimize the expectation value of the multilayer mean-field Hamiltonian $\langle\overline{\mathcal{H}}^{ML}\rangle$ with respect to mean-field parameters of each layer,

$$\frac{\partial\langle\overline{\mathcal{H}}^{ML}\rangle}{\partial b_m} = 0, \qquad \frac{\partial\langle\overline{\mathcal{H}}^{ML}\rangle}{\partial \lambda_m} = 0 \qquad \text{and} \qquad \frac{\partial\langle\overline{\mathcal{H}}^{ML}\rangle}{\partial \eta} = 0. \tag{5.34}$$

This results in a set of equations for $b_m$

$$\boxed{\begin{aligned}&\frac{V_0}{N_{f\|}}\sum_{\mathbf{k}_{\|}\sigma\alpha}\left\{\left(\Phi_{\sigma\alpha}(\mathbf{k}_{\|})\langle c^{\dagger}_{\mathbf{k}_{\|}m\sigma} f_{\mathbf{k}_{\|}m\alpha}\rangle + H.c.\right) + \left[\Phi^{z}_{\sigma\alpha}\left(\langle c^{\dagger}_{\mathbf{k}_{\|}m+1\sigma} f_{\mathbf{k}_{\|}m\alpha}\rangle - \langle c^{\dagger}_{\mathbf{k}_{\|}m-1\sigma} f_{\mathbf{k}_{\|}m\alpha}\rangle\right) + H.c.\right]\right\} \\ &+ \frac{2b_m}{N_{f\|}}\sum_{\mathbf{k}_{\|}\alpha}\left\{\epsilon^{f}_{m\mathbf{k}_{\|}}\langle f^{\dagger}_{\mathbf{k}_{\|}m\alpha} f_{\mathbf{k}_{\|}m\alpha}\rangle - t^{f}\left(\langle f^{\dagger}_{\mathbf{k}_{\|}m+1\alpha} f_{\mathbf{k}_{\|}m\alpha}\rangle + \langle f^{\dagger}_{\mathbf{k}_{\|}m-1\alpha} f_{\mathbf{k}_{\|}m\alpha}\rangle\right)\right\} + 2\lambda_m b_m = 0\end{aligned}} \tag{5.35}$$

for $\lambda_m$ as

$$\boxed{\frac{1}{N_{f\|}}\sum_{\mathbf{k}_{\|}\alpha}\langle f^{\dagger}_{\mathbf{k}_{\|}m\alpha} f_{\mathbf{k}_{\|}m\alpha}\rangle + b_m^2 - 1 = 0}. \tag{5.36}$$

and only one equation for $\eta$ as

$$\boxed{\sum_{\mathbf{k}_\parallel m\sigma}\langle c^\dagger_{\mathbf{k}_\parallel m\sigma} c_{\mathbf{k}_\parallel m\sigma}\rangle - \sum_{\mathbf{k}_\parallel m\alpha}\langle f^\dagger_{\mathbf{k}_\parallel m\alpha} f_{\mathbf{k}_\parallel m\alpha}\rangle = 0}. \tag{5.37}$$

For a multilayer system to be an insulator, physical chemical potential ($\mu$) as a macroscopic quantity should be inside the gap. Practically, this is imposed by the equation of $\eta$ which enforces the number of $c$ and $f$ electrons to be equal on average for all layers and sites.
To solve the self-consistency equations for $(b_m, \lambda_m, \eta)$, one need to know about certain thermal averages. For Fermionic operators, in general, one can find them as follows [42]

$$\begin{cases}\langle c^\dagger_\nu c_{\nu'}\rangle = \int_{-\infty}^{+\infty} d\omega A_{cc}(\nu',\nu,\omega)n_F(\omega) = \frac{1}{\pi}\int_{-\infty}^{+\infty} ImG^A_{cc}(\nu',\nu,\omega)n_F(\omega)d\omega \\ \langle f^\dagger_\nu f_{\nu'}\rangle = \int_{-\infty}^{+\infty} d\omega A_{ff}(\nu',\nu,\omega)n_F(\omega) = \frac{1}{\pi}\int_{-\infty}^{+\infty} ImG^A_{ff}(\nu',\nu,\omega)n_F(\omega)d\omega \\ \langle c^\dagger_\nu f_{\nu'}\rangle = \int_{-\infty}^{+\infty} d\omega A_{fc}(\nu',\nu,\omega)n_F(\omega) = \frac{1}{\pi}\int_{-\infty}^{+\infty} ImG^A_{fc}(\nu',\nu,\omega)n_F(\omega)d\omega \\ \langle f^\dagger_\nu c_{\nu'}\rangle = \int_{-\infty}^{+\infty} d\omega A_{cf}(\nu',\nu,\omega)n_F(\omega) = \frac{1}{\pi}\int_{-\infty}^{+\infty} ImG^A_{cf}(\nu',\nu,\omega)n_F(\omega)d\omega \end{cases}, \tag{5.38}$$

where $\nu$ and $\nu'$ are collection of quantum numbers. Hence, to complete the self-consistency loop one would need the matrix elements of the full advanced Green's function in the frequency domain.
Before going further, it is worthwhile to visualize the structure of the time translational invariant advanced Green's function of the multilayer system in the basis of (flavor) ⊗ (layer) ⊗ (spin);

$$\begin{aligned} G^A(\mathbf{k}_\parallel,t) &= i\Theta(-t)\langle\{\Psi_{\mathbf{k}_\parallel}(t),\Psi^\dagger_{\mathbf{k}_\parallel}(0)\}\rangle \\ &= i\Theta(-t)\langle\{\begin{bmatrix} c_{\mathbf{k}_\parallel 1\uparrow}(t) \\ \vdots \\ c_{\mathbf{k}_\parallel N_z\downarrow}(t) \\ \hline f_{\mathbf{k}_\parallel 1\uparrow}(t) \\ \vdots \\ f_{\mathbf{k}_\parallel N_z\downarrow}(t)\end{bmatrix}, \left[c^\dagger_{\mathbf{k}_\parallel 1\uparrow}(0)\cdots c^\dagger_{\mathbf{k}_\parallel N_z\downarrow}(0)\middle| f^\dagger_{\mathbf{k}_\parallel 1\uparrow}(0)\cdots f^\dagger_{\mathbf{k}_\parallel N_z\downarrow}(0)\right]\}\rangle \\ &= \left[\begin{array}{ccc|ccc} G^A_{cc}(\mathbf{k}_\parallel,11,\uparrow\uparrow;t) & \cdots & G^A_{cc}(\mathbf{k}_\parallel,1N_z,\uparrow\downarrow;t) & G^A_{cf}(\mathbf{k}_\parallel,11,\uparrow\uparrow;t) & \cdots & G^A_{cf}(\mathbf{k}_\parallel,1N_z,\uparrow\downarrow;t) \\ \vdots & \ddots & \vdots & \vdots & \ddots & \vdots \\ G^A_{cc}(\mathbf{k}_\parallel,N_z1,\downarrow\uparrow;t) & \cdots & G^A_{cc}(\mathbf{k}_\parallel,N_zN_z,\downarrow\downarrow;t) & G^A_{cf}(\mathbf{k}_\parallel,N_z1,\downarrow\uparrow;t) & \cdots & G^A_{cf}(\mathbf{k}_\parallel,N_zN_z,\downarrow\downarrow;t) \\ \hline G^A_{fc}(\mathbf{k}_\parallel,11,\uparrow\uparrow;t) & \cdots & G^A_{fc}(\mathbf{k}_\parallel,1N_z,\uparrow\downarrow;t) & G^A_{ff}(\mathbf{k}_\parallel,11,\uparrow\uparrow;t) & \cdots & G^A_{ff}(\mathbf{k}_\parallel,1N_z,\uparrow\downarrow;t) \\ \vdots & \ddots & \vdots & \vdots & \ddots & \vdots \\ G^A_{fc}(\mathbf{k}_\parallel,N_z1,\downarrow\uparrow;t) & \cdots & G^A_{fc}(\mathbf{k}_\parallel,N_zN_z,\downarrow\downarrow;t) & G^A_{ff}(\mathbf{k}_\parallel,N_z1,\downarrow\uparrow;t) & \cdots & G^A_{ff}(\mathbf{k}_\parallel,N_zN_z,\downarrow\downarrow;t) \end{array}\right]. \end{aligned} \tag{5.39}$$

By Fourier transforming $G^A(\mathbf{k}_\parallel,t)$ to the frequency domain one can obtain $G^A(\mathbf{k}_\parallel,\omega)$, which mimics the same structure.

To find the matrix elements of $G^A(\mathbf{k}_\parallel,\omega)$, the equation of motion method can be used in frequency

domain

$$\Big((\omega - i\delta)\mathbb{1} - \overline{\mathcal{H}}^{ML}(\mathbf{k}_{\parallel})\Big)G^{A}(\mathbf{k}_{\parallel}, \omega) = \mathbb{1} \tag{5.40}$$

and eventually, the problem reduces to the inversion of $\Big((\omega - i\delta)\mathbb{1} - \overline{\mathcal{H}}^{ML}(\mathbf{k}_{\parallel})\Big)$.

## 5.4 Results

### 5.4.1 TKI in a slab geometry with constant mean-field parameters

#### Band spectrum

In order to find the band spectrum of the TKI, one needs to diagonalize the multilayer Hamiltonian (5.33). The results shows a very narrow hybridization gap with surface states. For a set of parameters corresponding to a weak topological Kondo insulator, two Dirac cones appear inside the gap, very close to the chemical potential at the border of Brillouin zone ($k/k_{BZ} = \pm 1$).

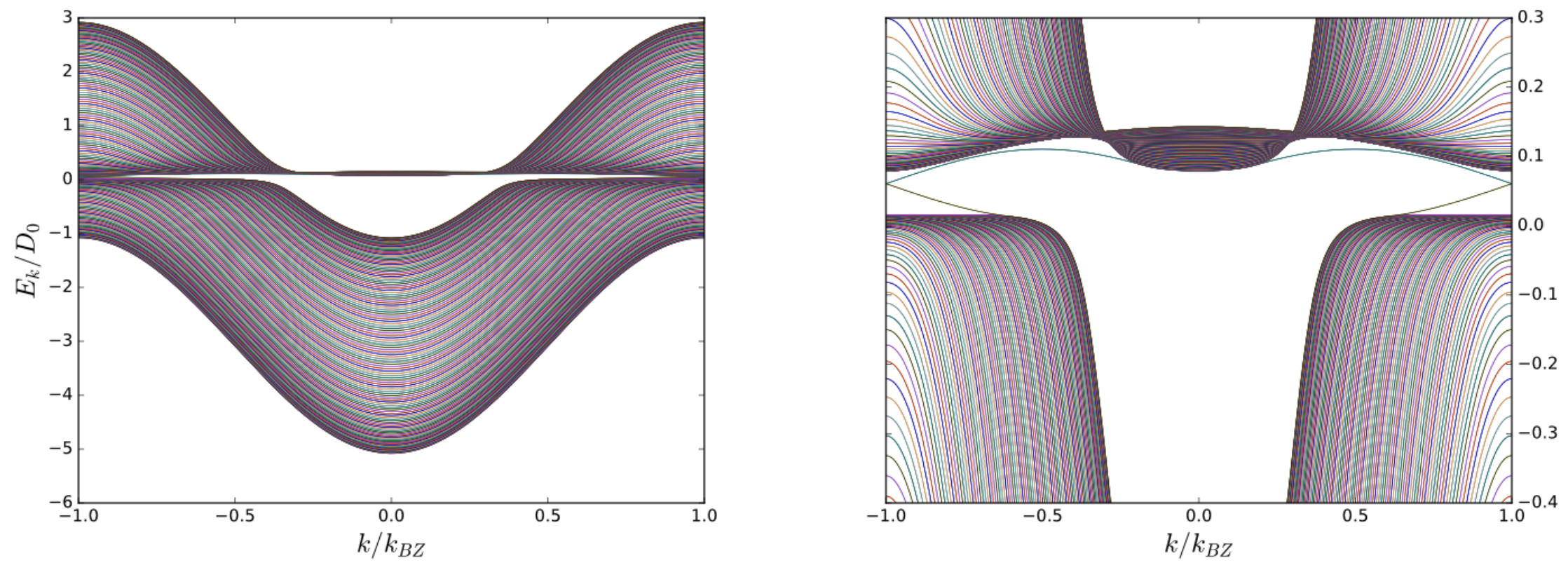

Figure 5.2: Left panel: Renormalized band spectrum of a weak TKI for $N_z = 140$ along the $(k_x, k_y) = (k, 0)$, with open boundary condition in $z$-axis. The constant mean-field parameters are $(b, \lambda/D_0, \eta/D_0) = (0.463, 0.162, 0.127)$, and the parameters chosen are $V_0/D_0 = 0.1$, $E_0^f/D_0 = -0.2$, $t^c/D_0 = 0.2$, $t^f/D_0 = -0.001$ and $\mu/D_0 = 0.063$. Right panel: A closer view of the gap, where the Dirac cones is clearly visible,

For another set of parameters, we obtain the strong topological insulators with a single Dirac cones at the center of Brillouin zone ($k/k_{BZ} = 0$).

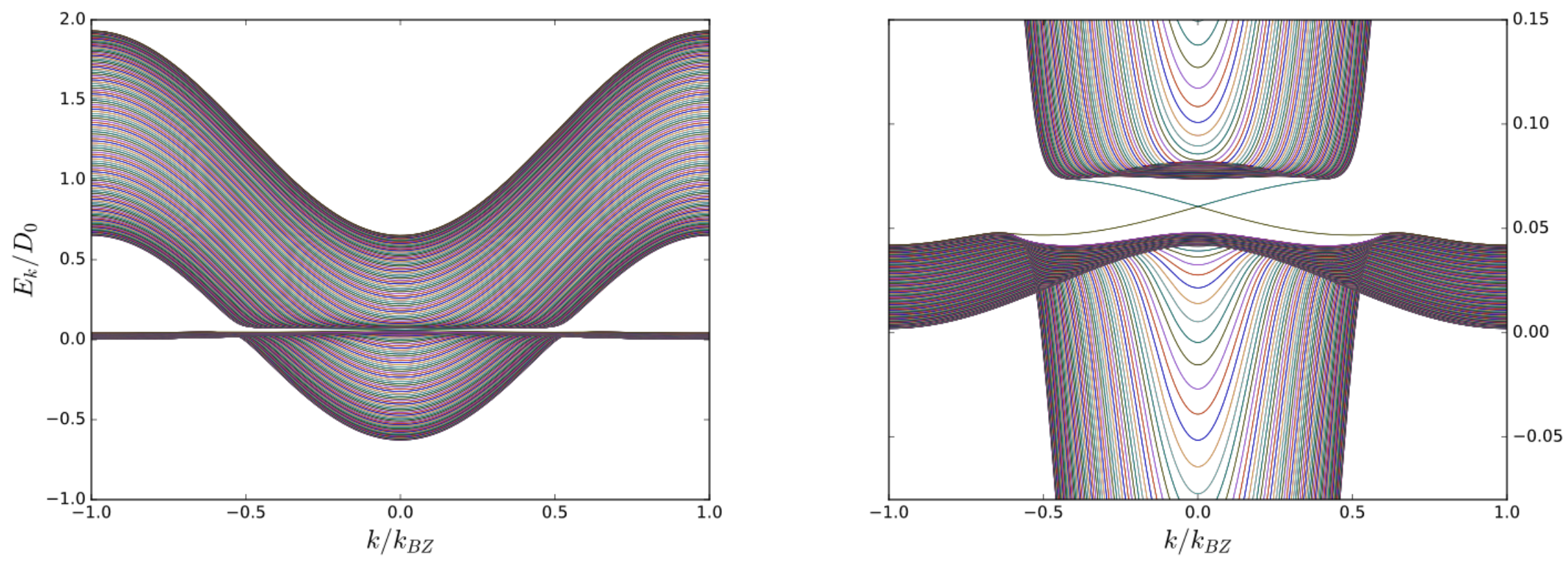

Figure 5.3: Left panel: Renormalized band spectrum of a strong TKI for $N_z = 140$ along the $(k_x, k_y) = (k, 0)$, with open boundary condition in $z$-axis. The constant mean-field parameters are $(b, \lambda/D_0, \eta/D_0) = (0.351, 0.146, 0.115)$, and the parameters chosen are $V_0/D_0 = 0.1$, $E_0^f/D_0 = -0.2$, $t^c/D_0 = -0.01$, $t^f/D_0 = 0.1$ and $\mu/D_0 = 0.063$. Right panel: A closer view of the gap, where the Dirac cone is clearly visible.

One should note that we used bulk mean-field parameters, which are constant throughout the lattice, to compute these band spectrum. Ignoring the effect of boundaries on the values of mean-field parameters allowed us to have large number of layers ($N_z = 140$) in the $z$-axis to find the dispersion without solving the inhomogeneous mean-field equations, which are numerically too costly. In the parameter regime explored, we observed that band crossing happens only when $sgn(t^c) = -sgn(t^f)$. These results can be compared with [11, 43].

### Bulk and edge states

Bulk state and edge state of the TKI can be obtained from diagonalization of the multilayer Hamiltonian (5.33). The result shows that the bulk state has finite value in the bulk and vanishing at the edge. The following result is only one of the possible bulk states.

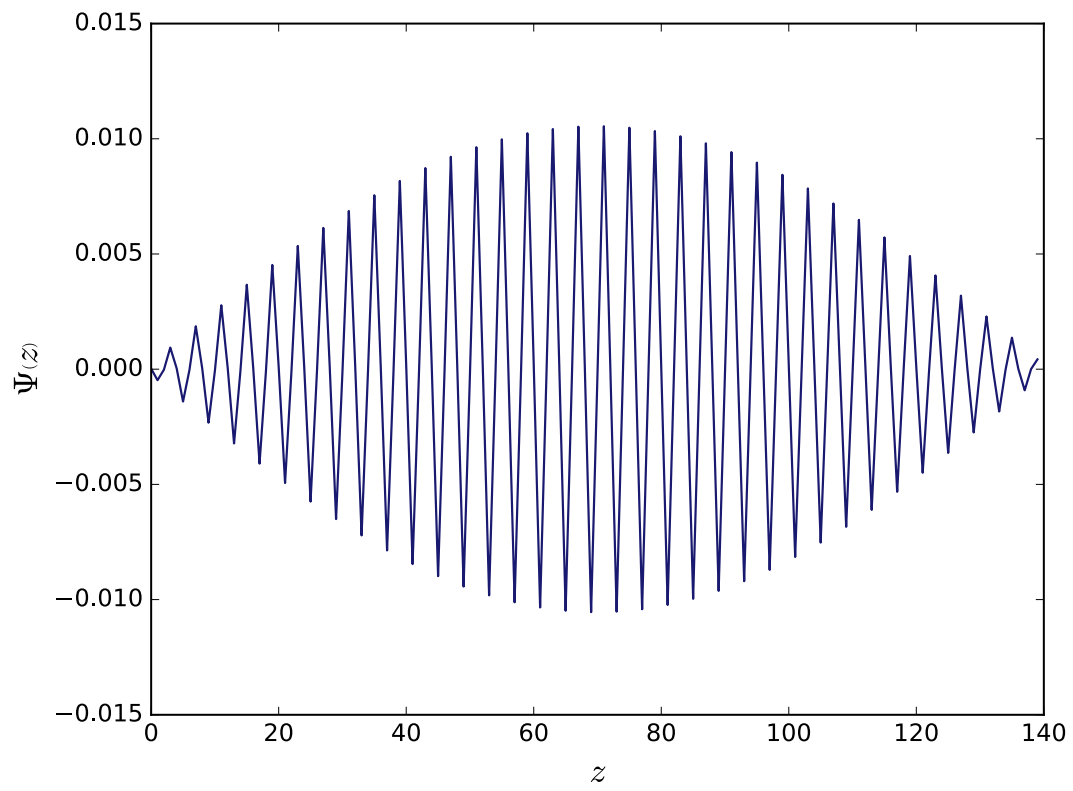

Figure 5.4: Wavefunction of the bulk state of a weak TKI with $N_z = 140$ along the $z$-axis corresponding to the energy eigenvalue $E_n/D_0 = -0.379$. The constant mean-field parameters are $(b, \lambda/D_0, \eta/D_0) = (0.463, 0.162, 0.127)$, and the parameters chosen are $V_0/D_0 = 0.1$, $E_0^f/D_0 = -0.2$, $t^c/D_0 = 0.2$, $t^f/D_0 = -0.001$ and $\mu/D_0 = 0.063$.

The results show that edge states are well-localized in the boundaries, relatively last 14% of the lattice sites, and vanishing in the bulk.

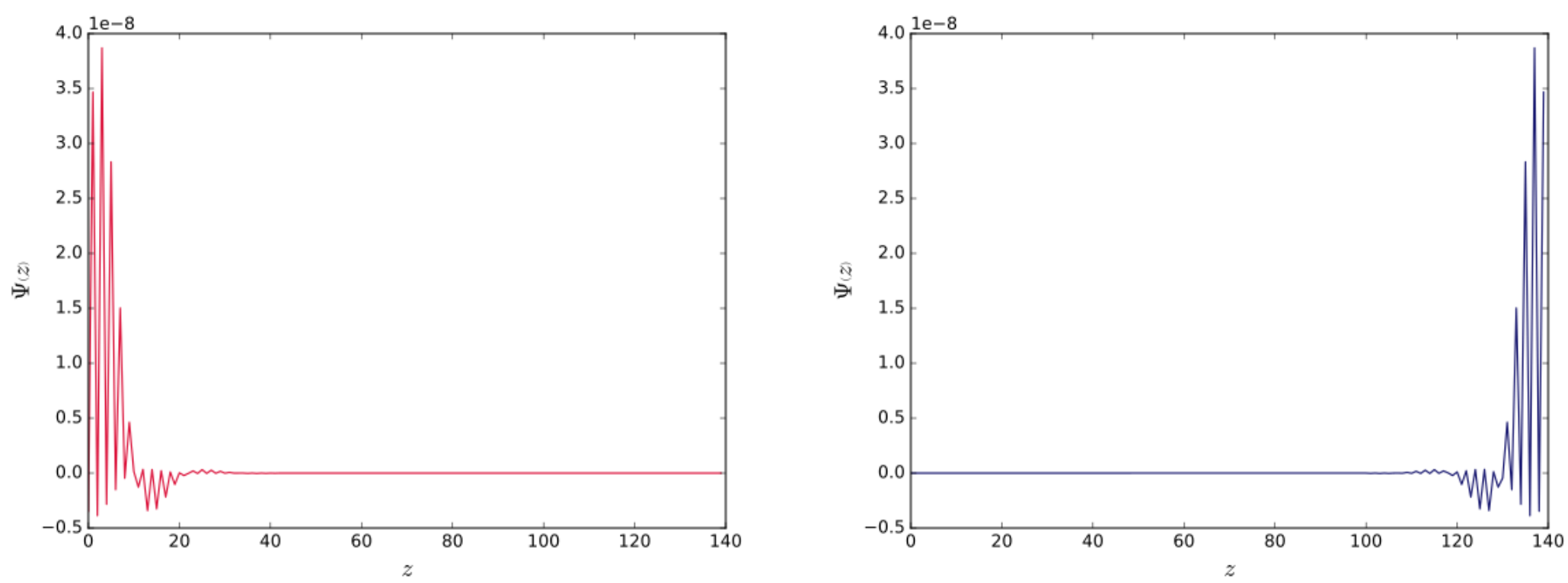

Figure 5.5: Wavefunction of the edge states of a weak TKI with $N_z = 140$ along the $z$-axis corresponding to the energy eigenvalue $E_n/D_0 = 0.066$. The constant mean-field parameters are $(b, \lambda/D_0, \eta/D_0) = (0.463, 0.162, 0.127)$, and the parameters chosen are $V_0/D_0 = 0.1$, $E_0^f/D_0 = -0.2$, $t^c/D_0 = 0.2$, $t^f/D_0 = -0.001$ and $\mu/D_0 = 0.063$.

### 5.4.2 Solution of the inhomogeneous slave-boson mean-field equations

#### Layer dependence of the mean-field parameters

The solutions of the inhomogeneous slave-boson mean-field equations show that the mean-field parameters $(b, \lambda)$ change layer by layer. They are almost constant at the bulk, with very small fluctuations, and a rise at the boundaries.

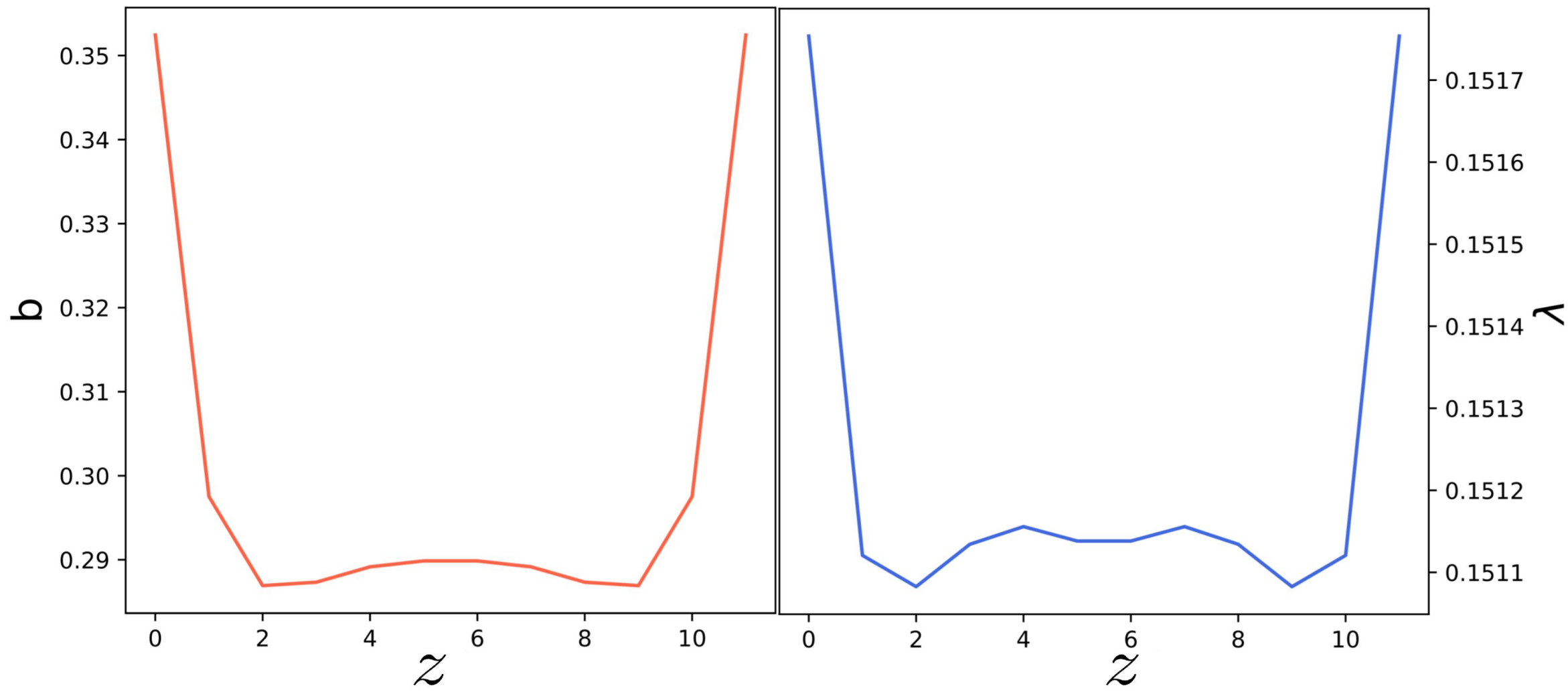

Figure 5.6: Layer dependence of the mean-field parameters $(b, \lambda)$ and constant $\eta/D_0 = 0.132$ at $T/D_0 = 10^{-4}$ for a TKI in a slab geometry with $N_z = 12$ planes in the $z$-axis. The parameters chosen are $V_0/D_0 = 0.1$, $E_0^f/D_0 = -0.2$, $t^c/D_0 = 0.1$, $t^f/D_0 = -0.001$, $\mu/D_0 = 0$ and $\delta = 10^{-2}$.

According to these results, due to the increase of the slave-boson $b$ at the boundaries, hybridization amplitude $(bV_0)$ and $f$ electron's hopping $(b^2 t^f)$ becomes stronger at the surface. There are also other works with different approaches related to the 2D and 3D TKI with a slab geometry performed by Thompson & Sachdev [44], Werner & Assad [45] and Alexandrov et al. [46].

#### Multilayer spectral functions with layer-dependent mean-field parameters

Single particle spectral function of the multilayer system can be obtained from equation (5.38). The result for the total spectral function of the conduction electrons of a weak TKI, shows a very narrow hybridization gap with smeared edge states in the Fermi level, at the border of Brillouin zone. On the other hand, the total spectral function of the $f$ electrons shows that they are mainly localized in the center of the upper band with brighter edge states at the borders. This means that $f$ electrons have more contribution to the edge states than the conduction electrons. To show this, we also computed the $f$ electron's edge state spectral function.

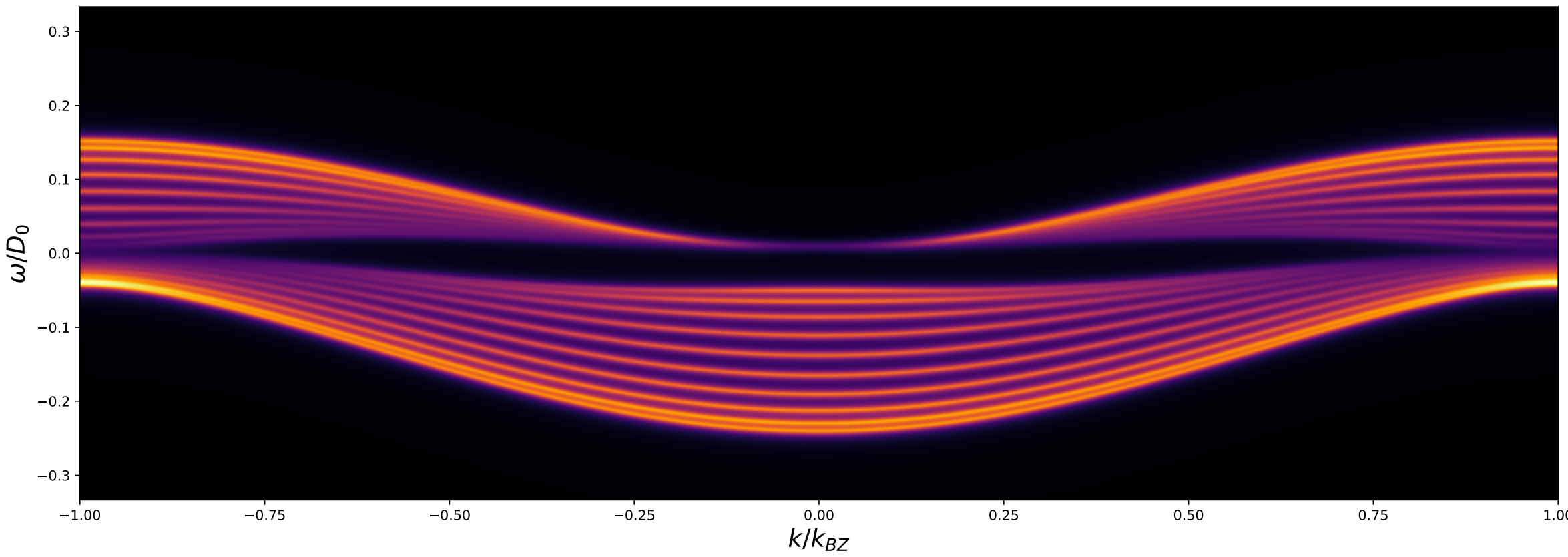

Figure 5.7: Intensity plot of the total spectral function of the conduction electrons at $T/D_0 = 10^{-4}$ for a TKI with layer-dependent mean-field parameters $(b, \lambda/D_0)$ and constant $\eta/D_0 = 0.132$, including $N_z = 12$ planes in the $z$-axis. The parameters chosen are $V_0/D_0 = 0.1$, $E_0^f/D_0 = -0.2$, $t^c/D_0 = 0.1$, $t^f/D_0 = -0.001$, $\mu/D_0 = 0$ and $\delta = 10^{-2}$.

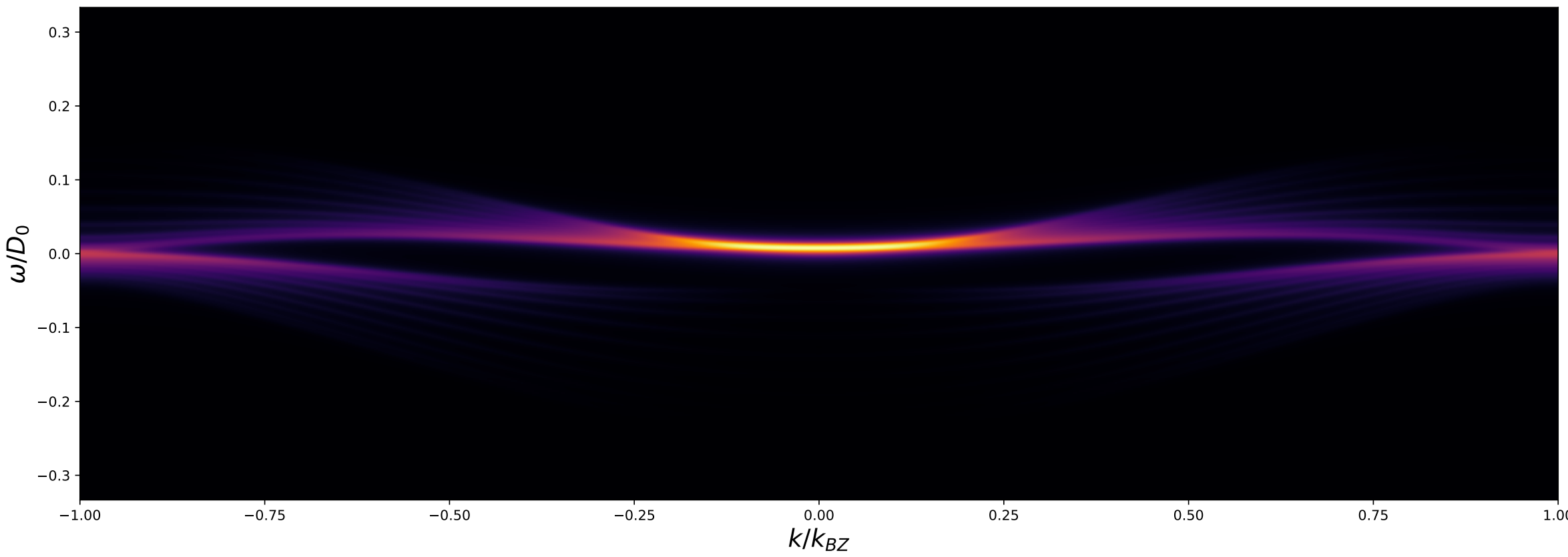

Figure 5.8: Intensity plot of the total spectral function of the $f$ electrons at $T/D_0 = 10^{-4}$ for a TKI with layer-dependent mean-field parameters $(b, \lambda/D_0)$ and constant $\eta/D_0 = 0.132$, including $N_z = 12$ planes in the $z$-axis. The parameters chosen are $V_0/D_0 = 0.1$, $E_0^f/D_0 = -0.2$, $t^c/D_0 = 0.1$, $t^f/D_0 = -0.001$, $\mu/D_0 = 0$ and $\delta = 10^{-2}$.

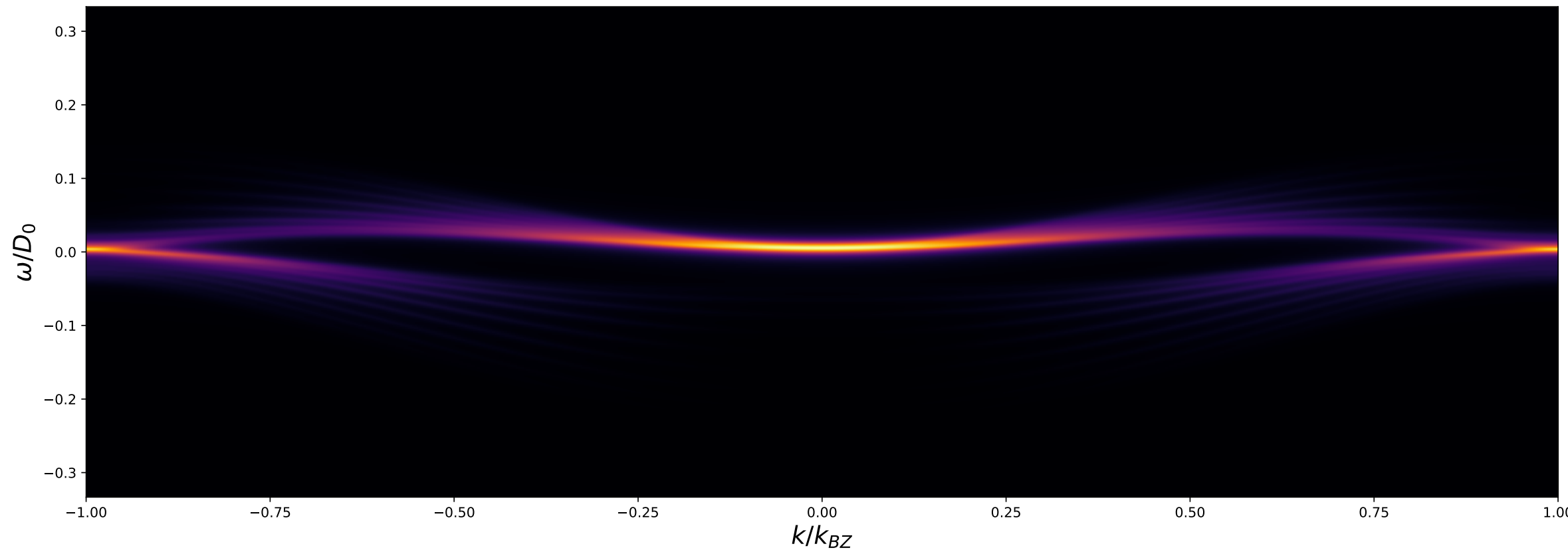

Figure 5.9: Intensity plot of the $f$ electrons edge state spectral function at $T/D_0 = 10^{-4}$ for a TKI with layer-dependent mean-field parameters $(b, \lambda/D_0)$ and constant $\eta/D_0 = 0.132$, including $N_z = 12$ planes in the $z$-axis. The parameters chosen are $V_0/D_0 = 0.1$, $E_0^f/D_0 = -0.2$, $t^c/D_0 = 0.1$, $t^f/D_0 = -0.001$, $\mu/D_0 = 0$ and $\delta = 10^{-2}$.

## CHAPTER 6

# Conclusion and outlook

In this work, we developed a self-consistent theory to study a 3D topological Kondo insulators in the bulk and the slab geometry, at the mean-field level. At the first step, following the works of Maxim Dzero and Piers Coleman [9, 11], spin-orbit coupling and crystal field effect were included in the infinite-$U$ Anderson lattice model. This led to a $k$-dependent, nonlocal, odd parity, time-reversal invariant hybridization form factor between $c$ and $f$ electrons, which have opposite parities. In comparison with the original infinite-$U$ Anderson lattice model, which has $k$-independent hybridization gap that cannot close, such a spin-orbit coupled form factor allows closing of the gap near the chemical potential.

At the second step, to develope a self-consistent thoery for the 3D bulk of the TKI, which has periodic boundary conditions in all axis, a homogeneous slave-boson mean-field theory was performed. This enabled us to realize a narrow hybridization gap between $c$ and $f$ bands in the bulk, while chemical potential lies inside the gap. It is worth mentioning that to find the mean-field parameters, self-consistent mean-field equations solved iteratively. These equations have a triple integral over momentum and a single integral over frequency. To solve these equations in a reasonable amount of time, we imposed an isotropy condition on the hybridization form factor. Thus, one way to improve the results of the bulk is to lift this condition, which may also increase the convergence of solutions.

Following this strategy, we developed a self-consistent theory for a 3D TKI in a slab geometry. By breaking periodic boundary condition in $z$-axis, we considered a multilayer system with layered-dependent mean-field parameters, where each layer is coupled to its nearest-neighbors by hoping and nonlocal hybridization. Self-consistency equations loop for such multilayer system grows very fast, proportional to double the number of layers, which made a slow convergence of the solutions. The results for 12 layers, while not ideal for this problem, are compatible with the results found with constant mean-field parameters. Generally, it would be interesting to explore the parameter ranges more thoroughly than done in this thesis, and for this reason a more efficient scheme for setting up computations would be valuable.

Finally, to improve and expand the presented results, finite quasiparticle life-time effects can be incorporated by taking bosonic fluctuations about the mean-field solution into account and by calculating the corresponding self-energies. This enables one to calculate characteristic, observable quantities, like the surface conductivity, for such system, including life-time effects. Moreover, one can also implement the Dynamical Mean Field Theory (DMFT), including the spin-orbit coupled hybridization form factor, which provides a more realistic picture from TKI even at high temperatures [47].

# APPENDIX A

# Simplification of the mean-field equation for the Bose amplitude $b$

To simplify the mean-field equation for $b$, one should substitute the $cf$ correlation (4.29) into equation (4.7) as

$$\frac{2}{\pi N}\sum_{\mathbf{k},\sigma,\alpha}\Phi_{\sigma\alpha}(\mathbf{k})\int_{-\infty}^{+\infty}d\omega\;\; Im\Big(G^{A}_{fc}(\mathbf{k},\omega;\alpha,\sigma)\Big)\,n_F(\omega) - \frac{2b}{\pi N}\sum_{\mathbf{k},\alpha}\xi^{f}_{\mathbf{k}}\int_{-\infty}^{+\infty}d\omega\;\; Im\Big(G^{A}_{ff}(\mathbf{k},\omega;\alpha,\alpha)\Big)\,n_F(\omega) + 2\lambda b = 0. \tag{A.1}$$

The imaginary part of the $G^{A}_{fc}$ can be calculated, using the identity

$$\frac{1}{x \pm i0^{+}} = \frac{\mathcal{P}}{x} \mp i\pi\delta(x), \tag{A.2}$$

and taking into account that the form factor matrix element $\Phi_{\sigma\alpha}(\mathbf{k})$ is real-valued for $\sigma_x$ and $\sigma_z$ and complex for $\sigma_y$, as follows

$$Im\Big(G^{A}_{fc}(\mathbf{k},\omega;\alpha,\sigma)\Big) = \begin{cases} \frac{\pi b\Phi_{\alpha\sigma}(\mathbf{k})}{\mathcal{W}_{\mathbf{k}}^{+}-\mathcal{W}_{\mathbf{k}}^{-}}\Big(\delta(\omega-\mathcal{W}_{\mathbf{k}}^{+}) - \delta(\omega-\mathcal{W}_{\mathbf{k}}^{-})\Big); & \text{for } \;\sigma_x \text{ and } \sigma_z \text{ elements} \\ \frac{(-i)b\Phi_{\alpha\sigma}(\mathbf{k})}{\mathcal{W}_{\mathbf{k}}^{+}-\mathcal{W}_{\mathbf{k}}^{-}}\Big(\frac{\mathcal{P}}{\omega-\mathcal{W}_{\mathbf{k}}^{+}} - \frac{\mathcal{P}}{\omega-\mathcal{W}_{\mathbf{k}}^{-}}\Big); & \text{for } \;\sigma_y \text{ elements} \end{cases}, \tag{A.3}$$

where the $(-i)$ factor is multiplied to enforce the imaginary operator on $\sigma_y$ matrix elements. Then the frequency integration of this, would be

$$\int_{-\infty}^{+\infty}d\omega\;\; Im\Big(G^{A}_{fc}(\mathbf{k},\omega;\alpha,\sigma)\Big)\,n_F(\omega) = \begin{cases} \frac{\pi b\Phi(\mathbf{k})}{\mathcal{W}_{\mathbf{k}}^{+}-\mathcal{W}_{\mathbf{k}}^{-}}\Big(n_F(\mathcal{W}_{\mathbf{k}}^{+}) - n_F(\mathcal{W}_{\mathbf{k}}^{-})\Big) \\ \frac{(-i)b\Phi(\mathbf{k})}{\mathcal{W}_{\mathbf{k}}^{+}-\mathcal{W}_{\mathbf{k}}^{-}}\left(\underbrace{\int_{-\infty}^{+\infty}d\omega\,\frac{n_F(\omega)}{\omega-\mathcal{W}_{\mathbf{k}}^{+}}}_{(*)} - \underbrace{\int_{-\infty}^{+\infty}d\omega\,\frac{n_F(\omega)}{\omega-\mathcal{W}_{\mathbf{k}}^{-}}}_{(**)}\right) \end{cases}. \tag{A.4}$$

The principal value integral $(*)$ and $(**)$, can be calculated with the residue theorem in complex plane.

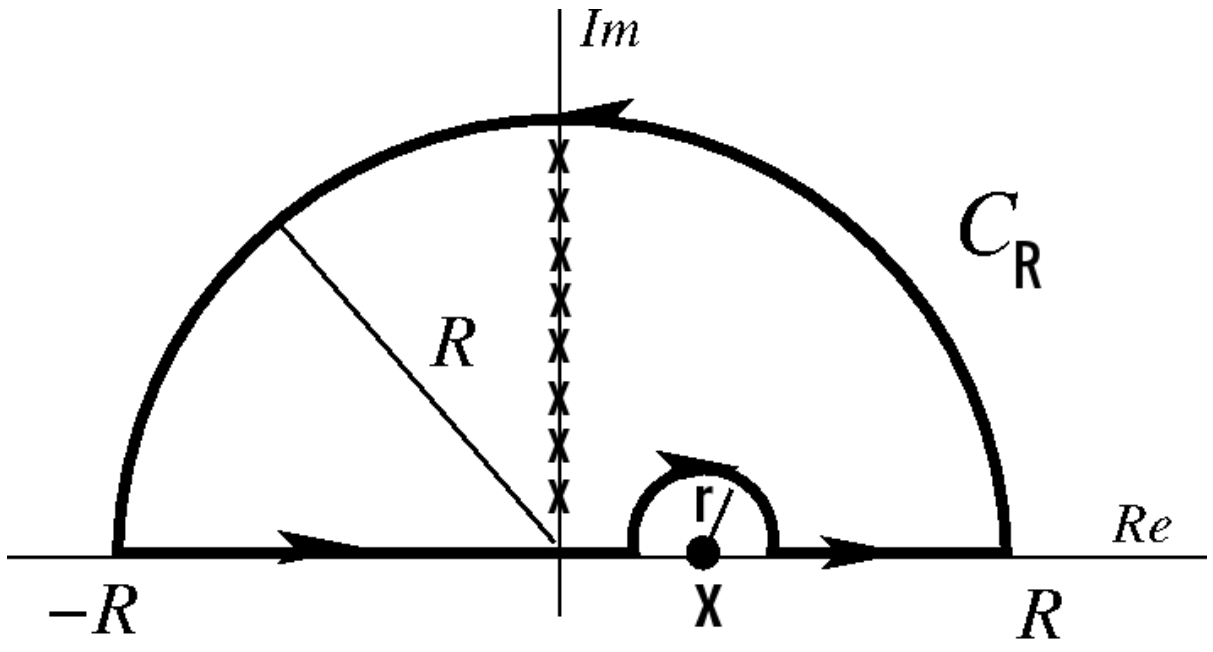

Figure A.1: Illustration of the integration contour and poles. Fermi function has infinite poles in the fermionic Matsubara frequencies, $\omega_n = \frac{\pi}{\beta}(2n+1)$ with $n > 0$, on the imaginary axis, and the function $g(z) = \frac{1}{z-x_0}$ has only one pole on the real axis, which is excluded from the contour.

$$\oint dz \underbrace{= \frac{n_F(z)}{z-x_0}}_{n_F(z)g(z)} = \underbrace{\int_{C_{R\to\infty}} dz \frac{n_F(z)}{z-x_0}}_{=0 \;\; (n_F(z)\to 0 \;\; \text{for} \;\; R\to\infty)} + \int_{-\infty}^{x_0-r} dx \frac{n_F(x)}{x-x_0} + \underbrace{\int_{C_{r\to 0}} dz \frac{n_F(z)}{z-x_0}}_{\text{shift} \;\; z\to z+x_0} + \int_{x_0+r}^{+\infty} dx \frac{n_F(x)}{x-x_0}$$

$$\to \int_{-\infty}^{+\infty} dx \frac{n_F(x)}{x-x_0} = \underbrace{\oint dz \; n_F(z)g(z)}_{(i)} - \underbrace{\int_{C_{r\to 0}} dz \frac{n_F(z+x_0)}{z}}_{(ii)} . \tag{A.5}$$

The integral $(i)$ can be calculated as

$$\oint dz \; n_F(z)g(z) = \frac{2\pi i}{\beta} \sum_{n>0} g(i\omega_n) = 2\pi i \;\; Res[g(z)]_{z=z_j} n_F(z_j) = 0 \tag{A.6}$$

where $z_j$ is the pole of the function $g(z)$, which is excluded from the contour; consequently, it is analytic in the whole contour and has no residue. [More elaborate calculation of such well-known integrals can be found in CITATION:Bruus].

The integral $(ii)$ is

$$\int_{C_{r\to 0}} dz \frac{n_F(z+x_0)}{z} = \int_{\pi}^{0} d\theta r i e^{i\theta} \frac{n_F(re^{i\theta}+x_0)}{re^{i\theta}} = -i \int_0^{\pi} d\theta \;\; n_F(re^{i\theta}+x_0) = -n_F(x_0)\pi i \qquad \text{for} \;\; r\to 0. \tag{A.7}$$

So the principal value integral (A.5) would be

$$\int_{-\infty}^{+\infty} dx \frac{n_F(x)}{x-x_0} = n_F(x_0)\pi i \tag{A.8}$$

Therefore, the integrals $(*)$ and $(**)$ become

$$\begin{cases} \int_{-\infty}^{+\infty} d\omega \frac{n_F(\omega)}{\omega - W_{\mathbf{k}}^{+}} = n_F(W_{\mathbf{k}}^{+})\pi i \\ \int_{-\infty}^{+\infty} d\omega \frac{n_F(\omega)}{\omega - W_{\mathbf{k}}^{-}} = n_F(W_{\mathbf{k}}^{-})\pi i \end{cases} \tag{A.9}$$

By substituting these results into the equation (A.4), one can find that the contribution of the integral for all cases is the same

$$\int_{-\infty}^{+\infty} d\omega \;\; Im\Big(G^{A}_{fc}(\mathbf{k},\omega;\alpha,\sigma)\Big)\, n_F(\omega) = \frac{\pi b \Phi_{\alpha\sigma}(\mathbf{k})}{\mathcal{W}_{\mathbf{k}}{}^{+} - \mathcal{W}_{\mathbf{k}}{}^{-}} \Big(n_F(\mathcal{W}_{\mathbf{k}}{}^{+}) - n_F(\mathcal{W}_{\mathbf{k}}{}^{-})\Big) \tag{A.10}$$

Now by substituting this result into the equation (A.1), one can find

$$\frac{2b}{N}\sum_{\mathbf{k}} \underbrace{\sum_{\sigma,\alpha} \Phi_{\sigma\alpha}(\mathbf{k})\Phi_{\alpha\sigma}(\mathbf{k})}_{=|S_{\mathbf{k}}|^2} \;\; \frac{n_F(\mathcal{W}_{\mathbf{k}}{}^{+}) - n_F(\mathcal{W}_{\mathbf{k}}{}^{-})}{\mathcal{W}_{\mathbf{k}}{}^{+} - \mathcal{W}_{\mathbf{k}}{}^{-}} - \frac{2b}{\pi N}\sum_{\mathbf{k},\alpha} \xi^{f}_{\mathbf{k}} \int_{-\infty}^{+\infty} d\omega \;\; Im\Big(G^{A}_{ff}(\mathbf{k},\omega;\alpha,\alpha)\Big)\, n_F(\omega) + 2\lambda b = 0.$$

$$\rightarrow \; \frac{1}{N}\sum_{\mathbf{k}} \; |S_{\mathbf{k}}|^2 \; \frac{n_F(\mathcal{W}_{\mathbf{k}}{}^{+}) - n_F(\mathcal{W}_{\mathbf{k}}{}^{-})}{\mathcal{W}_{\mathbf{k}}{}^{+} - \mathcal{W}_{\mathbf{k}}{}^{-}} - \frac{1}{\pi N}\sum_{\mathbf{k},\alpha} \xi^{f}_{\mathbf{k}} \int_{-\infty}^{+\infty} d\omega \;\; Im\Big(G^{A}_{ff}(\mathbf{k},\omega;\alpha,\alpha)\Big)\, n_F(\omega) + \lambda = 0. \tag{A.11}$$

# List of Figures

# Acknowledgements

First of all I would like to thank Prof. Johann Kroha for supervising this thesis and providing helpful guidance along the way. A special thanks go to my advanced seminar supervisor Dr. Andrea Severing who was the first one introduced me this interesting topic. I would also like to thank Prof. Simon Trebst for co-refereeing this thesis.

I am grateful of Ammar Nejati who passionately and patiently helped me and also for his friendly advices and invaluably constructive criticisms. I would like to express my gratitude to Zhong Yuan, Marvin Lenk and Francisco Meirinhos for the fruitful discussions, numerical helps and proof reading of this thesis.

Most importantly, I would like to express my wholehearted and profound thank to my parents and sister for their unconditional love and support without which this would not have been possible.